\documentclass[
aps,
pra,
twocolumn,
superscriptaddress,
floatfix,
longbibliography
]{revtex4-1}
\usepackage[dvipdfmx]{graphicx}
\usepackage{dcolumn}
\usepackage{bm}
\usepackage{amsmath}
\usepackage{amssymb}
\usepackage{color} 
\usepackage{xcolor}
\usepackage{mathrsfs} 

\usepackage{hyperref}
\hypersetup{
  colorlinks=true,
  linkcolor=[rgb]{0.00,0.60,0.0},
  citecolor=[rgb]{0.0,0.0,1.0},
  urlcolor=[rgb]{1.0,0.0,0.4},
  setpagesize=false
}

\begin{document}

\title{
  Discretized quantum adiabatic process for free fermions 
  and comparison with the imaginary-time evolution
}

\author{Tomonori Shirakawa}
\affiliation{Computational Materials Science Research Team, RIKEN Center for Computational Science (R-CCS), Kobe, Hyogo 650-0047, Japan}

\author{Kazuhiro Seki}
\affiliation{Computational Quantum Matter Research Team, RIKEN Center for Emergent Matter Science (CEMS), Wako, Saitama 351-0198, Japan}

\author{Seiji Yunoki}
\affiliation{Computational Materials Science Research Team, RIKEN Center for Computational Science (R-CCS), Kobe, Hyogo 650-0047, Japan}
\affiliation{Computational Quantum Matter Research Team, RIKEN Center for Emergent Matter Science (CEMS), Wako, Saitama 351-0198, Japan}
\affiliation{Computational Condensed Matter Physics Laboratory, RIKEN Cluster for Pioneering Research (CPR), Saitama 351-0198, Japan}

\date{\today}

\begin{abstract}
  Motivated by recent progress of quantum technologies 
  making small-scale programmable quantum computing possible, 
  here we study a discretized quantum adiabatic process 
  for a one-dimensional free-fermion system described by a variational wave function, i.e.,
  a parametrized quantum circuit. 
  The wave function is composed of $M$ layers of two elementary sets of time-evolution operators, 
  each set being decomposed into commutable local operators acting on neighboring sites. 
  The evolution time of each time-evolution operator 
  is treated as a variational parameter so as to minimize the expectation value of the energy. 
  We show that the exact ground state is reached by applying the layers of time-evolution operators
  as many as a quarter of the system size, implying that at least in this case, the state is exactly
  prepared in a quantum circuit with linear depth.  
  This is the minimum number $M_B$ of layers set by the limit of speed, i.e., the Lieb-Robinson bound, 
  for propagating quantum entanglement via the local time-evolution operators. 
  Indeed, we show the mutual information of the variational wave function that reveals a causality-cone like structure 
  in the propagation of quantum entanglement. 
  Quantities such as the energy $E$ and the entanglement entropy $S$ of
  the optimized variational wave function with the number $M$ of layers less than $M_B$
  are independent of the system size $L$ but fall into some universal functions of $M$, 
  indicating that the entanglement generated in this 
  variational ansatz with a finite $M$ is bounded, 
  as in the case of the matrix product states with a finite bond dimension.   
  Furthermore, in this case, we find that these two quantities behaves asymptotically 
  as $E/L-  \varepsilon_{\infty} \sim M^{-2}$ ($ \varepsilon_{\infty}$:
  the exact ground-state energy per site in the thermodynamic limit) 
  and $S\approx\frac{1}{3}\ln M$. 
  The development of the entanglement in the variational ansatz through the discretized 
  quantum adiabatic process is further manifested in 
  the progressive propagation of single-particle orbitals in the variational wave function.
  We also find that the optimized variational parameters 
  converge systematically to a smooth function of the discretized time, 
  which provides the optimum scheduling function in the quantum adiabatic process, with 
  the effective total evolution time of the variational ansatz to the exact ground state 
  being proportional to the system size $L$. This is a drastic improvement as compared to
  the evolution time proportional to $L^2$ for the continuous-time quantum adiabatic process with a linear scheduling,  
  and is attributed to diabaticity of the discretized quantum adiabatic process represented in the variational ansatz.
  Finally, we investigate the imaginary-time evolution counterpart of this variational wave function, 
  where the causality relation is absent due to the non-unitarity of the imaginary-time evolution operators, 
  and thus the norm of the wave function is no longer conserved. 
  We find that the convergence to the exact ground state is exponentially fast,
  despite that the system is at the critical point, 
  suggesting that implementation of the non-unitary imaginary-time evolution
  in a quantum circuit is highly promising to further shallow the circuit depth,
  provided that the local non-unitary operators are represented with a reasonable amount of unitary operators. 
\end{abstract}

\maketitle

\tableofcontents

\section{\label{sec:introduction}Introduction}

Currently realized and near-future expected quantum computing devices, 
called noisy intermediate-scale quantum (NISQ) devices~\cite{Preskill2018}, 
suffer various noise due to the poor gate fidelity and short coherent time 
so that the number of quantum gates as well as qubits reliably available
in quantum devices is severely limited. 
Therefore, it is highly desirable to find quantum algorithms working efficiently on such a limited condition.
One of the great challenges in quantum computing is to demonstrate 
quantum supremacy for practical calculations in quantum devices 
that can outperform the classical counterparts~\cite{Preskill2018,Boixo2018,Arute2019}.

Quantum simulations of quantum many-body systems, 
such as the Hubbard model and quantum chemistry systems,
have been anticipated to be the most promising application for quantum computers~\cite{Feynman1982}. 
One of the prominent algorithms specially in the NISQ era is the variational quantum 
eigensolver (VQE)~\cite{Peruzzo:2014aa,Yung2014}, 
a quantum-classical hybrid algorithm, 
in which a variational wave function describing a quantum state is represented by a quantum circuit 
composed of parametrized quantum gates. 
In the VQE, the energy and often the derivatives of the energy with respect to the variational parameters
are estimated on quantum computers and these quantities 
are used to optimize the variational wave function
by minimizing the variational energy on classical computers.

In this regard,
it has been pointed out that
the barren plateau phenomena occurs as a potentially
serious issue in the VQE method~\cite{McClean:2018aa}: 
If a random circuit is used, 
the derivatives of a VQE wave function with respect to the parameters vanish
as the number of qubits as well as quantum gates increases. 
Although this might not necessarily occur in general for any parametrized circuit, 
it is rather preferable to find a suitable circuit structure and reduce the 
number of parametrized gates necessary for representing a particular quantum state.

There have been several schemes proposed to improve and go beyond the plain VQE algorithm for 
quantum simulations of quantum many-body systems on NISQ devices. 
One of the strategies is to systematically reduce
the number of variational parameters in a parametrized quantum circuit, 
keeping the accuracy of the variational wave function. For example, 
the adaptive derivative-assembled pseudo-Trotter ansatz variational quantum eigensolver (ADAPT-VQE)~\cite{Grimsley:2019aa}  
employs the unitary coupled cluster ansatz with generalized single and double excitations from a single Slater determinant reference, 
but the parametrized quantum gates are additively selected, one at each iteration, in the collection of one- and two-body operator pools 
by searching the appropriate gate that gives the largest gradient, and hence the optimization is performed only 
for the selectively accumulated quantum gates. 
In a symmetry-adapted VQE scheme, the symmetry of the Hamiltonian is imposed in the variational wave function 
to reduced the number of parametrized gates at the expense of introducing a non-unitary projection operator 
that is treated partly as postprocessing on classical computers~\cite{Seki2020}. 
A non-orthogonal VQE scheme is a multireference version of the VQE algorithm where a generalized eigenvalue problem 
in a subspace spanned by a collection of the parametrized variational wave functions is solved, in addition to optimizing the 
variational parameters~\cite{Huggins2020}. A similar idea of expanding a quantum subspace within the VQE scheme is also 
proposed in Refs.~\cite{McClean2017,Colless2018,Ollitrault2019,Takeshita2020}.

Another strategy is to employ the imaginary-time evolution that is non-unitary. Recently, Motta {\it et al}. proposed 
a quantum imaginary-time evolution (QITE) algorithm~\cite{Motta:2020aa}, in which 
a local infinitesimally small imaginary-time evolution operator, including the normalization factor of the imaginary-time 
evolved quantum state, is mapped to a non-local unitary real-time evolution operator by solving a linear system of equations 
on classical computers to properly parametrize the non-local unitary operator that approximately reproduces the local non-unitary 
imaginary-time evolution operator~\cite{Motta:2020aa,Yeter-Aydeniz2020,Nishi2020,Gomes2020}. 
Note that the parameters in a non-local unitary operator here are determined by solving a linear 
system of equations, not by optimizing a cost function as in the VQE scheme. 
It is also interesting to note that in Ref.~\cite{Motta:2020aa} they used the QITE algorithm to expand a quantum subspace 
by constructing non-orthogonal Krylov-subspace basis states and proposed a quantum version of a Lanczos-like algorithm. 
In this regard, an imaginary-time evolution is not necessarily required to generate a Krylov subspace, as demonstrated in 
Refs.~\cite{Parrish2019,Stair2020} by using a real-time evolution.  
Very recently, a quantum version of the power method is proposed to generate a Krylov subspace~\cite{Seki2020B}.

Considering applications for the NISQ devices, it is crucial to find a way, based on some guiding principle, 
of designing a quantum circuit ansatz, ideally with linear depth or less, that can efficiently represent a quantum state of interest. 
It is easily shown mathematically that the imaginary-time evolution can 
yield the exact solution of a ground state in the limit of long-time evolution, provided that the imaginary-time 
evolution is treated exactly. 
The Lanczos method also guarantees to converge to a ground state with a desired accuracy as the dimension of a Krylov subspace is 
increased. Although these methods are well established classically, their quantum versions are still under 
development, as described above. 
There have been various circuit ansatzes proposed in the VQE scheme 
such as a unitary coupled cluster ansatz 
for mostly quantum chemistry application~\cite{Peruzzo:2014aa,Yung2014,Romero_2018,McArdle2020} 
and a hardware efficient ansatz~\cite{Kandala:2017aa}.
These ansatzes can represent any quantum state, in principle,
by increasing the number of gates and thus the circuit depth~\cite{Evangelista2019}, 
and have been implemented in the NISQ devices
with success particularly for small molecules~\cite{OMalley2016,Kandala:2017aa,Yaangchao2017,McCaskey:2019aa}.

Here, in this paper, we shall focus on a circuit ansatz realized
by discretizing a quantum adiabatic process from an initial product state to a final state 
corresponding to a ground state of a Hamiltonian to be solved~\cite{WenWeiHo,Mbeng1,Mbeng2,Wauters}. 
This is inspired by the quantum approximate optimization algorithm (QAOA) for combinatorial optimization 
problems that are represented as an Ising model~\cite{qaoa}.
In this paper, this circuit ansatz is called a discretized quantum adiabatic process (DQAP) ansatz. 
An advantage of a DQAP ansatz is the fact that a circuit constructed by the DQAP can yield the exact ground state 
without any parametrization in the continuous circuit limit because of the quantum adiabatic theorem~\cite{Ehrenfest1916,Born1928,Schwinger1937,Kato1950}.
However, the convergence of a DQAP ansatz with the finite number of discretization steps is unknown, in general,
and this is the main issue addressed in this paper.

We thereby study a DQAP ansatz for free fermions on a one-dimensional lattice at half filling.  
This is an ideal system to analyze a DQAP ansatz because a quantum state evolved by the DQAP with an initial state described by 
a single Slater determinant state can still be described by a single Slater determinant state, and therefore one can keep track of 
each occupied single-particle orbital in the Slater determinant state during the DQAP. 
In the DQAP ansatz considered here, we first prepare as the initial state a product state
of local bonding states formed on neighboring sites, which can be described by a single Slater determinant state. 
We then let the state evolve forward via the DQAP by repeatedly applying layers of two elementary sets of local time-evolution operators 
(see Fig.~\ref{fig:1d:ansatz}), 
where the evolution time in each layer of the time-evolution operators is treated as a variational parameter so as to minimize 
the variational energy.  
We examine how the state described by the DQAP ansatz 
evolves with increasing the number $M$ of layers by monitoring the variational energy, single-particle orbitals 
in the Slater determinant state, the entanglement entropy, and the mutual information. 

We find that the exact ground state is attained
by applying the layers of time-evolution operators as many as a quarter of the system 
size, which is the minimum number $M_B$ of layers necessary to 
entangle the entire system by the local time-evolution operators, corresponding to the Lieb-Robinson bound for the 
propagation of quantum information~\cite{Lieb1972}.
In contrast, the DQAP ansatz with the number $M$ of layers less than $M_B$, thus not describing the exact ground state,  
represents another series of quantum states in that physical quantities such as the energy and the entanglement entropy 
evaluated for these states are independent of the system size but scale with $M$, 
which indicates that the entanglement carried by 
the DQAP ansatz with a finite number of layers is bounded,  
as in the case of the matrix product states with a finite bond dimension~\cite{PerezGarcia2007,Verstraete2008}.
We also find that the optimized variational parameters in the DQAP ansatz converge systematically 
to a smooth function of the discretized time, 
which thus provides the optimized scheduling function for the quantum adiabatic process.
Furthermore, we show that the effective total evolution time of the optimized DQAP ansatz with $M_B$ layers of the local 
time-evolution operators, thus representing the exact ground state, is proportional to the system size $L$. 
This is in sharp contrast to the case of the continuous-time quantum adiabatic process with a linear scheduling, where 
the total evolution time necessary to reach the exact ground state with a given accuracy is proportional to $L^2$. 
Indeed, we find that the intermediate states in the DQAP ansatz
cannot represent the ground state of the instantaneous Hamiltonian,
suggesting that diabaticity of the transition processes in the DQAP ansatz
is essential for the quadratic speedup of the total evolution time. 
For comparison, we also investigate the imaginary-time evolution of the DQAP ansatz, which can still be described by 
a single Slater determinant state. 
We find that the convergence to the ground state is exponentially fast with respect to the number of layers of 
the local imaginary-time evolution operators~\cite{Zanca2017}, despite that the system is at the critical point where the one-particle density matrix 
decays algebraically with distance.

The rest of this paper is organized as follows.
We first describe the free-fermion model and establish the notation used throughout this paper in Sec.~\ref{sec:model}, 
and introduce the DQAP ansatz in Sec.~\ref{sec:dqap}.
We then provide the analytical formulas for various quantities such as the one-particle density matrix, entanglement entropy, 
and mutual information, and also explain the optimization method 
in Secs.~\ref{sec:tool}-- \ref{sec:ent}. 
The numerical results for the DQAP ansatz are given in Sec.~\ref{sec:result} 
and these results are compared with those for the imaginary-time counterpart in Sec.~\ref{sec:res:imag}. 
We then conclude the paper with a brief discussion in Sec.~\ref{sec:summary}.
To make our paper self-contained,
the details of the derivation for the free-fermion formulas are provided in Appendix~\ref{sec:ff}.
The numerical details of the optimized parameters are discussed in Appendix~\ref{sec:random} 
and details of the entanglement entropy with respect to the one-particle density matrix 
are explained in Appendix~\ref{sec:separation}.
The continuous-time quantum adiabatic process
with a linear scheduling is analyzed in Appendix~\ref{sec:conttime}
and a geometrically optimal scheduling derived by the quantum adiabatic brachistochrone (QAB)
is discussed in Appendix~\ref{sec:qab}.
Throughout the paper, we set the reduced Planck's constant $\hbar=1$.

\section{\label{sec:method}Model and method}

In this section, we first define the free-fermion model with matrix notation in Sec.~\ref{sec:model}, 
and introduce a DQAP to construct a variational ansatz in Sec.~\ref{sec:dqap}. 
We then summarize the analytical formulas for the variational ansatz in the free-fermion case in Sec.~\ref{sec:tool}. 
The optimization method to optimize the variational parameters is described in Sec.~\ref{sec:opt}.  
To discuss the entanglement property, 
we also derive the analytical formulas of the reduced density matrix, 
entanglement entropy, and mutual information for a free-fermion wave function 
in Sec.~\ref{sec:ent}.

\subsection{\label{sec:model}Model}

The free-fermion system considered in this paper is described by the following Hamiltonian:  
\begin{equation}
  \hat{\mathcal{H}} = \sum_{x=1}^L \sum_{x^{\prime}=1}^{L}
  t_{xx^{\prime}} \hat{c}_{x}^{\dagger} \hat{c}_{x^{\prime}}, 
  \label{eq:ham}
\end{equation}
where $\hat{c}_x$ ($\hat{c}_x^{\dagger}$) denotes the annihilation (creation) operator
of a fermion at site $x\in \{ 1, 2, \cdots, L \}$. 
For convenience, we represent the Hamiltonian in Eq.~(\ref{eq:ham}) as 
\begin{equation}
  \hat{\mathcal{H}} = \hat{\bm c}^{\dagger} {\bm T} \hat{\bm c},
  \label{eq:ham:quadform}
\end{equation}
where $\hat{\bm c}^{\dagger}$ ($\hat{\bm c}$) is an $L$-dimensional row (column) vector 
of the fermion operators given by
\begin{equation}
  \hat{\bm c}^{\dagger} =
  (
  \hat{c}_1^{\dagger}\ \hat{c}_2^{\dagger}\ \cdots\ \hat{c}_L^{\dagger}
  ), \ 
  \hat{\bm c} =
  \left(
  \begin{array}{c}
    \hat{c}_1 \\
    \hat{c}_2 \\
    \vdots \\
    \hat{c}_L \\
  \end{array}
  \right),
  \label{eq:vec:c}
\end{equation}
and ${\bm T}$ is an $L \times L$ matrix
whose elements are given by $\left[ {\bm T} \right]_{xx^{\prime}} = t_{xx^{\prime}}$.

Let ${\bm U}$ be an $L\times L$ unitary matrix that diagonalizes the matrix ${\bm T}$ as
\begin{equation}
  {\bm U}^{\dagger} {\bm T} {\bm U} = {\bm E},
\end{equation}
where ${\bm E}$ is the $L\times L$ diagonal matrix whose diagonal elements
are the eigenvalues of ${\bm T}$: ${\bm E} = {\rm diag}( E_1, E_2, \cdots, E_L )$.
Here, we assume $E_n$ ($n=1,2,\cdots,L$) in ascending order, i.e.,  
$E_1 \leq E_2 \leq \cdots \leq E_L$.
Using the unitary matrix ${\bm U}$, one can define the new fermion operators
\begin{equation}
  \hat{\bm a}^{\dagger} = ( \hat{a}_1^{\dagger}\ \hat{a}_2^{\dagger}\ \cdots \hat{a}_L^{\dagger} ),\
  \hat{\bm a} = \left(
  \begin{array}{c}
    \hat{a}_1 \\
    \hat{a}_2 \\
    \vdots \\
    \hat{a}_L \\
  \end{array}
  \right)
\end{equation}
given by
\begin{equation}
  \hat{\bm a}^{\dagger} = \hat{\bm c}^{\dagger} {\bm U}, \ 
  \hat{\bm a} = {\bm U}^{\dagger} \hat{\bm c},
\end{equation}
and the Hamiltonian $\hat{\mathcal{H}}$
in Eq.~(\ref{eq:ham}) is represented as
\begin{equation}
  \hat{\mathcal{H}} = \hat{\bm a}^{\dagger} {\bm E} \hat{\bm a} = \sum_{n=1}^L E_n \hat{a}_n^{\dagger} \hat{a}_n.
  \label{eq:ham:mod}
\end{equation}
The ground state of the Hamiltonian $\hat{\mathcal{H}}$ with $N$ fermions is 
a state with the lowest $N$ energy levels in Eq.~(\ref{eq:ham:mod}) being occupied, i.e.,
\begin{equation}
  \vert \psi \rangle = \prod_{n=1}^N \hat{a}_n^{\dagger} \vert 0 \rangle,
  \label{eq:gs:a}
  \end{equation}
where $\vert 0\rangle$ is the vacuum of fermions.
Using the original fermion operator $\hat{\bm c}^{\dagger}$,
the ground state in Eq.~(\ref{eq:gs:a}) is now expressed as
\begin{equation}
  \vert \psi \rangle = \prod_{n=1}^N [ \hat{\bm c}^{\dagger} \mbox{\boldmath{$\Psi$}} ]_{n} \vert 0 \rangle
  \label{eq:psi}
\end{equation}
where $\mbox{\boldmath{$\Psi$}}$ is
an $L \times N$ matrix obtained by extracting the first $N$ columns from ${\bm U}$.
$[ \hat{\bm c}^{\dagger} \mbox{\boldmath{$\Psi$}} ]_n$
indicates the $n$th element of the $N$ dimensional row vector $\hat{\bm c}^{\dagger} \mbox{\boldmath{$\Psi$}}$.

It is important to note that 
the row index $x$ of $[\mbox{\boldmath{$\Psi$}}]_{xn}$ indicates
the site index while the column index $n$ is the index labeling the
single-particle state with the single-particle energy $E_n$ obtained
by diagonalizing the matrix ${\bm T}$. 
$[ \hat{\bm c}^{\dagger} \mbox{\boldmath{$\Psi$}} ]_n$ in Eq.~(\ref{eq:psi}) thus 
corresponds to the $n$th single-particle orbital that composes the Slater determinant state of the ground state $\vert \psi \rangle$. 
Hereafter, we simply refer to the column vectors of $\mbox{\boldmath{$\Psi$}}$ as single-particle orbitals.

\subsection{\label{sec:dqap}Variational ansatz based on a discretized quantum adiabatic process}

The quantum adiabatic process is a quantum process following the quantum adiabatic 
theorem~\cite{Ehrenfest1916,Born1928,Schwinger1937,Kato1950},  
in which a slowly driving system in time from an initial eigenstate, the ground state of Hamiltonian $\hat{\mathcal{H}}_{\rm i}$, 
stays in the instantaneous eigenstate of the time evolving Hamiltonian $\hat{\mathcal{H}}(\tau)$ at time $\tau$ 
and finally reaches to the ground state of Hamiltonian $\hat{\mathcal{H}}_{\rm f}$. 
Here the time evolving Hamiltonian $\hat{\mathcal{H}}(\tau)$ is expressed as 
\begin{equation}
  \hat{\mathcal{H}}(\tau) = s_{\rm i}(\tau) \hat{\mathcal{H}}_{\rm i} + s_{\rm f} (\tau) \hat{\mathcal{H}}_{\rm f},
  \label{eq:hqa}
\end{equation}
with $s_{\rm i}(\tau)$ and $s_{\rm f}(\tau)$ being the scheduling functions
that are smooth and satisfy the conditions: 
\begin{equation}
\begin{split}
&s_{\rm i}(\tau_{\rm i}) = s_{\rm f}(\tau_{\rm f}) = 1, \\
&s_{\rm i}(\tau_{\rm f}) =s_{\rm f}(\tau_{\rm i}) =0,
\end{split}
\label{eq:cond_qap}
\end{equation}
where $\tau_{\rm i}$ and $\tau_{\rm f}$
denote the initial and final times of the process, respectively. 
The final state  $\vert \psi (\tau_{\rm f}) \rangle$ at $\tau=\tau_{\rm f}$ after the time evolution is thus given as
\begin{equation}
  \vert \psi (\tau_{\rm f}) \rangle = \hat{\mathcal{U}}(\tau_{\rm f},\tau_{\rm i}) \vert \psi_{\rm i} \rangle,
  \label{eq:qap}
\end{equation}
where $\hat{\mathcal{U}}(\tau,\tau_{\rm i})$ is
the time-evolution operator obtained by solving the Schr\"odinger's equation
\begin{equation}
  {\rm i}\frac{\partial}{\partial \tau} \hat{\mathcal{U}} (\tau,\tau_{\rm i})=
  \hat{\mathcal{H}}(\tau) \hat{\mathcal{U}}(\tau,\tau_{\rm i}),
  \label{eq:ut}
\end{equation}
with $\hat{\mathcal{U}}(\tau_{\rm i},\tau_{\rm i}) = 1$ and
$\vert \psi_{\rm i} \rangle$ is the ground state of $\hat{\mathcal{H}}_{\rm i}$.

It is well known that for a sufficiently long time $\tau_{\rm f}-\tau_{\rm i}$,
implying a slow driving dynamics, 
the initial state $ \vert \psi (\tau_{\rm i}) \rangle=\vert \psi_{\rm i} \rangle$ is adiabatically transformed
into the ground state of the final Hamiltonian $\hat{\mathcal{H}}_{\rm f}$
through this adiabatic process
if there is a finite energy gap between the ground state and the excited states
of $\hat{\mathcal{H}}(\tau)$ for all $\tau$~\cite{Sarandy:2004aa,Albash2018}. 
A quantum adiabatic process is a real-time dynamics governed by 
the unitary time-evolution operator in Eq.~(\ref{eq:qap}).
This should be contrasted with the case of the imaginary-time evolution
where the imaginary-time evolution operator is no longer unitary. 
Since all operations in quantum computers are composed of unitary gate operations, 
a quantum adiabatic process would be a natural principle to follow in constructing a circuit ansatz 
for obtaining a ground state of a Hamiltonian in a quantum circuit.

We shall now consider the case where the final Hamiltonian 
$\hat{\mathcal{H}}_{\rm f}$ in Eq.~(\ref{eq:hqa}) is composed of a 
set of terms $\hat{\mathcal{V}}_p$ ($p=1,2,\cdots,P$),
\begin{equation}
  \hat{\mathcal{H}}_{\rm f} = \sum_{p=1}^P \hat{\mathcal{V}}_p,
   \label{eq:hp}  
\end{equation}
such that, in general,
$[ \hat{\mathcal{V}}_p,  \hat{\mathcal{V}}_{p'}]\ne0$ when $p\ne p'$. 
Here, we assume that each $\hat{\mathcal{V}}_p$ consists of a set of operators
\begin{equation}
\hat{\mathcal{V}}_p = \sum_{q=1}^{Q_p} \hat{\mathcal{O}}_q^{(p)}, \label{eq:oq}
\end{equation}
where all operators $\hat{\mathcal{O}}_q^{(p)}$ commute with each other for given $p$: 
\begin{equation}
  [ \hat{\mathcal{O}}_{q}^{(p)}, \hat{\mathcal{O}}_{q^{\prime}}^{(p)} ] = 0.
\end{equation}
In addition, we assume that the initial Hamiltonian
$\hat{\mathcal{H}}_{\rm i}$ in Eq.~(\ref{eq:hqa}) is given by one of $\hat{\mathcal{V}}_p$'s in $\hat{\mathcal{H}}_{\rm f}$ 
and here we consider 
\begin{equation}
  \hat{\mathcal{H}}_{\rm i} = \hat{\mathcal{V}}_{1}.
\end{equation}

Then, the time-evolution operator is written in the following form:
\begin{equation}
  \begin{split}
  \hat{\mathcal{U}}(\tau_{\rm f},\tau_{\rm i}) 
  = & T_{\tau} {\rm e}^{-{\rm i}\int_{\tau_{\rm i}}^{\tau_{\rm f}}\hat{\mathcal H}(\tau) d\tau} \\
  = & \lim_{M \to \infty}
  \prod_{m=M}^1 \hat{\mathcal{U}}_{\rm d} (\mbox{\boldmath{$\theta$}}_m) \\
  = & \lim_{M \to \infty} \hat{\mathcal{U}}_{\rm d} (\mbox{\boldmath{$\theta$}}_M) \hat{\mathcal{U}}_{\rm d}(\mbox{\boldmath{$\theta$}}_{M-1}) \cdots \hat{\mathcal{U}}_{\rm d}(\mbox{\boldmath{$\theta$}}_1),
  \end{split}
\end{equation}
with
\begin{equation}
  \begin{split}
    \hat{\mathcal{U}}_{\rm d}(\mbox{\boldmath{$\theta$}}_m) = & \prod_{p=1}^{P} {\rm e}^{-{\rm i}\theta_p^{(m)} \hat{\mathcal{V}}_p} \\
    = & \ {\rm e}^{-{\rm i}\theta_1^{(m)} \hat{\mathcal{V}}_1}
    {\rm e}^{-{\rm i}\theta_2^{(m)} \hat{\mathcal{V}}_2} \cdots
    {\rm e}^{-{\rm i}\theta_P^{(m)} \hat{\mathcal{V}}_P},
  \end{split}
\end{equation}
where 
$\mbox{\boldmath{$\theta$}}_m = \{ \theta_p^{(m)} \}_{p=1}^P$
should be chosen as  
\begin{equation}
  \theta_p^{(m)} =
  \left\{
  \begin{array}{ll}
    \left[ s_{\rm i}(\tau_m) + s_{\rm f}(\tau_m) \right] \delta \tau & \text{for }p=1 \\
    s_{\rm f}(\tau_m) \delta \tau & \text{for }p\neq 1, \\
  \end{array}
  \right.
\end{equation}
with
\begin{align}
  \delta \tau & = (\tau_{\rm f}-\tau_{\rm i})/M, \\
  \tau_m & = \tau_{\rm i} + m \delta \tau,
\end{align}
to reproduce $\hat{\mathcal{U}}(\tau_{\rm f},\tau_{\rm i})$ in the limit of $M\to\infty$. 
This is the most naive discretization procedure of time in the time-evolution operator $\hat{\mathcal{U}}(\tau_{\rm f},\tau_{\rm i}) $ 
and the ground state of $\hat{\mathcal{H}}_{\rm f}$ is obtained by operating $\hat{\mathcal{U}}(\tau_{\rm f},\tau_{\rm i})$ 
to the initial state $\vert \psi_{\rm i} \rangle$ as in Eq.~(\ref{eq:qap}).

The simplest scheduling functions $s_{\rm i}(\tau)$ and $s_{\rm f}(\tau)$ that satisfy the conditions given in Eqs.~(\ref{eq:cond_qap}) 
are a linear scheduling, i.e.,  
\begin{equation}
\begin{split}
&s_{\rm i} (\tau) =  1 - \frac{\tau-\tau_{\rm i}}{T} , \\
&s_{\rm f} (\tau) = \frac{\tau-\tau_{\rm i}}{T},
\end{split}
\label{eq:lin_sche}
\end{equation}
where $T=\tau_{\rm f} - \tau_{\rm i }$~\cite{Albash2018,Kadowaki1998}. In this case, 
\begin{equation}
  \theta_p^{(m)} =
  \left\{
  \begin{array}{ll}
     \delta \tau & \text{for }p=1 \\
    \frac{m}{M} \delta \tau & \text{for }p\neq 1. \\
  \end{array}
  \right. 
  \label{eq:theta_lin}
\end{equation}

Inspired by the quantum adiabatic process described above, here we instead consider, 
as a variational ansatz for the ground state of $\hat{\mathcal{H}}_{\rm f}$, 
the following state with a finite value of $M$:  
\begin{equation}
  \begin{split}
  \vert \psi_M (\mbox{\boldmath{$\theta$}}) \rangle = & \prod_{m=M}^1 \hat{\mathcal{U}}_{\rm d}(\mbox{\boldmath{$\theta$}}_m) \vert \psi_{\rm i} \rangle
  \\
  = & \ 
  \hat{\mathcal{U}}_{\rm d}(\mbox{\boldmath{$\theta$}}_M)
  \hat{\mathcal{U}}_{\rm d}(\mbox{\boldmath{$\theta$}}_{M-1})
  \cdots
  \hat{\mathcal{U}}_{\rm d}(\mbox{\boldmath{$\theta$}}_{1})  
  \vert \psi_{\rm i} \rangle,
  \end{split}
  \label{eq:dqap}
\end{equation}
where $\mbox{\boldmath{$\theta$}} = \{ \mbox{\boldmath{$\theta$}}_m \}_{m=1}^M$
are assumed to be variational parameters determined by minimizing the variational energy. 
This variational state $\vert \psi_M (\mbox{\boldmath{$\theta$}}) \rangle$
in Eq.~(\ref{eq:dqap}) is referred to as a DQAP ansatz.

Let us now illustrate the DQAP ansatz for the free-fermion system 
given in Eq.~(\ref{eq:ham}).
For simplicity, we assume that the system is one dimensional 
and the final Hamiltonian $\hat{\mathcal{H}}_{\rm f}$ is given by 
\begin{equation}
    \hat{\mathcal{H}}_{\rm f} =  - t \sum_{x=1}^{L-1} ( \hat{c}_{x+1}^{\dagger} \hat{c}_x + \hat{c}_{x}^{\dagger} \hat{c}_{x+1} ) 
     - t \gamma ( \hat{c}_{1}^{\dagger} \hat{c}_L + \hat{c}_{L}^{\dagger} \hat{c}_1 ),
  \label{eq:1d:ham}
\end{equation}
where
$\gamma$ sets the boundary conditions: 
$\gamma = 1$ for the periodic boundary conditions (PBCs) and
$\gamma = -1$ for the anti-periodic boundary conditions (APBCs).
We also assume that the number $L$ of sites is even and the number $N$ of fermions is at half filling, i.e., $N=L/2$.
In what follows, we set $t = 1$ as a unit of the energy.

\begin{figure}
  \includegraphics[width=\hsize]{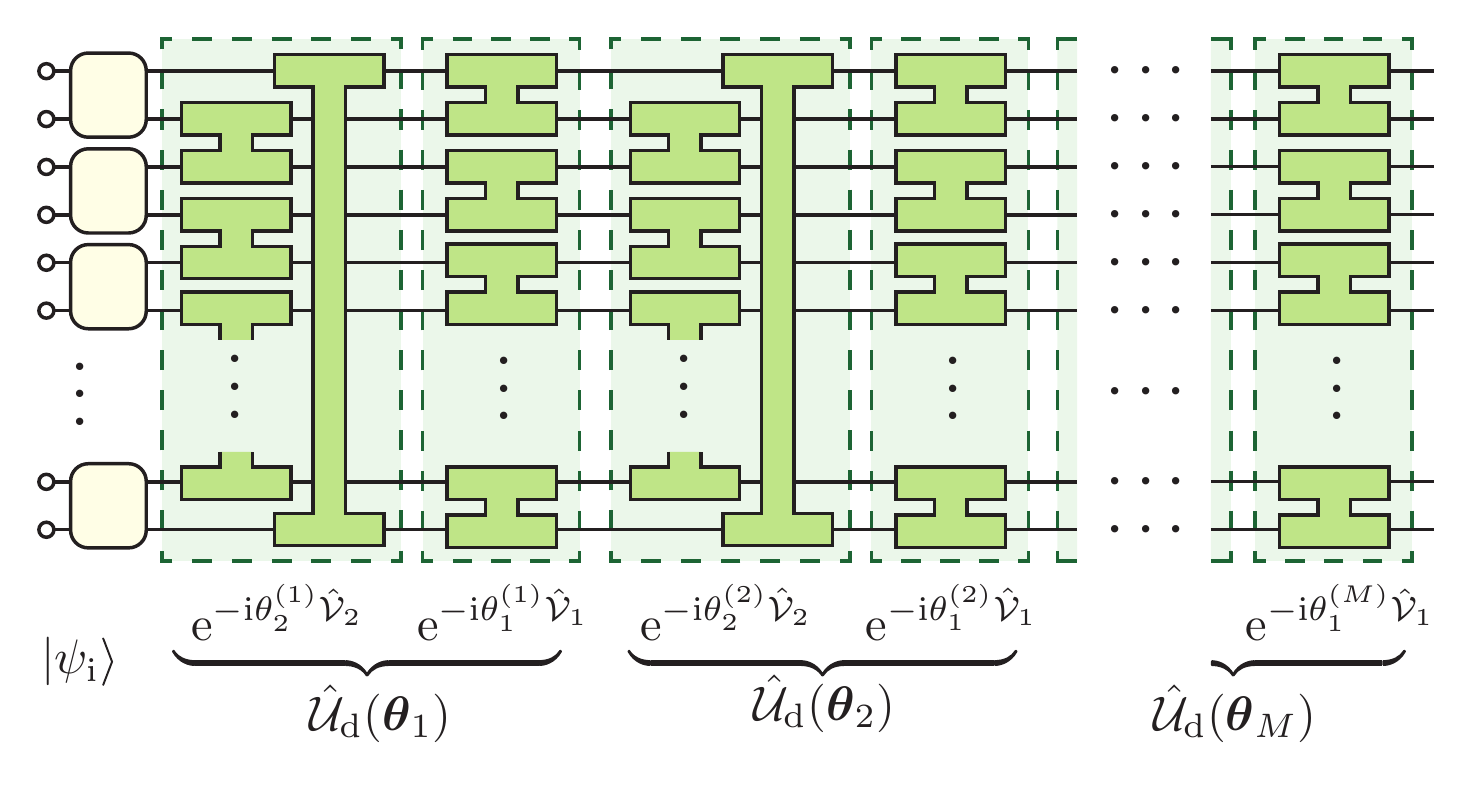}
  \caption{
  Schematic representation of the discretized quantum adiabatic process (DQAP) ansatz 
  $\vert \psi_M (\mbox{\boldmath{$\theta$}}) \rangle$, defined in Eq.~(\ref{eq:1d:dqap}), 
    for the one-dimensional free-fermion system.  
    Horizontal black solid lines indicate the local quantum states at sites $x $ (i.e., qubits).
    Green blocks denote the local time-evolution operators of the form 
    $\exp [{\rm i}\theta t (  \hat{c}_{x}^{\dagger} \hat{c}_{x^{\prime}} + \hat{c}_{x^{\prime}}^{\dagger} \hat{c}_{x} )]$.
    The boundary terms (blocks connecting the top and bottom black solid lines)
    involve generally non-local operations in the qubit representation.  
    However, they can be eliminated for the special cases discussed in the text. 
    The initial state $\vert \psi_{\rm i} \rangle$ is given in Eq.~(\ref{eq:1d:init}). 
    Light yellow squares indicate the local bonding states
    $(\hat{c}_{2x-1}^{\dagger} + \hat{c}_{2x}^{\dagger}) \vert 0 \rangle /\sqrt{2}$.
    }
  \label{fig:1d:ansatz}
\end{figure}

For this system, $\hat{\mathcal{V}}_p$ ($p=1,2$) is given by 
\begin{equation}
  \hat{\mathcal{V}}_{1} = - t \sum_{x=1}^{L/2} ( \hat{c}_{2x}^{\dagger} \hat{c}_{2x-1} + \hat{c}_{2x-1}^{\dagger} \hat{c}_{2x})
  \label{eq:1d:v:bond1}
\end{equation}
and 
\begin{equation}
  \begin{split}
    \hat{\mathcal{V}}_{2} = & - t \sum_{x=1}^{L/2-1} ( \hat{c}_{2x+1}^{\dagger} \hat{c}_{2x} + \hat{c}_{2x}^{\dagger} \hat{c}_{2x+1}) \\
    & - t \gamma (\hat{c}_{1}^{\dagger} \hat{c}_{L} + \hat{c}_{L}^{\dagger} \hat{c}_1).
  \end{split}
  \label{eq:1d:v:bond2}
\end{equation}
The initial state $\vert \psi_{\rm i} \rangle$ is the ground state of $\hat{\mathcal{V}}_{1}$
given by
\begin{equation}
  \vert \psi_{\rm i} \rangle = \prod_{x=1}^{L/2} \frac{1}{\sqrt{2}}
  (\hat{c}_{2x-1}^{\dagger} + \hat{c}_{2x}^{\dagger})  \vert 0 \rangle.
  \label{eq:1d:init}
\end{equation}
Here, the state
\begin{equation}
  \frac{1}{\sqrt{2}}( \hat{c}_{2x-1}^{\dagger} + \hat{c}_{2x}^{\dagger} ) \vert 0 \rangle
  \label{eq:local:bonding}
\end{equation}
is the local bonding state formed between sites $2x-1$ and $2x$.
The form of Eq.~(\ref{eq:1d:init}) suggests that
the initial state $ \vert \psi_{\rm i} \rangle$ is a product state
of local states, as is expected from the assumption in Eq.~(\ref{eq:oq}). 
Using Eqs.~(\ref{eq:1d:v:bond1})--(\ref{eq:1d:init}),
the DQAP ansatz is written as
\begin{equation}
  \vert \psi_M (\mbox{\boldmath{$\theta$}}) \rangle = \prod_{m=M}^1
  ( {\rm e}^{-{\rm i} \theta_1^{(m)} \hat{\mathcal{V}}_1} {\rm e}^{-{\rm i} \theta_2^{(m)} \hat{\mathcal{V}}_2 } ) \vert \psi_{\rm i} \rangle.
  \label{eq:1d:dqap}
\end{equation}
The schematic representation of this state is shown in Fig.~\ref{fig:1d:ansatz}.

We shall now discuss how this DQAP ansatz can be described in the qubit representation for quantum computing. 
First of all, a fermion system can always be mapped in 
the qubit representation through, e.g, the Jordan-Wigner transformation~\cite{Jordan1928},
\begin{equation}
\begin{split}
\hat{c}_x^\dag =& \hat{\sigma}_x^+ \hat{K}(x) \\
\hat{c}_x =&  \hat{K}^\dag(x) \hat{\sigma}_x^- 
\label{eq:JW}
\end{split}
\end{equation}
where 
\begin{equation}
\hat{K}(x) = {\rm e}^{-{\rm i}\frac{\pi}{2}\sum_{x'<x}\left(\hat{\rm Z}_{x'} + 1 \right)},
\end{equation}
$\hat{\sigma}_x^\pm=(\hat{\rm X}_x \pm {\rm i} \hat{\rm Y}_x)/2$, and 
$\{\hat{\rm X}_x, \hat{\rm Y}_x, \hat{\rm Z}_x \}$ are the Pauli operators (i.e., gates) at qubit $x$. 
Notice that 
$[\hat{\sigma}_x^{\pm},\hat{K}^{(\dag)}(x)]=0$ and
from Eq.~(\ref{eq:JW}) $\hat{c}_x^\dag \hat{c}_x = \frac{1}{2}(\hat{\rm Z}_x+1)$. 
With this transformation, any local fermion operator acting up to nearest-neighbor sites 
in a one-dimensional system can be represented by Pauli operators 
without introducing the sign factors due to the Jordan-Wigner string $\hat{K}(x)$, 
suggesting that the fermion representation is trivially equivalent to the qubit representation, 
except for the boundary terms. 
Indeed, the sign factors at the boundary are also canceled if an APBC (PBC) is imposed when $N$ is even (odd).  
 
\begin{figure}
  \includegraphics[width=\hsize]{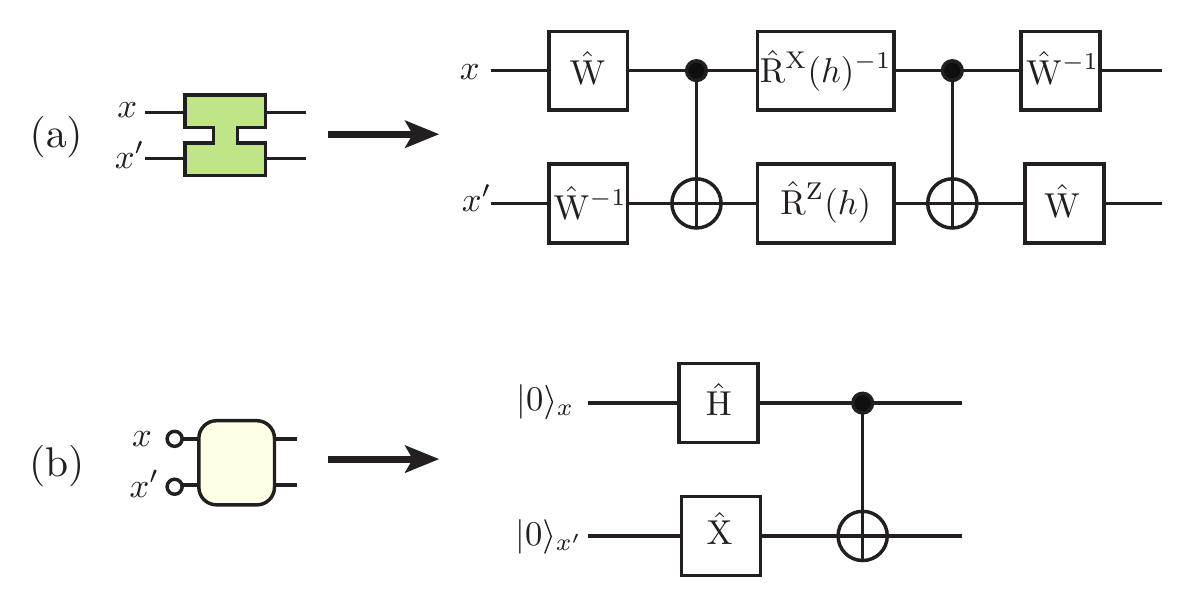}
  \caption{
    (a) Quantum circuit representing the local time-evolution operator 
    $\exp [{\rm i}\theta t (\hat{\sigma}_x^+ \hat{\sigma}_{x^\prime}^- +\hat{\sigma}_{x^\prime}^- \hat{\sigma}_x^+)]$, 
    composed of two CNOT gates and six single-qubit unitary gates.
    $\hat{\rm W} = {\rm e}^{{\rm i}\pi \hat{\rm X}/4}$, $\hat{\rm R}^{\rm X}(h) = {\rm e}^{-{\rm i}h \hat{\rm X}}$, and 
    $\hat{\rm R}^{\rm Z}(h) = {\rm e}^{-{\rm i}h\hat{\rm Z}}$
    with $h=\theta t$. 
    In the control-NOT gate $\hat{\rm C}_x(\hat{\rm X}_{x^{\prime}})$, 
    a black circle denotes the control qubit and an open plus circle
    indicates the $\hat{\rm X}$ operation. 
    (b) Preparation of the initial bonding state in a quantum circuit. 
    }
  \label{fig:ucircuit}
\end{figure}

In this case, the one-dimensional free-fermion system in Eq.~(\ref{eq:1d:ham}) under both PBCs and APBCs 
is mapped onto the spin-1/2 $XY$ model,
\begin{equation}
  \hat{\mathcal{H}}_{\rm spin} = - t
  \sum_{x=1}^{L} (
  \hat{\sigma}^+_{x+1} \hat{\sigma}^-_{x} +
  \hat{\sigma}^-_{x+1} \hat{\sigma}^+_{x} ),
\end{equation}
with $\hat{\sigma}^{\pm}_{L+1} = \hat{\sigma}^{\pm}_{1}$ (i.e., PBCs). 
Here, $\hat{\sigma}^+_{x}$ ($\hat{\sigma}^-_{x}$) represents
the local operator to flip the qubit state from $\vert 1 \rangle_x$ ($\vert 0 \rangle_x$)
to $\vert 0 \rangle_x$ ($\vert 1 \rangle_x$), 
but not the other way around, where
$\vert \sigma \rangle_x$ ($\sigma = 0,1$) denotes the local state at qubit $x$ in the Pauli $z$ basis. 
As shown in Fig.~\ref{fig:ucircuit}(a), the local time-evolution operator 
$\exp [{\rm i}\theta t (\hat{\sigma}_x^+ \hat{\sigma}_{x^\prime}^- +\hat{\sigma}_{x^\prime}^- \hat{\sigma}_x^+)]$ 
can be implemented in a quantum circuit~\cite{Vidal2004,Shende2004,Coffey2008}. 
It should also be noted that 
the condition of $N$ being even (odd) for APBCs (PBCs) 
corresponds to the closed shell condition in the free-fermion system.

With the Jordan-Wigner transformation, the fermion vacuum state $\vert 0\rangle$ is mapped to 
$\prod_{x=1}^L|1\rangle_x$, and therefore  
the initial state $\vert \psi_{\rm i} \rangle$ in Eq.~(\ref{eq:1d:init}) 
can be mapped in the qubit representation onto 
\begin{equation}
  \prod_{x=1}^{L/2} \frac{1}{\sqrt{2}} \left( \vert 0 \rangle_{2x-1} \vert 1 \rangle_{2x}
  + \vert 1 \rangle_{2x-1} \vert 0 \rangle_{2x} \right), 
  \label{eq:qint}
\end{equation}
i.e., a product state of spin-triplet states. 
As shown in Fig.~\ref{fig:ucircuit}(b), each spin-triplet state  
$\frac{1}{\sqrt{2}} ( \vert 0 \rangle_{2x-1} \vert 1 \rangle_{2x}
  + \vert 1 \rangle_{2x-1} \vert 0 \rangle_{2x} )$ can be generated as 
\begin{equation}
  \hat{\rm C}_{2x-1}(\hat{\rm X}_{2x})
  \hat{\rm H}_{2x-1} \hat{\rm X}_{2x} \vert 0 \rangle_{2x-1} \vert 0 \rangle_{2x}.
\end{equation}
Here, $\hat{\rm C}_{x}(\hat{\rm X}_{x^{\prime}})$ denotes the control-NOT gate
acting on qubit $x^{\prime}$ with the control qubit at qubit $x$, and
$\hat{\rm H}_x$ indicates the Hadamard gate acting on qubit $x$.

Finally, we briefly note that for a more general fermion system in higher spatial dimensions 
with a long-range hopping, 
the phase factors due to the Jordan-Wigner strings cannot be canceled  
and yield many-body interactions in the qubit representation. 
In principle, these many-body interactions can be treated as two-qubit operations 
by using, for example, the perturbative gadgets~\cite{PhysRevA.77.062329}. 
However, these techniques introduce additional sources of errors. 
Therefore, we leave the general cases for a future study 
and focus here on the one-dimensional system.

\subsection{\label{sec:tool}Useful properties of the DQAP ansatz for free fermions}

The DQAP ansatz for the free-fermion system
introduced in the previous section generally has the following form:
\begin{align}
  \vert \psi (\mbox{\boldmath{$\theta$}}) \rangle = & 
  \prod_{k=K}^1
  {\rm e}^{- {\rm i} \hat{\mathcal{W}}_k \theta_k} \vert \psi_0 \rangle
  \nonumber \\
  = & {\rm e}^{-{\rm i} \hat{\mathcal{W}}_K \theta_K}
  \cdots
  {\rm e}^{-{\rm i}\hat{\mathcal{W}}_2 \theta_{2}}
  {\rm e}^{-{\rm i}\hat{\mathcal{W}}_1 \theta_{1}}
  \vert \psi_0 \rangle,
  \label{eq:ffdqap:general}
\end{align}
where $\hat{\mathcal{W}}_k$ ($k=1,2,\cdots,K$) is a Hermitian single-particle
operator given by
\begin{equation}
  \hat{\mathcal{W}}_k = \hat{\bm c}^{\dagger} {\bm W}_k \hat{\bm c},
  \label{eq:ff:op:w}
\end{equation}
and $\vert \psi_0 \rangle$ is a ground state of an $N$-fermion system defined by the following single-particle Hamiltonian: 
\begin{equation}
  \hat{\mathcal{W}}_0 = \hat{\bm c}^{\dagger} {\bm W}_0 \hat{\bm c}, 
\end{equation}
thus representing a Slater determinant state of $N$ fermions.
We can now easily show that $\vert \psi (\mbox{\boldmath{$\theta$}}) \rangle$ in Eq.~(\ref{eq:ffdqap:general}) 
is more compactly written as 
\begin{equation}
  \vert \psi (\mbox{\boldmath{$\theta$}}) \rangle
  = \prod_{n=1}^N [ \hat{\bm c}^{\dagger} \mbox{\boldmath{$\Psi$}}_K ]_n \vert 0 \rangle,
  \label{eq:tsps:psi}
\end{equation}
where
\begin{equation}
  \mbox{\boldmath{$\Psi$}}_K = \prod_{k=K}^1 {\rm e}^{-{\rm i} \theta_k {\bm W}_k} \mbox{\boldmath{$\Psi$}}_0,
  \label{eq:mat:psik}
\end{equation}
and $\mbox{\boldmath{$\Psi$}}_0$ is an $L\times N$ matrix 
such that the $n$th column of $\mbox{\boldmath{$\Psi$}}_0$ 
is the eigenstate of ${\bm W}_0$ with the $n$th lowest eigenvalue. 
The derivation of Eq.~(\ref{eq:tsps:psi}) is given in Appendix~\ref{sec:ff}.
Equations~(\ref{eq:tsps:psi}) and (\ref{eq:mat:psik}) imply that 
a state initially prepared as a single Slater determinant state evolves in time, realized by repeatedly applying the unitary time-evolution 
operators, into a state that can still be represented as a single Slater determinant state. 
Therefore, we can even discuss the time evolution of each constituent single-particle orbital in the Slater determinant state.

It is also readily shown that the overlap between
two $N$-fermion states $\vert \psi \rangle$ and $\vert \phi \rangle$ is calculated as 
\begin{equation}
  \langle \psi \vert \phi \rangle = {\rm det} [ \mbox{\boldmath{$\Psi$}}^{\dagger} \mbox{\boldmath{$\Phi$}} ],
  \label{eq:ff:overlap}
\end{equation}
where $\vert \psi \rangle$ is an $N$-fermion state given in Eq.~(\ref{eq:psi}) but for any $\mbox{\boldmath{$\Psi$}}$ and 
\begin{equation}
  \vert \phi \rangle = \prod_{n=1}^N [ \hat{\bm c}^{\dagger} \mbox{\boldmath{$\Phi$}} ]_n \vert 0 \rangle
  \label{eq:phi}
\end{equation}
with $\mbox{\boldmath{$\Phi$}} $ being an $L\times N$ matrix. 
We can also show the following useful formula:  
\begin{align}
  {\mathscr G}_{xx^\prime}
  = \frac{\langle \psi \vert \hat{c}_x^{\dagger} \hat{c}_{x^{\prime}} \vert
    \phi \rangle}{\langle \psi \vert \phi \rangle}
  = &{\rm tr} [
    \mbox{\boldmath{$\Phi$}} (\mbox{\boldmath{$\Psi$}}^{\dagger} \mbox{\boldmath{$\Phi$}})^{-1}
    \mbox{\boldmath{$\Psi$}}^{\dagger} \mbox{\boldmath{$\delta$}}_{xx^{\prime}} ]
  \nonumber \\
  = &
  [\mbox{\boldmath{$\Phi$}} (\mbox{\boldmath{$\Psi$}}^{\dagger} \mbox{\boldmath{$\Phi$}})^{-1}
     \mbox{\boldmath{$\Psi$}}^{\dagger} ]_{x^{\prime}x},
  \label{eq:ff:expectation}
\end{align}
where ${\rm tr}[ {\bm A}]$ indicates the trace of a matrix ${\bm A}$ and 
$\mbox{\boldmath{$\delta$}}_{xx^{\prime}}$ is an $L\times L$ matrix
whose elements are given by
$\left[ \mbox{\boldmath{$\delta$}}_{xx^{\prime}} \right]_{x_1x_2}= \delta_{x x_1} \delta_{x^{\prime}x_2}$. 
We can furthermore derive that, for example,  
\begin{equation}
 \frac{\langle \psi \vert \hat{c}_x^{\dagger}  \hat{c}_y^{\dagger} \hat{c}_{y^{\prime}}  \hat{c}_{x^{\prime}} \vert
  \phi \rangle}{\langle \psi \vert \phi \rangle}
  = {\mathscr G}_{xx^\prime} {\mathscr G}_{yy^\prime} - {\mathscr G}_{xy^\prime} {\mathscr G}_{yx^\prime},  
  \label{eq:wick}
\end{equation} 
which is simply the Wick's theorem.

\subsection{\label{sec:opt}Optimization method}

In this paper, we employ the natural gradient method to optimize 
the variational parameters in the variational wave function.
Here we briefly summarize this optimization method for the DQAP ansatz.

The natural gradient method was originally introduced in the context of machine learning~\cite{Amari,Amari1998}.  
However, essentially the same method has also been independently proposed to optimize a many-body 
variational wave function~\cite{SRmethod} and has been successfully applied to various systems in quantum chemistry and 
condensed-matter physics~\cite{Casula2003,Yunoki2006}. 
This method has also been proposed recently in the context of quantum computing as a way 
to optimize a parametrized quantum circuit~\cite{McArdle:2019aa} and  
is nicely summarized in Ref.~\cite{Stokes2020}.

It is now well known that there are several ways to derive this optimization method~\cite{BeccaSorella}.
A simple way is the formulation based on an infinitesimal imaginary-time evolution 
in the variational parameter space. 
In this case, we determine the new variational parameters
$\mbox{\boldmath{$\theta$}}_{\rm new} = \mbox{\boldmath{$\theta$}} + \delta \mbox{\boldmath{$\theta$}}$
so as to satisfy 
\begin{equation}
  \vert \psi (\mbox{\boldmath{$\theta$}}+\delta \mbox{\boldmath{$\theta$}}) \rangle
  \propto ( 1 - \delta \beta \hat{\mathcal{H}} ) \vert \psi (\mbox{\boldmath{$\theta$}}) \rangle, 
\end{equation}
where $\vert \psi(\mbox{\boldmath{$\theta$}}) \rangle$ is given in Eq.~(\ref{eq:ffdqap:general}) with 
$K$ variational parameters $\mbox{\boldmath{$\theta$}}=\{\theta_1,\theta_2,\cdots,\theta_K \}$, 
$\hat{\mathcal{H}}$ is the Hamiltonian to be solved, in our case, given in Eq.~(\ref{eq:ham}), and 
$\delta \beta$ is an infinitesimal real number. 
Assuming that the variational parameters are all real,
$\delta \mbox{\boldmath{$\theta$}}$ is then determined as 
\begin{equation}
  \delta \mbox{\boldmath{$\theta$}} = \underset{\delta \mbox{\scriptsize\boldmath{$\theta$}}}{\rm arg min}
  \left[
    d^2( \vert \psi(\mbox{\boldmath{$\theta$}}+\delta \mbox{\boldmath{$\theta$}}) \rangle,
    (1 - \delta \beta \hat{\mathcal{H}} ) \vert \psi (\mbox{\boldmath{$\theta$}}) \rangle )
  \right], 
\end{equation}
where $d( \vert \psi \rangle, \vert \phi \rangle)$ is a distance
between two quantum states $\vert \psi \rangle$ and $\vert \phi \rangle$ and is given by 
\begin{equation}
  d^2( \vert \psi \rangle, \vert \phi \rangle) = 1 -
  \frac{\langle \psi \vert \phi \rangle \langle \phi \vert \psi \rangle}
       {\langle \psi \vert \psi \rangle \langle \phi \vert \phi \rangle},
\end{equation}
i.e., essentially the same as the fidelity, assuming that the two states $\vert \psi \rangle$ and $\vert \phi \rangle$ are 
not generally normalized. 

Expanding $d^2( \vert \psi(\mbox{\boldmath{$\theta$}}+\delta \mbox{\boldmath{$\theta$}}) \rangle, (1 - \delta \beta \hat{\mathcal{H}} ) \vert \psi (\mbox{\boldmath{$\theta$}}) \rangle )$
up to the second order of $\delta \mbox{\boldmath{$\theta$}}$ and $\delta \beta$,
we obtain the following quadratic form:
\begin{align}
 & d^2( \vert \psi(\mbox{\boldmath{$\theta$}}+\delta \mbox{\boldmath{$\theta$}}) \rangle, (1 - \delta \beta \hat{\mathcal{H}} ) \vert \psi (\mbox{\boldmath{$\theta$}}) \rangle ) \nonumber \\
  \approx &
  \delta \mbox{\boldmath{$\theta$}}^t {\bm S} \delta \mbox{\boldmath{$\theta$}} + \delta \beta
  ( \delta \mbox{\boldmath{$\theta$}}^t {\bm f} + {\bm f}^{\dagger} \delta \mbox{\boldmath{$\theta$}} )
  +
  \delta \beta^2 \bar{E}^2,
  \label{eq:sto}
\end{align}
where $\delta \mbox{\boldmath{$\theta$}}$ on the right hand side is a $K$-dimensional column vector 
with the $k$th element being $\delta\theta_k$, 
 ${\bm S}$ is a $K\times K$ matrix given by
\begin{equation}
  \begin{split}
  \left[ {\bm S} \right]_{k k^{\prime}} = & 
  \frac{\langle \partial_k \psi(\mbox{\boldmath{$\theta$}}) \vert \partial_{k^{\prime}} \psi(\mbox{\boldmath{$\theta$}}) \rangle}
       {\langle \psi(\mbox{\boldmath{$\theta$}}) \vert \psi(\mbox{\boldmath{$\theta$}}) \rangle} \\
       & -
  \frac{\langle \partial_k \psi(\mbox{\boldmath{$\theta$}}) \vert \psi(\mbox{\boldmath{$\theta$}}) \rangle}
       {\langle \psi(\mbox{\boldmath{$\theta$}}) \vert \psi(\mbox{\boldmath{$\theta$}}) \rangle}
  \frac{\langle \psi(\mbox{\boldmath{$\theta$}}) \vert \partial_{k^{\prime}} \psi(\mbox{\boldmath{$\theta$}}) \rangle}
       {\langle \psi(\mbox{\boldmath{$\theta$}}) \vert \psi(\mbox{\boldmath{$\theta$}}) \rangle},
  \end{split}
  \label{eq:smat}
\end{equation}
with 
\begin{equation}
  \partial_k = \frac{\partial}{\partial \theta_k}, 
\end{equation}
${\bm f}$ is a $K$-dimensional column vector given by
\begin{equation}
  \begin{split}
    \left[ {\bm f} \right]_{k} =
    & 
    \frac{\langle \partial_k \psi(\mbox{\boldmath{$\theta$}}) \vert \hat{\mathcal{H}} \vert \psi(\mbox{\boldmath{$\theta$}}) \rangle}
         {\langle \psi (\mbox{\boldmath{$\theta$}}) \vert \psi (\mbox{\boldmath{$\theta$}}) \rangle} \\
    & -
    \frac{\langle \partial_k \psi(\mbox{\boldmath{$\theta$}}) \vert \psi(\mbox{\boldmath{$\theta$}}) \rangle}
         {\langle \psi (\mbox{\boldmath{$\theta$}}) \vert \psi (\mbox{\boldmath{$\theta$}}) \rangle}
    \frac{\langle \psi(\mbox{\boldmath{$\theta$}}) \vert \hat{\mathcal{H}} \vert \psi(\mbox{\boldmath{$\theta$}}) \rangle}
         {\langle \psi (\mbox{\boldmath{$\theta$}}) \vert \psi (\mbox{\boldmath{$\theta$}}) \rangle},
  \end{split}
\end{equation}
and $\bar{E}^2$ is the variance of the Hamiltonian $\mathcal{\hat{H}}$ given by
\begin{equation}
  \begin{split}
  \bar{E}^2 = 
  \frac{ \langle \psi (\mbox{\boldmath{$\theta$}}) \vert \hat{\mathcal{H}}^2 \vert \psi (\mbox{\boldmath{$\theta$}}) \rangle }
       { \langle \psi (\mbox{\boldmath{$\theta$}}) \vert \psi (\mbox{\boldmath{$\theta$}}) \rangle} 
        - 
  \left( \frac{ \langle \psi (\mbox{\boldmath{$\theta$}}) \vert \hat{\mathcal{H}} \vert \psi (\mbox{\boldmath{$\theta$}}) \rangle }
       { \langle \psi (\mbox{\boldmath{$\theta$}}) \vert \psi (\mbox{\boldmath{$\theta$}}) \rangle} \right)^2. 
  \end{split}
\end{equation}
Here, we assume that $\vert \psi (\mbox{\boldmath{$\theta$}}) \rangle$ is not normalized and thus these formulas can be used 
in general cases. Note also that ${\bm S}$ is Hermitian, i.e., ${\bm S}^\dag = {\bm S}$.

The stationary point of the quadratic equation given in Eq.~(\ref{eq:sto}) is 
now easily obtained by solving the following linear equation: 
\begin{equation}
  ( {\bm S} + {\bm S}^{\ast} ) \delta \mbox{\boldmath{$\theta$}} =
  - \delta \beta ( {\bm f} + {\bm f}^{\ast} ).
  \label{eq:update}
\end{equation}
Notice that since $ {\bm S} + {\bm S}^{\ast} $ and ${\bm f} + {\bm f}^{\ast}$ are both real, the solution 
$\delta \mbox{\boldmath{$\theta$}}^t = (\delta\theta_1,\delta\theta_2,\cdots,\delta\theta_K)$ is guaranteed to also be real. 
We can thereby obtain the new variational parameters 
$\mbox{\boldmath{$\theta$}}_{\rm new} =  \mbox{\boldmath{$\theta$}} + \delta \mbox{\boldmath{$\theta$}}$ 
by solving the above linear equation, in which $\delta\beta$ is learning rate and can be chosen properly.

We can now easily show that
\begin{equation}
  \begin{split}
\delta E
&=\frac{\langle \psi(\mbox{\boldmath{$\theta$}}_{\rm new}) \vert \hat{\mathcal{H}} \vert \psi(\mbox{\boldmath{$\theta$}}_{\rm new}) \rangle}
         {\langle \psi (\mbox{\boldmath{$\theta$}}_{\rm new}) \vert \psi (\mbox{\boldmath{$\theta$}}_{\rm new}) \rangle}  
-
\frac{\langle \psi(\mbox{\boldmath{$\theta$}}) \vert \hat{\mathcal{H}} \vert \psi(\mbox{\boldmath{$\theta$}}) \rangle}
         {\langle \psi (\mbox{\boldmath{$\theta$}}) \vert \psi (\mbox{\boldmath{$\theta$}}) \rangle} \\
&\approx 
\sum_k \delta\theta_k \left(  \left[ {\bm f} \right]_{k} + \left[ {\bm f} \right]_{k}^* \right) \\
&=
- \frac{1}{\delta\beta} \delta \mbox{\boldmath{$\theta$}}^t ( {\bm S} + {\bm S}^{\ast} ) \delta \mbox{\boldmath{$\theta$}}. 
  \end{split}
\end{equation}
Since $ {\bm S} + {\bm S}^{\ast} = 2 {\rm Re} [ {\bm S} ]$ is a positive semidefinite matrix~\cite{Seki2020}, $\delta E\le 0$ as long as $\delta\beta > 0$. 
We should also note that if we expand the following quantity: 
\begin{equation}
  d^2(\vert \psi (\mbox{\boldmath{$\theta$}}) \rangle, \vert \psi(\mbox{\boldmath{$\theta$}}+\delta \mbox{\boldmath{$\theta$}})\rangle)
  = \delta \mbox{\boldmath{$\theta$}}^t {\bm S} \delta \mbox{\boldmath{$\theta$}} + O(\delta \theta_k^3),
\end{equation}
the matrix ${\bm S}$ defined in Eq.~(\ref{eq:smat}) naturally appears.  
Therefore, ${\bm S}$ can be regarded as a metric tensor
for the distance $d(\vert \psi \rangle, \vert \phi \rangle)$ in the parameter space $\mbox{\boldmath{$\theta$}}$.

Using Eqs.~(\ref{eq:ff:overlap}) and (\ref{eq:ff:expectation}), 
we can explicitly derive the forms of ${\bm S}$ and ${\bm f}$, respectively,
for the variational state given in Eq.~(\ref{eq:ffdqap:general}) as 
\begin{equation}
  \begin{split}
  \left[ {\bm S} \right]_{kk^{\prime}} = &
       {\rm tr} \left[
         (\partial_k \mbox{\boldmath{$\Psi$}}_K^{\dagger}) (\partial_{k^{\prime}} \mbox{\boldmath{$\Psi$}}_K) \right] \\
  & - {\rm tr} \left[
    (\partial_k \mbox{\boldmath{$\Psi$}}_K^{\dagger})
     \mbox{\boldmath{$\Psi$}}_K 
     \mbox{\boldmath{$\Psi$}}_K^{\dagger} (\partial_{k^{\prime}} \mbox{\boldmath{$\Psi$}}_K) \right]
  \end{split}
  \label{eq:s_kk}
\end{equation}
and 
\begin{equation}
  \begin{split}
  \left[ {\bm f} \right]_{k} = &
       {\rm tr} \left[ (\partial_k \mbox{\boldmath{$\Psi$}}_K^{\dagger}) {\bm T} \mbox{\boldmath{$\Psi$}}_K \right] \\
       & - {\rm tr} \left[ (\partial_k \mbox{\boldmath{$\Psi$}}_K^{\dagger}) \mbox{\boldmath{$\Psi$}}_K
         \mbox{\boldmath{$\Psi$}}_K^{\dagger} {\bm T} \mbox{\boldmath{$\Psi$}}_K \right],
  \end{split}
  \label{eq:f_kk}
\end{equation}
where we have used that $ \mbox{\boldmath{$\Psi$}}_K^{\dagger} \mbox{\boldmath{$\Psi$}}_K ={\bf I}_{N}$ 
and ${\bf I}_{N}$ is the $N$-dimensional unit matrix. This condition corresponds to the fact that the single-particle orbitals 
in $ \mbox{\boldmath{$\Psi$}}_K$ are orthonormalized, and the generalization to the case where they are not orthonormalized 
is described in Sec.~\ref{sec:res:imag}.
$\partial_k \mbox{\boldmath{$\Psi$}}_K$ is an $L \times N$ matrix 
defined as the first derivative of $\mbox{\boldmath{$\Psi$}}_K$ given in Eq.~(\ref{eq:mat:psik}) 
with respect to the $k$th variational parameter $\theta_k$, i.e., 
\begin{equation}
  \partial_k \mbox{\boldmath{$\Psi$}}_K
  = -{\rm i}\left( \prod_{l=K}^{k+1} {\rm e}^{-{\rm i} \theta_{l} {\bm W}_{l} } \right)
  {\bm W}_k
  \left( \prod_{l=k}^{1} {\rm e}^{-{\rm i} \theta_{l} {\bm W}_{l}} \right)
  \mbox{\boldmath{$\Psi$}}_0.
\end{equation}

Finally, notice that the update formula given in Eq.~(\ref{eq:update})
can be regarded as an extension of the steepest descent algorithm 
that corresponds to the case when the metric tensor ${\bm S}$ is
the unit matrix. This indicates that the optimization method described here cannot exceed 
the limitation of the locality of the search space in general. 
However, we find that this is not a problem in our case 
since we can easily obtain the optimal results without any difficulty. 
The details of this point are found in Appendix~\ref{sec:random}.

\subsection{\label{sec:ent}Entanglement entropy for free fermions}

The entanglement von Neumann entropy is a measure to quantify the quantum entanglement 
between a subspace and its complement of a quantum state, 
and has been used to characterize various quantum states. 
The formula is quite simplified for the free-fermion systems and 
here we briefly outline how the entanglement entropy is calculated.

Let $\mathbb{A} = \{ x_1, x_2, \cdots, x_{L_A} \}$ be a subset of sites 
(the number of sites in $\mathbb{A}$ being $L_A$) that 
are picked out of the all sites $\mathbb{U} = \{ 1, 2, \cdots,x, \cdots, L \}$.
Let $\mathbb{B}$ be the complementary subspace of $\mathbb{A}$: $\mathbb{B} = \overline{\mathbb{A}}$.
We also assume that $\vert \psi \rangle$ is a normalized quantum state and 
can be represented by using the basis on $\mathbb{U}$. 
The reduced density matrix $\hat{\rho}_{\mathbb{A}}$ of subspace $\mathbb{A}$ is given by
\begin{equation}
  \hat{\rho}_{\mathbb{A}} = {\rm Tr}_{\mathbb{B}} [ \vert \psi \rangle \langle \psi \vert ],
\end{equation}
where ${\rm Tr}_{\mathbb{B}}$ indicates the trace over all bases defined on subspace $\mathbb{B}$.
The entanglement entropy $S_{\mathbb{A}}$
of subspace $\mathbb{A}$ is defined by using
this reduced density matrix $\hat{\rho}_{\mathbb{A}}$ as 
\begin{equation}
  S_{\mathbb{A}} = - {\rm Tr}_{\mathbb{A}} \hat{\rho}_{\mathbb{A}} \ln \hat{\rho}_{\mathbb{A}}, 
  \label{eq:def:ent}
\end{equation}
where ${\rm Tr}_{\mathbb{A}}$ is the trace over all bases defined on subspace $\mathbb{A}$.

Notice first that the expectation value of any physical quantity $\hat{\mathcal{O}}_{\mathbb{A}}$ defined on subspace $\mathbb{A}$
can be obtained by using the reduced density matrix $\hat{\rho}_{\mathbb{A}}$ as  
\begin{equation}
  \langle \psi \vert \hat{\mathcal{O}}_{\mathbb{A}} \vert \psi \rangle = {\rm Tr}_{\mathbb{A}} [ \hat{\rho}_{\mathbb{A}} \hat{\mathcal{O}}_{\mathbb{A}} ].
  \label{eq:dmat:identity}
\end{equation}
For the fermion system, $\hat{\mathcal{O}}_{\mathbb{A}}$ is generally composed of 
a product of $\hat{c}_{x}$ and $\hat{c}_{x}^{\dagger}$ with $x \in \mathbb{A}$. 
Therefore, when $\vert \psi \rangle$ is a single-particle state, 
we can use the Wick's theorem [see, for example, Eq.~(\ref{eq:wick})].
This implies that $\hat{\rho}_{\mathbb{A}}$ can be written as 
\begin{equation}
  \hat{\rho}_{\mathbb{A}} = {\rm e}^{-\hat{\bm c}_{\mathbb{A}}^{\dagger} \mbox{\scriptsize\boldmath{$\Gamma$}} \hat{\bm c}_{\mathbb{A}}}
  / {\rm Tr}_{\mathbb{A}} [ {\rm e}^{-\hat{\bm c}_{\mathbb{A}}^{\dagger} \mbox{\scriptsize\boldmath{$\Gamma$}} \hat{\bm c}_{\mathbb{A}}} ],
  \label{eq:dmat}
\end{equation}
where $\hat{\bm c}_{\mathbb{A}}^{\dagger}$ and $\hat{\bm c}_{\mathbb{A}}$
are similar to those in Eqs.~(\ref{eq:vec:c}) but the elements here are fermion operators 
$\hat{c}_x^{\dagger}$ and $\hat{c}_x$ with $x \in \mathbb{A}$,  
and $\mbox{\boldmath{$\Gamma$}}$ is an $L_A\times L_A$ Hermitian matrix~\cite{ChungPeschel}.
Indeed, one can derive the matrix $\mbox{\boldmath{$\Gamma$}}$
directly from a given single Slater determinant state $\vert \psi \rangle$~\cite{ChungPeschel,Cheong2004}.
Here, we shall follow a different route~\cite{Peschel_2003}.

Since the Wick's theorem can decompose the expectation value of any operator into a product of one-particle density matrices,  
we can determine $\mbox{\boldmath{$\Gamma$}}$ by equating
the expectation values of the single-particle operator 
$ \hat{c}_{x}^{\dagger} \hat{c}_{x^{\prime}} $, i.e.,
$\langle \psi \vert \hat{c}_{x}^{\dagger} \hat{c}_{x^{\prime}}  \vert \psi \rangle$ 
with ${\rm Tr}_{\mathbb{A}} [ \hat{\rho}_{\mathbb{A}}  \hat{c}_{x}^{\dagger} \hat{c}_{x^{\prime}}  ]$. 
To this end, let us introduce the following $L_A\times L_A $ one-particle density matrix:
\begin{equation}
  {\bm D}_{\mathbb{A}} =
  \langle \psi \vert \hat{\bm c}_{\mathbb{A}}^{\ast} \hat{\bm c}_{\mathbb{A}}^t \vert \psi \rangle,
  \label{eq:D_A}
\end{equation}
where $\hat{\bm c}_{\mathbb{A}}^{\ast}$ ($\hat{\bm c}_{\mathbb{A}}^t$) is 
the matrix transpose of $\hat{\bm c}_{\mathbb{A}}^{\dagger}$ ($\hat{\bm c}_{\mathbb{A}}$) 
and $\vert \psi \rangle$ is a single Slater determinant state given by the form of Eq.~(\ref{eq:psi}). 
Using Eq.~(\ref{eq:ff:expectation}), each element of ${\bm D}_{\mathbb{A}}$ can be obtained as 
\begin{equation}
  \left[ {\bm D}_{\mathbb{A}} \right]_{xx^{\prime}} =
       {\rm tr} \left[ ( \mbox{\boldmath{$\Psi$}}^{\dagger} \mbox{\boldmath{$\Psi$}} )^{-1}
         \mbox{\boldmath{$\Psi$}}^{\dagger} \mbox{\boldmath{$\delta$}}_{xx^{\prime}}
         \mbox{\boldmath{$\Psi$}} \right]
         =\left[
           \mbox{\boldmath{$\Psi$}}
           ( \mbox{\boldmath{$\Psi$}}^{\dagger} \mbox{\boldmath{$\Psi$}} )^{-1}
           \mbox{\boldmath{$\Psi$}}^{\dagger}         
         \right]_{x^{\prime}x}
\end{equation}
for $x,x^{\prime} \in \mathbb{A}$. 
Since ${\bm D}_{\mathbb{A}}$ is an $L_A\times L_A$ Hermitian matrix,
we can then diagonalize this matrix ${\bm D}_{\mathbb{A}}$ as
\begin{equation}
  {\bm D}_{\mathbb{A}} = {\bm V} \mbox{\boldmath{$\Delta$}} {\bm V}^{\dagger}, 
  \label{eq:dmat:rep1}
\end{equation}
where ${\bm V}$ denotes the unitary matrix composed of the
eigenvectors of matrix ${\bm D}_{\mathbb{A}}$ and
$\mbox{\boldmath{$\Delta$}}$ is the diagonal matrix whose diagonal
elements correspond to the eigenvalues $\delta_l$ ($l=1,2,\cdots,L_A$)
of matrix ${\bm D}_{\mathbb{A}}$: $\mbox{\boldmath{$\Delta$}} = {\rm diag}(\delta_1, \delta_2, \cdots, \delta_{L_A})$.

Let us also define the following $L_A\times L_A$ matrix: 
\begin{equation}
  {\bm D}^{\prime}_{\mathbb{A}} = {\rm Tr}_{\mathbb{A}} [ \hat{\rho}_{\mathbb{A}} \hat{\bm c}_{\mathbb{A}}^{\ast} \hat{\bm c}_{\mathbb{A}}^t ],
\end{equation}
assuming that $\hat{\rho}_{\mathbb{A}}$ is given in Eq.~(\ref{eq:dmat}). 
We then obtain that 
\begin{equation}
  {\bm D}^{\prime}_{\mathbb{A}} = {\bm U}_{\mathbb{A}}^{\ast} ( {\bf I}_{L_A} + {\rm e}^{\mbox{\scriptsize\boldmath{$\Lambda$}}} )^{-1} {\bm U}_{\mathbb{A}}^t,
  \label{eq:dmat:rep2}
\end{equation}
where ${\bf I}_{L_A}$ is the $L_A$-dimensional unit matrix, ${\bm U}_{\mathbb{A}}$ is the unitary matrix composed
of the eigenvectors of matrix $\mbox{\boldmath{$\Gamma$}}$,
and $\mbox{\boldmath{$\Lambda$}}$ is the diagonal matrix
whose diagonal elements correspond to the eigenvalues $\lambda_l$ ($l=1,2,\cdots,L_A$)
of matrix $\mbox{\boldmath{$\Gamma$}}$:
$\mbox{\boldmath{$\Lambda$}} = {\rm diag}(\lambda_1, \lambda_2, \cdots, \lambda_{L_A})$. 
Because of Eq.~(\ref{eq:dmat:identity}), we now impose that 
${\bm D}_{\mathbb{A}} = {\bm D}_{\mathbb{A}}^{\prime}$.
Comparing Eqs.~(\ref{eq:dmat:rep1}) and (\ref{eq:dmat:rep2}),
we obtain that 
\begin{equation}
  {\bm U}_{\mathbb{A}} = {\bm V}^{\ast}
  \label{eq:dmat:evec}
\end{equation}
and
\begin{equation}
  \mbox{\boldmath{$\Lambda$}} =
  \ln ( {\bf I}_{L_A} - \mbox{\boldmath{$\Delta$}} ) - \ln \mbox{\boldmath{$\Delta$}}.
  \label{eq:dmat:eval}
\end{equation}
Therefore, we finally find that 
\begin{equation}
  \mbox{\boldmath{$\Gamma$}} =
       {\bm V}^{\ast} ( \ln ( {\bf I}_{L_A} - \mbox{\boldmath{$\Delta$}} ) - \ln \mbox{\boldmath{$\Delta$}} )
       {\bm V}^t. 
\end{equation}

Giving the form of the reduced density matrix $\hat{\rho}_{\mathbb{A}}$ in Eq.~(\ref{eq:dmat}),
the entanglement entropy $S_{\mathbb{A}}$ defined in Eq.~(\ref{eq:def:ent}) can now be written as  
\begin{equation}
  S_{\mathbb{A}} = {\rm tr} \left[
    \mbox{\boldmath{$\Lambda$}}
    ( {\bf I}_{L_A} + {\rm e}^{\mbox{\scriptsize\boldmath{$\Lambda$}}} )^{-1}
    + \ln ( {\bf I}_{L_A} + {\rm e}^{-\mbox{\scriptsize\boldmath{$\Lambda$}}} )
    \right].
  \label{eq:ent:orig}
\end{equation}
Using Eq.~(\ref{eq:dmat:eval}), 
we can show that the entanglement entropy $S_{\mathbb{A}}$ is simply given as 
\begin{equation}
  S_{\mathbb{A}} = - {\rm tr}
  \left[
    ( {\bf I}_{L_A} - \mbox{\boldmath{$\Delta$}} )
    \ln 
    ( {\bf I}_{L_A} - \mbox{\boldmath{$\Delta$}} )
    +
    \mbox{\boldmath{$\Delta$}}
    \ln
    \mbox{\boldmath{$\Delta$}}
  \right]. 
  \label{eq:ent}
\end{equation}
Therefore, the entanglement entropy $S_{\mathbb{A}}$ is determined solely 
from the eigenvalues of the one-particle density matrix ${\bm D}_{\mathbb{A}}$.
Note that the eigenvalues of ${\bm D}_{\mathbb{A}}$ are bounded
as $0 \leq \delta_l \leq 1$. 
In the context of quantum chemistry, the eigenvectors of ${\bm D}_{\mathbb{A}}$
are called the natural orbitals 
and the eigenvalue $\delta_l$ corresponds to the density of each natural orbital. 
Since the $L_A\times L_A$ matrices inside the trace in Eq.~(\ref{eq:ent}) are all 
diagonal, we can discuss separately the individual contribution of the natural orbitals 
to the entanglement entropy $S_{\mathbb{A}}$. 
For example, the contribution to $S_{\mathbb{A}}$ is maximum when $\delta_l = 0.5$, 
while it is minimum when $\delta_l = 0$ or $\delta_l = 1$.
This implies that when $\delta_l = 0.5$, 
the corresponding natural orbital in subspace $\mathbb{A}$ is highly hybridized 
with orbitals in subspace $\mathbb{B}$,
giving an intuition of the quantum entanglement in the free-fermion system.

As described above, the entanglement entropy $S_{\mathbb{A}}$ is a measure to quantify the
quantum entanglement between subspaces $\mathbb{A}$ and $\mathbb{B} = \overline{\mathbb{A}}$.
Instead, it is often required to discuss how the quantum state is entangled
between a subspace $\mathbb{A} \subset \mathbb{U}$ and another subspace 
$\mathbb{B} \subset \mathbb{U}$ with $\mathbb{A} \cap \mathbb{B} = \emptyset$ and 
$\mathbb{A} \cup \mathbb{B} \ne \mathbb{U}$. 
One of the quantities for this purpose is
the mutual information $I_{\mathbb{A},\mathbb{B}}$ defined by
\begin{equation}
  I_{\mathbb{A},\mathbb{B}} = S_{\mathbb{A}} + S_{\mathbb{B}} - S_{\mathbb{A}\cup \mathbb{B}}.
\end{equation}
We consider a special case when $\mathbb{A} = \{ x \}$ and $\mathbb{B} = \{ x^{\prime} \}$, and 
the mutual information $I_{x,x^{\prime}}$ for this special case is 
\begin{equation}
  I_{x,x^{\prime}} = S_{\{ x \}} + S_{ \{ x^{\prime} \} } - S_{\{ x,x^{\prime} \} }.
  \label{eq:2p:minf}
\end{equation}

There are several remarks on $I_{x,x^{\prime}}$. 
First, of all, 
${\bm D}_{ \{ x \} } = D_x = \langle \psi \vert \hat{c}_{x}^{\dagger} \hat{c}_{x} \vert \psi \rangle$,
which is the density of fermions at site $x$.
Therefore, if the system is homogenous, 
$D_x$ is independent of $x$ and $D_x = N/L$. 
When the system is at half filling, $D_x = 0.5$  
and thus $S_{ \{ x \} } = \ln 2$,
which is the maximum value of the entanglement entropy for a single site. 
Second, $I_{x,x^{\prime}}$ is determined by $S_{ \{ x,x^{\prime} \}}$,
which can be calculated from the eigenvalues of the one-particle density matrix 
\begin{equation}
  {\bm D}_{ \{ x, x^{\prime} \} } =
  \left(
  \begin{array}{cc}
    \langle \psi \vert \hat{c}_{x}^{\dagger} \hat{c}_{x} \vert \psi \rangle &
    \langle \psi \vert \hat{c}_{x}^{\dagger} \hat{c}_{x^{\prime}} \vert \psi \rangle \\
    \langle \psi \vert \hat{c}_{x^{\prime}}^{\dagger} \hat{c}_{x} \vert \psi \rangle &
    \langle \psi \vert \hat{c}_{x^{\prime}}^{\dagger} \hat{c}_{x^{\prime}} \vert \psi \rangle \\
  \end{array}
  \right).
  \label{eq:2p:dmat}
\end{equation}
Since the diagonal term is $0.5$ when the system is homogenous at half filling, 
the off-diagonal elements determine the value of $I_{x,x^{\prime}}$. 
For example, if $\vert \langle \psi \vert \hat{c}_{x}^{\dagger} \hat{c}_{x^{\prime}} \vert \psi \rangle \vert= 0.5$,
we find that $S_{ \{ x, x^{\prime} \} } = 0$ and thus 
$I_{x,x^{\prime}} = 2 \ln 2$. In contrast, if $\langle \psi \vert \hat{c}_{x}^{\dagger} \hat{c}_{x^{\prime}} \vert \psi \rangle = 0$,
we find that $S_{ \{ x, x^{\prime} \} } = 2 \ln 2$ and thus $I_{x,x^{\prime}} = 0$.

\section{\label{sec:result}Numerical Results}

Here, we show the results of numerical simulations 
for the one-dimensional free-fermion system described in Eq.~(\ref{eq:1d:ham}) 
and examine how the DQAP ansatz $\vert \psi_M (\mbox{\boldmath{$\theta$}}) \rangle$ given in 
Eq.~(\ref{eq:1d:dqap}) approaches the exact ground state with 
increasing the number $M$ of layers of the local time-evolution operators (see Fig.~\ref{fig:1d:ansatz}). 
We focus on the fermion density at half filling, i.e., $N=L/2$ and use the natural gradient method described in Sec.~\ref{sec:opt} 
to optimize the variational parameters $\mbox{\boldmath{$\theta$}}$ in the DQAP ansatz 
$\vert \psi_M (\mbox{\boldmath{$\theta$}}) \rangle$.

\subsection{\label{sec:res:ene}Convergence of ground state energy}

\begin{figure}
  \includegraphics[width=\hsize]{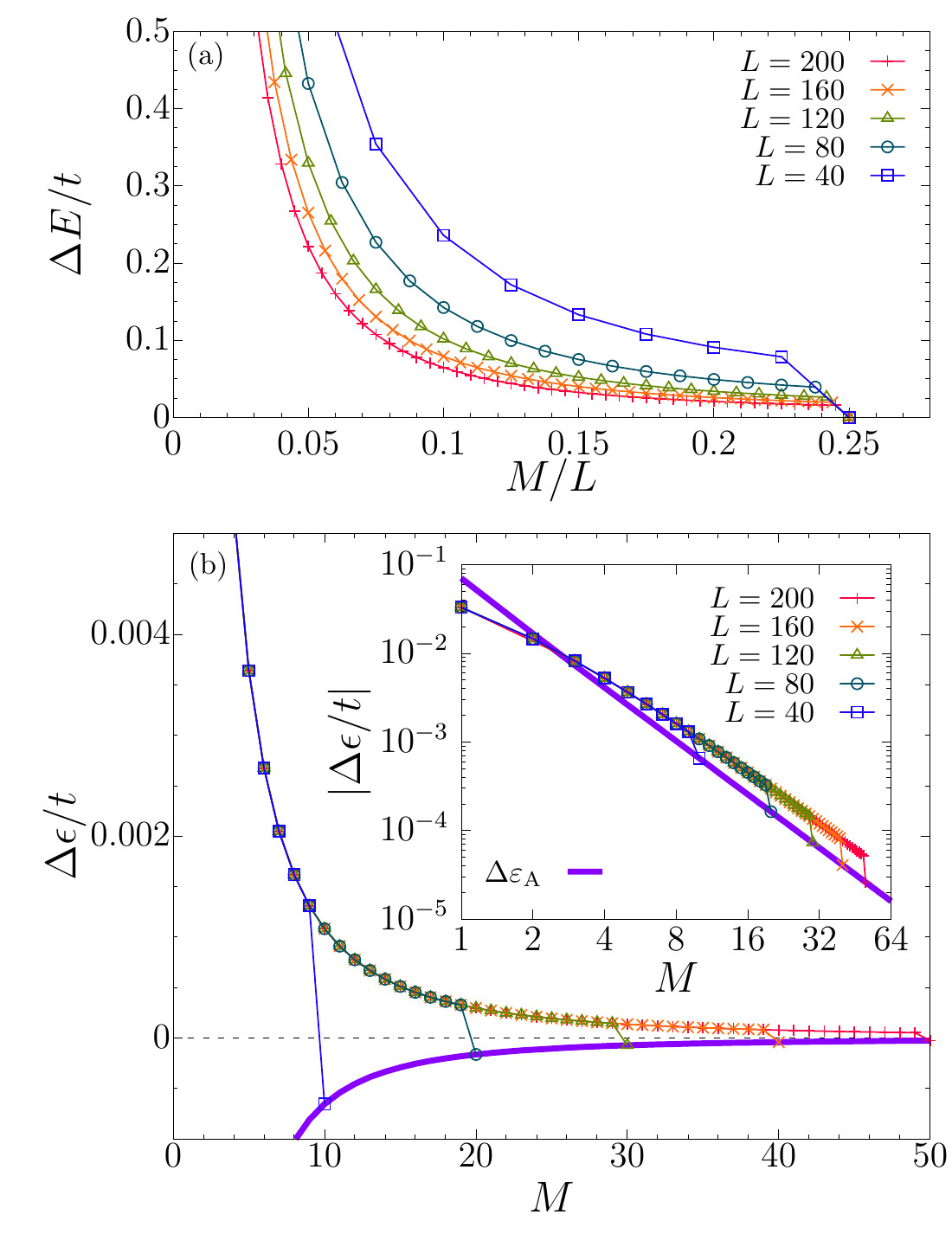}
  \caption{(a) Energy difference $\Delta E = E_M(L) - E_{\rm exact}(L)$
    between the variational energy $E_M(L)$ 
    and the exact energy $E_{\rm exact}(L)$
    as a function of $M/L$ for various system sizes $L$. 
    (b) Energy difference
    $\Delta \varepsilon = E_M(L)/L - \varepsilon_{\infty}$ as a function of $M$ for various system sizes $L$, 
    where $\varepsilon_{\infty} = \lim_{L\to \infty} E_{\rm exact}(L)/L = - 2 \vert t \vert /\pi$ is the exact energy per site
    in the thermodynamic limit. 
    Purple line indicates $\Delta \varepsilon_{\rm A}= E_{\rm exact}(L)/L - \varepsilon_{\infty}$ with $L=4M$.  
    Inset: Logarithmic plot of $\vert \Delta \varepsilon \vert$.
    The results are obtained under APBCs. 
  }
  \label{fig:energy}
\end{figure}

We optimize the variational parameters 
$\mbox{\boldmath{$\theta$}} = \{ \theta_1^{(1)}, \theta_2^{(1)}, \cdots, \theta_1^{(M)}, \theta_2^{(M)} \}$
in the DQAP ansatz $\vert \psi_M (\mbox{\boldmath{$\theta$}}) \rangle$ given in Eq.~(\ref{eq:1d:dqap})
so as to minimize the variational energy 
\begin{equation}
  E_M(L) = \langle \psi_M (\mbox{\boldmath{$\theta$}}) \vert \hat{\mathcal{H}} \vert \psi_M (\mbox{\boldmath{$\theta$}}) \rangle 
  \label{eq:expE}
\end{equation}
for a given system size $L$. 
In order to check the convergence of the variational energy, 
Fig.~\ref{fig:energy}(a) shows the energy difference $\Delta E = E_M(L) - E_{\rm exact}(L)$
from the exact energy $E_{\rm exact}(L)$ 
for various system sizes $L$ 
as a function of $M$. 
Here, we use $L = 4 n_L$ ($n_L$: integer) 
with APBCs and thus the closed shell condition is satisfied for the ground state (see Sec.~\ref{sec:dqap}). 
As shown in Fig.~\ref{fig:energy}(a), $\Delta E$ monotonically decreases with increasing $M$ 
and we obtain that $\Delta E = 0$ within the machine precision exactly at $M = L/4$ for all values of $L$ studied~\cite{pbc}.

\begin{figure*}
  \includegraphics[width=0.9\hsize]{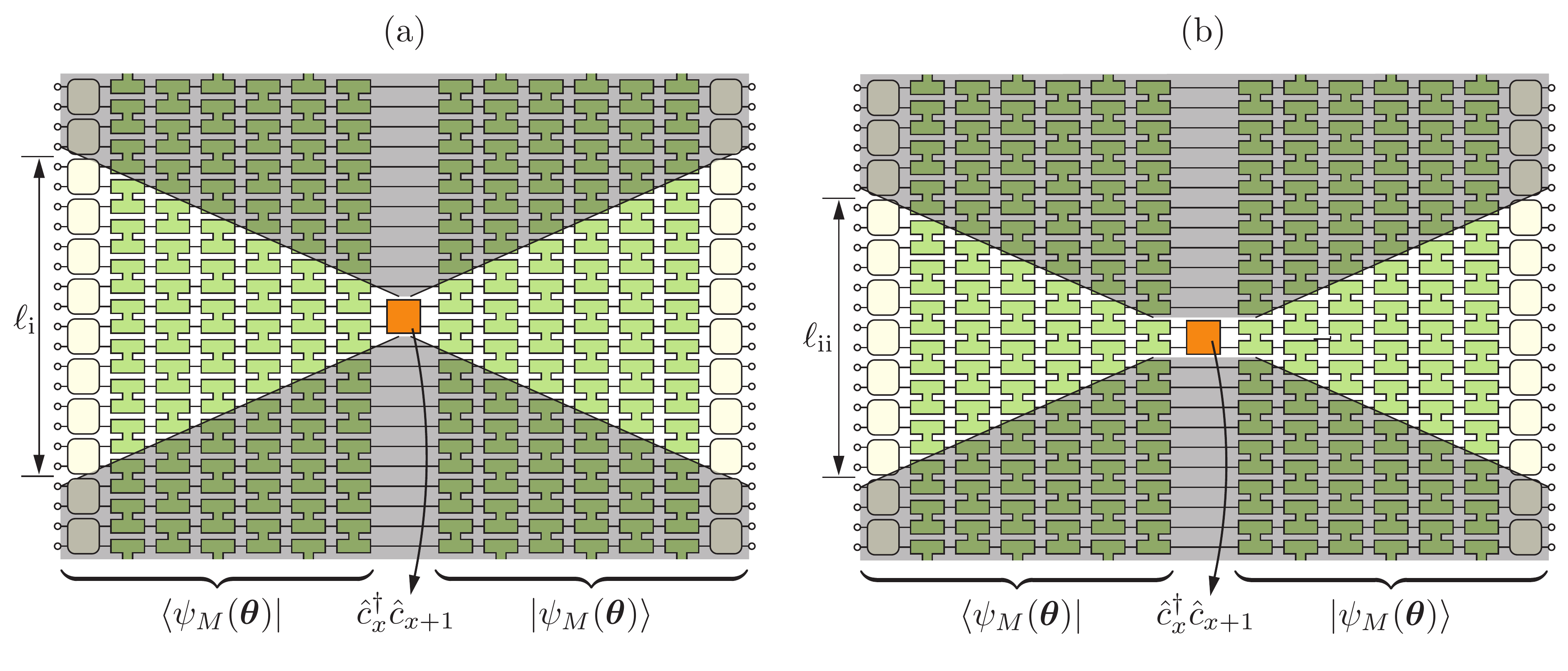}
  \caption{
    Schematic figures of a causality structure for the propagation of quantum entanglement 
    via the local unitary time-evolution operators when 
    $\langle \psi_M(\mbox{\boldmath{$\theta$}}) \vert \hat{c}_x^{\dagger} \hat{c}_{x+1} \vert \psi_M(\mbox{\boldmath{$\theta$}}) \rangle$ 
    is evaluated. 
    Each green block with ``H" shape indicates the local unitary time-evolution operator 
    and light yellow objects at the left and right ends indicate the local bonding states 
    composing the initial state $\vert \psi_{\rm i} \rangle$. 
    The operator $\hat{c}_x^{\dagger} \hat{c}_{x+1}$, indicated by orange squares at the center, 
    acts (a) over two neighboring local time-evolution operators and 
    (b) only on a single local time-evolution operator.  
    The local unitary time-evolution operators in the shaded regions do not contribute to the expectation value 
    because these unitary operators are canceled when $\langle \psi_M(\mbox{\boldmath{$\theta$}}) \vert$ and 
    $\vert \psi_M(\mbox{\boldmath{$\theta$}}) \rangle$ 
    are multiplied form the left and right sides, respectively. 
    $\ell_{\rm i}$ ($\ell_{\rm ii}$) is the number of sites (i.e., qubits) that are relevant to 
    the expectation value. $\ell_{\rm i} = 16$ in (a) and $\ell_{\rm ii}=14$ in (b), where $M=3$. 
  }
  \label{fig:causality}
\end{figure*}

To better understand this observation, 
let us examine closely how the expectation value of the energy for the DQAP ansatz $\vert \psi_M (\mbox{\boldmath{$\theta$}}) \rangle$ 
in Eq.~(\ref{eq:expE}) is evaluated. 
For this purpose, we should notice that the energy expectation value is essentially given simply by the sum of terms 
$\langle \psi_M(\mbox{\boldmath{$\theta$}}) \vert \hat{c}_{x}^{\dagger} \hat{c}_{x+1} \vert \psi_M (\mbox{\boldmath{$\theta$}}) \rangle$ 
(and also $\langle \psi_M(\mbox{\boldmath{$\theta$}}) \vert \hat{c}_{x+1}^{\dagger} \hat{c}_{x} \vert \psi_M (\mbox{\boldmath{$\theta$}}) \rangle$ but it is basically the same as $\langle \psi_M(\mbox{\boldmath{$\theta$}}) \vert \hat{c}_{x}^{\dagger} \hat{c}_{x+1} \vert \psi_M (\mbox{\boldmath{$\theta$}}) \rangle$ for the purpose of the discussion here) 
over all $x$'s. Therefore, it is adequate to consider each term separately.   
Because of the form of construction for the DQAP ansatz $\vert \psi_M (\mbox{\boldmath{$\theta$}}) \rangle$, there are two different 
cases of 
$\langle \psi_M(\mbox{\boldmath{$\theta$}}) \vert \hat{c}_{x}^{\dagger} \hat{c}_{x+1} \vert \psi_M (\mbox{\boldmath{$\theta$}}) \rangle$: 
the operator $\hat{c}_x^{\dagger} \hat{c}_{x+1}$ acts (i) over two neighboring local time-evolution operators 
${\rm e}^{{\rm i}t\theta_1^{(M)} ( \hat{c}_{x-1}^{\dagger} \hat{c}_{x} + \hat{c}_{x}^{\dagger} \hat{c}_{x-1} ) }$ and 
${\rm e}^{{\rm i}t\theta_1^{(M)} ( \hat{c}_{x+1}^{\dagger} \hat{c}_{x+2} + \hat{c}_{x+2}^{\dagger} \hat{c}_{x+1} ) }$, 
as schematically shown in Fig.~\ref{fig:causality}(a), and acts (ii) only on a single local time-evolution operator 
${\rm e}^{{\rm i}t\theta_1^{(M)} ( \hat{c}_{x}^{\dagger} \hat{c}_{x+1} + \hat{c}_{x+1}^{\dagger} \hat{c}_{x} ) }$, 
as shown in Fig.~\ref{fig:causality}(b).

Let us first consider case (i). In this case, 
\begin{equation}
  \begin{split}
    & \langle \psi_M(\mbox{\boldmath{$\theta$}}) \vert \hat{c}_{x}^{\dagger} \hat{c}_{x+1} \vert \psi_M(\mbox{\boldmath{$\theta$}}) \rangle \\
   =  & \langle \psi_{\rm i} \vert
  \prod_{m=1}^M ( {\rm e}^{{\rm i}\theta_2^{(m)} \hat{\mathcal{V}}_2^{(m)}} {\rm e}^{{\rm i} \theta_1^{(m)} \hat{\mathcal{V}}_1^{(m)}} )   \hat{c}_{x}^{\dagger} \hat{c}_{x+1} \\
  & \times 
  \prod_{m=M}^1 ( {\rm e}^{-{\rm i}\theta_1^{(m)} \hat{\mathcal{V}}_1^{(m)}} {\rm e}^{-{\rm i} \theta_2^{(m)} \hat{\mathcal{V}}_2^{(m)}} )
  \vert \psi_{\rm i} \rangle,
  \end{split}
  \label{eq:evcc}
\end{equation}
where
\begin{equation}
  \hat{\mathcal{V}}_1^{(m)} = - t \sum_{y=x-2(M-m)-1}^{x+2(M-m)+1} (\hat{c}_{y}^{\dagger} \hat{c}_{y+1} + \hat{c}_{y+1}^{\dagger} \hat{c}_{y})
\end{equation}
and
\begin{equation}
  \hat{\mathcal{V}}_2^{(m)} = - t \sum_{y=x-2(M-m)-2}^{x+2(M-m)+2} (\hat{c}_{y}^{\dagger} \hat{c}_{y+1} + \hat{c}_{y+1}^{\dagger} \hat{c}_{y} ),
\end{equation}
assuming that $L \ge 4M+2$. 
Namely, one can eliminate many of the local unitary time-evolution operators in the expectation value 
due to the cancellations of the left and right sides of the product, as illustrated in Fig.~\ref{fig:causality}(a). 
The number $\ell_{\rm i}$ of sites (i.e., qubits) that contribute to the local
expectation value $\langle \psi_M (\mbox{\boldmath{$\theta$}}) \vert \hat{c}_{x}^{\dagger} \hat{c}_{x+1} \vert \psi_M (\mbox{\boldmath{$\theta$}}) \rangle$ is linearly dependent on $M$: $\ell_{\rm i} = 4M + 4$~\cite{note1}. 
This implies that the propagation of quantum entanglement via the {\it local} time-evolution operators is bounded in space 
and this boundary forms a causality-cone like structure shown schematically in Fig.~\ref{fig:causality}(a). 
This upper limit on the propagation 
speed is known as the Lieb-Robinson bound~\cite{Lieb1972}. 

In case (ii), we can also evaluate the local
expectation value $\langle \psi_M (\mbox{\boldmath{$\theta$}}) \vert \hat{c}_{x}^{\dagger} \hat{c}_{x+1} \vert \psi_M (\mbox{\boldmath{$\theta$}}) \rangle$ in the same manner as in Eq.~(\ref{eq:evcc}) except that now 
\begin{equation}
  \hat{\mathcal{V}}_1^{(m)} = - t \sum_{y=x-2(M-m)}^{x+2(M-m)} (\hat{c}_{y}^{\dagger} \hat{c}_{y+1} + \hat{c}_{y+1}^{\dagger} \hat{c}_{y})
\end{equation}
and
\begin{equation}
  \hat{\mathcal{V}}_2^{(m)} = - t \sum_{y=x-2(M-m)-1}^{x+2(M-m)+1} (\hat{c}_{y}^{\dagger} \hat{c}_{y+1} + \hat{c}_{y+1}^{\dagger} \hat{c}_{y} ),\label{eq:v2bound}
\end{equation}
assuming that $L \ge 4M$. Therefore, in this case, 
the number $\ell_{\rm ii}$ of sites that contribute to the local expectation value 
$\langle \psi_M (\mbox{\boldmath{$\theta$}}) \vert \hat{c}_{x}^{\dagger} \hat{c}_{x+1} \vert \psi_M (\mbox{\boldmath{$\theta$}}) \rangle$ 
is also linearly dependent on $M$: $\ell_{\rm ii} = 4M+2$~\cite{note1}. 
This sets the boundaries of a causality-cone like structure in 
Fig.~\ref{fig:causality}(b), within which the quantum entanglement is developed.

To reach the exact ground state energy for a given
system size $L$, $\ell_{\rm i}$ and $\ell_{\rm ii} $ 
have to be equal to or exceed the system size $L$, which corresponds to 
\begin{equation}
M\ge \lceil (L-2)/4\rceil ,
\end{equation}
with $\lceil z\rceil$ being the smallest integer greater than or equal to $z$. 
This condition is independent of the boundary conditions because Eqs.~(\ref{eq:evcc})--(\ref{eq:v2bound}) 
do not depend on the boundary conditions.
The fact that the exact ground state energy is obtained 
exactly when $M=L/4$ found in Fig.~\ref{fig:energy}(a) (recall that we choose $L=4n_L$ with $n_L$ integer, there) implies 
that the exact ground state is constructed in the DQAP ansatz $\vert \psi_M (\mbox{\boldmath{$\theta$}}) \rangle$ 
with the shortest possible depth 
$M_B=\lceil (L-2)/4\rceil$  set by the Lieb-Robinson bound.
The same conclusion is also reached in the case when the PBC is employed~\cite{pbc}.

We should also note that as indicated in Fig.~\ref{fig:causality}, 
the causality-cone like structure of the local unitary time-evolution operators 
contributing to the local expectation value does not depend on system size $L$.  
As a consequence, we expect that the optimized variational energy per site would 
not depend on $L$ as along as $L>4M+2\,(=\ell_{\rm ii})$. 
Indeed, as shown in Fig.~\ref{fig:energy}(b), 
the optimized variational energies per site with a given value of $M$ are exactly the same for different values of $L$ 
until $M$ reaches to the boundary at $M=L/4$ for APBCs~\cite{conv}, where the variational energy abruptly changes to the exact value 
for the system size $L$.
Moreover, we find that the optimized variational energy per site for $M < (L-2)/4$ under PBCs
is identical to that for the same $M$ (but $M<L/4$) under APBCs.
We should note that 
a similar analysis for the transverse-field Ising model has also been reported in Ref.~\cite{Mbeng1}.

\subsection{\label{sec:res:psi}Time-evolution of single-particle orbitals }

We now explore how the DQAP ansatz $\vert \psi_M (\mbox{\boldmath{$\theta$}}) \rangle$ is evolved 
by applying the local time-evolution operators. 
Following the argument in Sec.~\ref{sec:tool}, 
the DQAP ansatz $\vert \psi_M (\mbox{\boldmath{$\theta$}}) \rangle$ given in Eq.~(\ref{eq:1d:dqap})
can be written as 
\begin{equation}
  \vert \psi_M (\mbox{\boldmath{$\theta$}}) \rangle
  = \prod_{n=1}^N [ \hat{\bm c}^{\dagger} \mbox{\boldmath{$\Psi$}}_M ]_n \vert 0 \rangle,
  \label{eq:tsps}
\end{equation}
where
\begin{equation}
  \mbox{\boldmath{$\Psi$}}_M = \prod_{m=M}^{1}
  ( {\rm e}^{-{\rm i}\theta_1^{(m)}{\bm V}_1} {\rm e}^{-{\rm i}\theta_2^{(m)}{\bm V}_2})
  \mbox{\boldmath{$\Psi$}}_{\rm i}. 
  \label{eq:1d:psi}
\end{equation} 
Here, ${\bm V}_1$ and ${\bm V}_2$ are $L\times L$ matrices representing  
$\hat{\mathcal{V}}_1$ and $\hat{\mathcal{V}}_2$
given in Eqs.~(\ref{eq:1d:v:bond1}) and (\ref{eq:1d:v:bond2}), respectively, i.e.,  
\begin{equation}
  \hat{\mathcal{V}}_1 = \hat{\bm c}^{\dagger} {\bm V}_1 \hat{\bm c}, \ 
  \hat{\mathcal{V}}_2 = \hat{\bm c}^{\dagger} {\bm V}_2 \hat{\bm c},
\end{equation}
with 
\begin{equation}
  {\bm V}_1 = \left(
  \begin{array}{ccccccc}
    0 & -t & 0 & 0 & \cdots & 0 & 0 \\
    -t & 0 & 0 & 0 & \cdots & 0 & 0 \\
    0 & 0 & 0 & -t & \cdots & 0 & 0 \\
    0 & 0 & -t & 0 & \cdots & 0 & 0 \\
    \vdots & \vdots & \vdots & \vdots & \ddots & \vdots & \vdots \\
    0 & 0 & 0 & 0 & \cdots & 0 & -t \\
    0 & 0 & 0 & 0 & \cdots & -t & 0 \\
  \end{array}
  \right)
  \label{eq:matv1}
\end{equation}
and 
\begin{equation}
  {\bm V}_2 = \left(
  \begin{array}{ccccccc}
    0 & 0 & 0 & 0 & \cdots & 0 & -\gamma t \\
    0 & 0 & -t & 0 & \cdots & 0 & 0 \\
    0 & -t & 0 & 0 & \cdots & 0 & 0 \\
    0 & 0 & 0 & 0 & \cdots & 0 & 0 \\
    \vdots & \vdots & \vdots & \vdots & \ddots & \vdots & \vdots \\
    0 & 0 & 0 & 0 & \cdots & 0 & 0 \\
    -\gamma t & 0 & 0 & 0 & \cdots & 0 & 0 \\
  \end{array}
  \right), 
  \label{eq:matv2}
\end{equation}
and 
$\mbox{\boldmath{$\Psi$}}_{\rm i}$ is the $L\times N$ matrix 
representing the initial state $\vert \psi_{\rm i}\rangle$ in Eq.~(\ref{eq:1d:init}), i.e., 
\begin{equation}
\vert \psi_{\rm i} \rangle = \prod_{n=1}^N [\hat{\bm c}^{\dagger} \mbox{\boldmath{$\Psi$}}_{\rm i} ]_{n} \vert 0 \rangle,
\end{equation}
with 
\begin{equation}
  \mbox{\boldmath{$\Psi$}}_{\rm i} =
  \frac{1}{\sqrt{2}}
  \left(
  \begin{array}{cccc}
    1 & 0 & \cdots & 0 \\
    1 & 0 & \cdots & 0 \\
    0 & 1 & \cdots & 0 \\
    0 & 1 & \cdots & 0 \\
    \vdots & \vdots & \ddots & \vdots \\
    0 & 0 & \cdots & 1 \\
    0 & 0 & \cdots & 1 \\
  \end{array}
  \right)
  \label{eq:matint}
\end{equation}
and the number $N$ of fermions being $L/2$.

It should be noted that since ${\bm V}_1$ and ${\bm V}_2$ are both block diagonal matrices 
with each block being a $2 \times 2$ matrix, 
these can easily be exponentiated as 
\begin{equation}
  \begin{split}
    & {\rm e}^{-{\rm i}\theta{\bm V}_1} \\
    = & \left(
  \begin{array}{ccccccc}
    \cos \theta t & {\rm i} \sin \theta t & 0 & 0 & \cdots & 0 & 0 \\
    {\rm i} \sin \theta t & \cos \theta t & 0 & 0 & \cdots & 0 & 0 \\
    0 & 0 & \cos \theta t & {\rm i} \sin \theta t & \cdots & 0 & 0 \\
    0 & 0 & {\rm i} \sin \theta t & \cos \theta t & \cdots & 0 & 0 \\
    \vdots & \vdots & \vdots & \vdots & \ddots & \vdots & \vdots \\
    0 & 0 & 0 & 0 & \cdots & \cos \theta t & {\rm i} \sin \theta t \\
    0 & 0 & 0 & 0 & \cdots & {\rm i} \sin \theta t & \cos \theta t \\
  \end{array}
  \right)
  \end{split}
  \label{eq:mat_v1}
\end{equation}
and 
\begin{equation}
  \begin{split}
    & {\rm e}^{-{\rm i}\theta{\bm V}_2} \\
    = & \left(
    \begin{array}{ccccccc}
      \cos \theta t & 0 & 0 & 0 & \cdots & 0 &  {\rm i} \gamma \sin \theta t \\
      0 & \cos \theta t & {\rm i} \sin \theta t & 0 & \cdots & 0 & 0 \\
      0 & {\rm i} \sin \theta t & \cos \theta t & 0 & \cdots & 0 & 0 \\
      0 & 0 & 0 & \cos \theta t & \cdots & 0 & 0 \\
      \vdots & \vdots & \vdots & \vdots & \ddots & \vdots & \vdots \\
      0 & 0 & 0 & 0 & \cdots & \cos \theta t & 0 \\
      {\rm i} \gamma \sin \theta t & 0 & 0 & 0 & \cdots & 0 & \cos \theta t \\
    \end{array}
    \right).
  \end{split}
  \label{eq:mat_v2}
\end{equation}
It is also apparent from Eq.~(\ref{eq:matint}) that each column vector in $\mbox{\boldmath{$\Psi$}}_{\rm i} $ 
corresponds to a single-particle orbital, representing the local bonding state 
$\frac{1}{\sqrt{2}}( \hat{c}^{\dagger}_{2x-1} + \hat{c}^{\dagger}_{2x} )\vert 0\rangle$ 
in this case given in Eq.~(\ref{eq:local:bonding}), which constitutes
the Slater determinant state $\vert \psi_{\rm i}\rangle$ for $N$ fermions. 
Therefore, we can now clearly understand that the time-evolved state 
$\vert \psi_M (\mbox{\boldmath{$\theta$}}) \rangle$ from a state initially prepared as a single Slater determinant state 
$\vert \psi_{\rm i}\rangle$ can still be represented as a single Slater determinant state, in which each single-particle orbital 
is given by each column vector of $\mbox{\boldmath{$\Psi$}}_M$. 
We can thus examine the time evolution of each single-particle orbital in the Slater determinant state, which is described 
by Eqs.~(\ref{eq:1d:psi}), (\ref{eq:mat_v1}), and (\ref{eq:mat_v2}).

For this purpose, we first introduce the following $L\times N$ matrix: 
\begin{equation}
  \mbox{\boldmath{$\Phi$}}_m = \prod_{m^{\prime}=m}^{1}
       {\rm e}^{-{\rm i}\theta_1^{(m^{\prime})} {\bm V}_1}
       {\rm e}^{-{\rm i}\theta_2^{(m^{\prime})} {\bm V}_2}
       \mbox{\boldmath{$\Psi$}}_{\rm i}
       \label{eq:phim}
\end{equation}
for $m=0,1,2,\cdots, M$, where the variational parameters 
$\mbox{\boldmath{$\theta$}} = \{ \theta_1^{(1)}, \theta_2^{(1)}, \cdots, \theta_1^{(M)}, \theta_2^{(M)} \}$ 
are determined so as to minimize the variational energy for $\vert \psi_M (\mbox{\boldmath{$\theta$}}) \rangle$, 
and thus $\mbox{\boldmath{$\Phi$}}_M = \mbox{\boldmath{$\Psi$}}_M$. We also set that 
$\mbox{\boldmath{$\Phi$}}_0 = \mbox{\boldmath{$\Psi$}}_{\rm i}$. 
There are two elemental properties of $\mbox{\boldmath{$\Phi$}}_m$. First, 
the single-particle orbitals in $\mbox{\boldmath{$\Phi$}}_m$ are mutually orthonormalized. 
This is simply because of the consequence of the unitary evolution:
\begin{equation}
  \mbox{\boldmath{$\Phi$}}_m^{\dagger}
  \mbox{\boldmath{$\Phi$}}_m =
  \mbox{\boldmath{$\Psi$}}_{\rm i}^{\dagger}
  \mbox{\boldmath{$\Psi$}}_{\rm i} =
  {\bf I}_{N}. 
  \label{eq:psi:orth}
\end{equation}
Second, it is apparent by construction in Eq.~(\ref{eq:phim}) 
that, apart from the phase factor due to the boundary conditions (i.e., in the case of APBCs), 
a single-particle orbital in $\mbox{\boldmath{$\Phi$}}_m$ is transformed to other single-particle orbitals 
by the translation of two lattice spaces, i.e., 
$[\mbox{\boldmath{$\Phi$}}_m]_{x,n} =
[\mbox{\boldmath{$\Phi$}}_m]_{x\pm2,n\pm1}$, where 
$n=0$ and $N+1$ correspond to $N$ and $1$, respectively. 
Therefore, the single-particle orbitals in $\mbox{\boldmath{$\Phi$}}_m$ are associated with the Wannier orbitals with a unit 
cell of two lattice spaces.

Let us now introduce the spatial extent $d_m$ (in unit of lattice constant)
of a single-particle orbital in $ \mbox{\boldmath{$\Phi$}}_m$,  
i.e., $d_m$ being the number of consecutive nonzero elements in each column of $ \mbox{\boldmath{$\Phi$}}_m$.
It is obvious form Eq.~(\ref{eq:matint}) that $d_0=2$ for $ \mbox{\boldmath{$\Phi$}}_0=\mbox{\boldmath{$\Psi$}}_{\rm i}$. 
Without knowing the explicit values of the variational parameters $\mbox{\boldmath{$\theta$}}$, 
we can readily show that the spatial extent of a single-particle orbital increases by four each time applying matrices 
${\rm e}^{-{\rm i}\theta_2^{(m)}{\bm V}_2}$ and ${\rm e}^{-{\rm i}\theta_1^{(m)} {\bm V}_1}$ 
given in Eqs.~(\ref{eq:mat_v2}) and (\ref{eq:mat_v1}),
respectively, i.e., $d_m=d_{m-1}+4$. 
Therefore, the spatial extent of a single-particle orbital in $\mbox{\boldmath{$\Phi$}}_m$ is generally given as 
$d_m = 4m + 2$ for our initial matrix $\mbox{\boldmath{$\Phi$}}_0 = \mbox{\boldmath{$\Psi$}}_{\rm i}$. 
Consequently, the spatial extent $d_m$ of a single-particle orbital in $\mbox{\boldmath{$\Phi$}}_m$
exceeds (reaches) the system size $L$ at $m = L/4$
[$m=(L-2)/4$] for APBCs (PBCs), where we choose $L=4n_L$ ($L=4n_L+2$) with $n_L$ integer. 
In other words, for the single-particle orbitals in $ \mbox{\boldmath{$\Psi$}}_M = \mbox{\boldmath{$\Phi$}}_M$ to 
extend over the entire system, the smallest number $M$ of layers in $\mbox{\boldmath{$\Psi$}}_M$ is
$L/4$ [$(L-2)/4$] for 
APBCs (PBCs), which is in good accordance with the results in Fig.~\ref{fig:energy} and the discussion in Sec.~\ref{sec:res:ene}. 
This is understood because the spatial extent $d_m$ of the single-particle orbitals essentially sets the limit of 
the propagation of quantum entanglement in the DQAP state.

\begin{figure}
  \includegraphics[width=1.05\columnwidth]{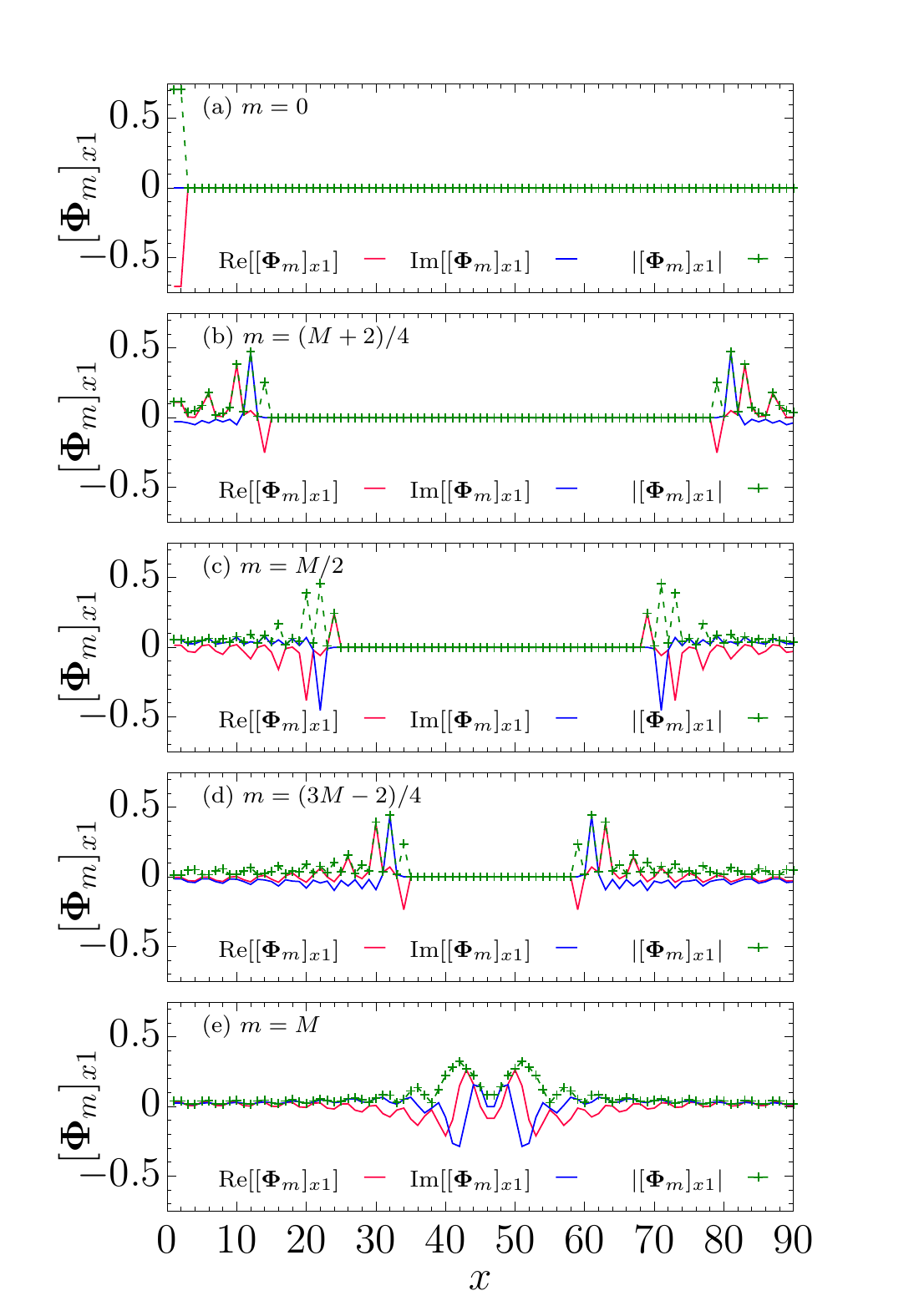}
  \caption{
  Time evolution of the first single-particle orbital $[ \mbox{\boldmath{$\Phi$}}_m ]_{x1}$ at 
    (a) $m=0$, (b) $m=(M+2)/4$, (c) $m=M/2$, (d) $m=(3M-2)/4$, and (e) $m=M$. 
    The DQAP ansatz $\vert \psi_M (\mbox{\boldmath{$\theta$}})\rangle$ is optimized 
    for $L=90$ and $N=45$ under PBC with $M=(L-2)/4$, 
    thus representing the exact ground state. 
  }
  \label{fig:orbital}
\end{figure}

Figure~\ref{fig:orbital} shows the numerical results of the time evolution of a single-particle orbital 
in  $\mbox{\boldmath{$\Psi$}}_M$ of the DQAP ansatz $\vert \psi_M (\mbox{\boldmath{$\theta$}}) \rangle$, 
for which the variational parameters $\mbox{\boldmath{$\theta$}}$ are optimized 
for $L=90$ with $M=(L-2)/4$ under PBCs, thus representing the exact ground state. 
Initially, the single-particle orbital is spatially localized at sites $x=1$ and 2, and 
propagates gradually in time (i.e., increasing $m$) by splitting a wave into the two opposite directions, 
finally reaching each other at $m=M$ when the spatial extent $d_m$ of the single-particle orbital 
becomes as large as the system size $L$. 

We should note here that the single-particle orbitals in $ \mbox{\boldmath{$\Psi$}}_M$  
are not uniquely determined. 
Instead, an $L\times L$ matrix $\mbox{\boldmath{$\Xi$}}_M$ given by
\begin{equation}
  \mbox{\boldmath{$\Xi$}}_M = \mbox{\boldmath{$\Psi$}}_M \mbox{\boldmath{$\Psi$}}_M^{\dagger}
\end{equation}
is invariant for all sets of single-particle orbitals which represent the exact ground state.
The expectation value of any physical operator evaluated for 
$\vert \psi_M (\mbox{\boldmath{$\theta$}}) \rangle$ is the same, despite that 
$ \mbox{\boldmath{$\Psi$}}_M$ is not uniquely determined, as long as $\mbox{\boldmath{$\Xi$}}_M$ 
is the same for different $\mbox{\boldmath{$\Psi$}}_M$. This can be easily proved from Eq.~(\ref{eq:ff:expectation}) 
because the single-particle orbitals are orthonormalized, i.e., 
$\mbox{\boldmath{$\Psi$}}_M^{\dagger} \mbox{\boldmath{$\Psi$}}_M = {\bf I}_{N}$.
It is also apparent that $\mbox{\boldmath{$\Xi$}}_M$ is invariant under the transformation
\begin{equation}
  \mbox{\boldmath{$\Psi$}}_M \to \mbox{\boldmath{$\Psi$}}_M^{\prime} = \mbox{\boldmath{$\Psi$}}_M {\bm Q}, 
  \label{eq:untq}
\end{equation}
where ${\bm Q}$ is an $N \times N$ unitary matrix.
Starting with different initial variational parameters, the numerical optimization of the variational parameters in 
$\vert \psi_M (\mbox{\boldmath{$\theta$}}) \rangle$ 
might find different sets of optimized variational parameters and thus 
different $\mbox{\boldmath{$\Psi$}}_M$'s.  
We indeed obtain several sets of single-particle orbitals with different single-particle orbital shapes, which 
nonetheless constitute the exact ground state, and all of them give the same value of $\mbox{\boldmath{$\Xi$}}_M$. 
However, we note that all these sets of single-particle orbitals are time evolved 
as those shown in Fig.~\ref{fig:orbital}, and they extend over the entire system at $m=M=(L-2)/4$ for PBCs.

We shall now consider the number of independent matrix elements 
in an $L\times N$ complex matrix $\mbox{\boldmath{$\Psi$}}$ when the exact ground state is constructed in the form   
$\vert \psi \rangle = \prod_{n=1}^N [ \hat{\bm c}^{\dagger} \mbox{\boldmath{$\Psi$}} ]_{n} \vert \psi_{\rm i} \rangle$ 
where $|\psi_{\rm i}\rangle$ is given in Eq.~(\ref{eq:1d:init}). 
To be specific, we assume that $L=4n_L+2$ ($n_L$: integer) with PBCs at half filling, i.e., $N=L/2$. 
We should first recall that $ [\mbox{\boldmath{$\Psi$}} ]_{x,n}$ represents the
$n$th single-particle orbital at site $x$.  
Because a single-particle orbital can be mapped to other single-particle orbitals by the translation of two lattice spaces, 
i.e., $[\mbox{\boldmath{$\Psi$}} ]_{x,n} = [\mbox{\boldmath{$\Psi$}} ]_{x\pm2,n\pm1}$, 
there are $L$ independent complex elements in $\mbox{\boldmath{$\Psi$}}$. 
In addition, there exists the reflection symmetry at the center of bond, i.e., 
$[\mbox{\boldmath{$\Psi$}} ]_{x,n} = [\mbox{\boldmath{$\Psi$}} ]_{-x+4n-1,n}$, 
which reduces the number of independent complex elements in $\mbox{\boldmath{$\Psi$}}$ down to $L/2$. 
Furthermore, the orthonormality of the single-particle orbitals, i.e., 
$\mbox{\boldmath{$\Psi$}}^{\dagger}\mbox{\boldmath{$\Psi$}} = {\bf I}_{N}$, 
yields $n_L+1$ independent equations and thus there are $3n_L+1$ independent {\it real} elements in $\mbox{\boldmath{$\Psi$}}$.

Next, we shall consider the transformation of $\mbox{\boldmath{$\Psi$}}$ by ${\bm Q}$, i.e., 
$\mbox{\boldmath{$\Psi$}} \to \mbox{\boldmath{$\Psi$}}^{\prime} = \mbox{\boldmath{$\Psi$}} {\bm Q}$, 
as discussed above in Eq.~(\ref{eq:untq}). 
Assuming that $\mbox{\boldmath{$\Psi$}}^{\prime}$ has the same translational and reflection symmetries 
as in $\mbox{\boldmath{$\Psi$}}$, 
we can show that the matrix elements of ${\bm Q}$ are also related to each other, 
similarly to the matrix elements of $\mbox{\boldmath{$\Psi$}}$. 
Thus, the independent complex matrix elements in ${\bm Q}$ is $n_L+1$. 
In addition, the unitarity of ${\bm Q}$ yields $n_L+1$ independent equations 
and therefore there are $n_L+1$ independent {\it real} elements in ${\bm Q}$. 
This suggests that, among $3n_L+1$ independent real elements in $\mbox{\boldmath{$\Psi$}}$,
$n_L+1$ elements are redundant. 
Therefore, there are $2n_L = (L-2)/2$ independent real elements that represent $\mbox{\boldmath{$\Psi$}}$. 
It is interesting to note that this number coincides with the number of the variational parameters 
$\mbox{\boldmath{$\theta$}} = \{ \theta_1^{(1)}, \theta_2^{(1)}, \cdots, \theta_1^{(M)}, \theta_2^{(M)} \}$ in 
the DQAP ansatz $\vert \psi_M (\mbox{\boldmath{$\theta$}}) \rangle$ with $M=(L-2)/4$, 
which corresponds to the shorted possible depth of $\vert \psi_M (\mbox{\boldmath{$\theta$}}) \rangle$ to represent 
the exact ground state, as discussed in Sec.~\ref{sec:res:ene}. 

In the case of APBCs with $L=4n_L$, exactly the same argument follows except that 
now there are $n_L$ independent equations 
generated due to the orthonormality of the single-particle orbitals in  
$\mbox{\boldmath{$\Psi$}} $ and the unitary matrix ${\bm Q}$ contains $n_L$ independent real 
elements. Therefore, there are $2n_L=L/2$ independent real elements in $\mbox{\boldmath{$\Psi$}}$. 
This number also coincides with the minimum number of variational parameters $\mbox{\boldmath{$\theta$}}$ in 
the DQAP ansatz $\vert \psi_M (\mbox{\boldmath{$\theta$}}) \rangle$ with $M=L/4$ that can represent the exact ground state.

\subsection{\label{sec:res:eent}Entanglement entropy}

\begin{figure}[htbp]
  \includegraphics[width=\hsize]{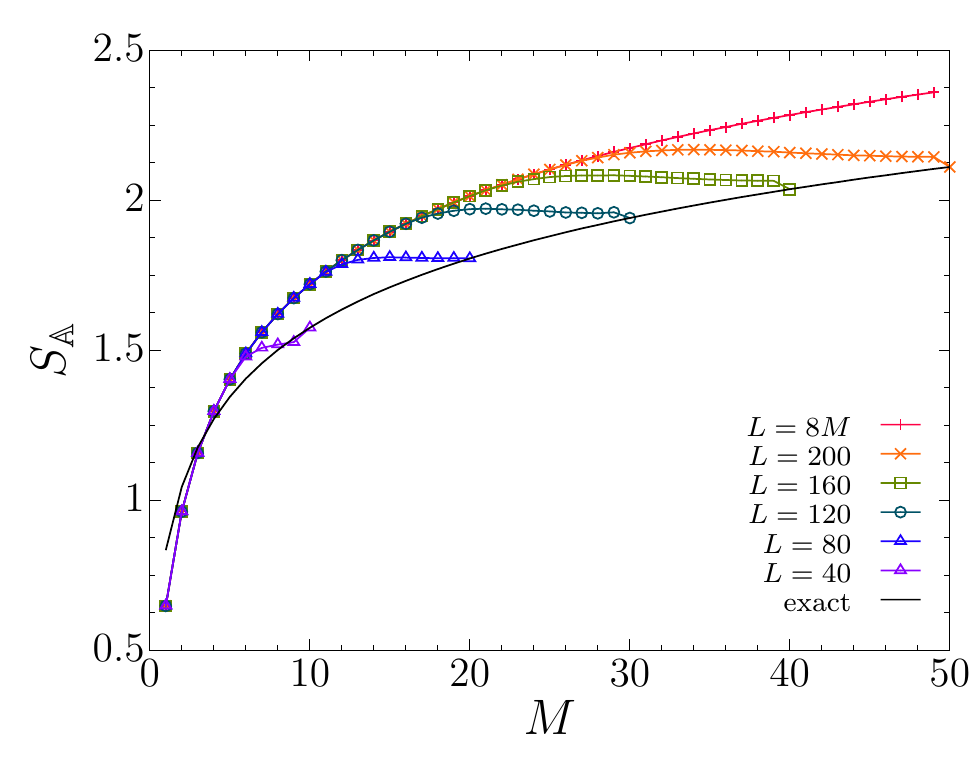}
  \caption{
    Entanglement entropy $S_{\mathbb{A}}$ as a function of the number $M$ of layers in the DQAP ansatz 
    $\vert \psi_M (\mbox{\boldmath{$\theta$}}) \rangle$ for different system sizes $L$ under APBCs at half filling. 
    The bipartition is assumed to be the half of the system.
    For comparison, the results for the case where $L$ and $M$ are both varied with keeping the ratio 
    $L=8M$ (i.e., $L_A=4M$) are also shown.
    Black solid line indicates the entanglement entropy of the exact ground state 
    for the system size $L$, which is plotted at $M=L/4$. }
  \label{fig:ent}
\end{figure}

\begin{figure}[htbp]
  \includegraphics[width=0.8\hsize]{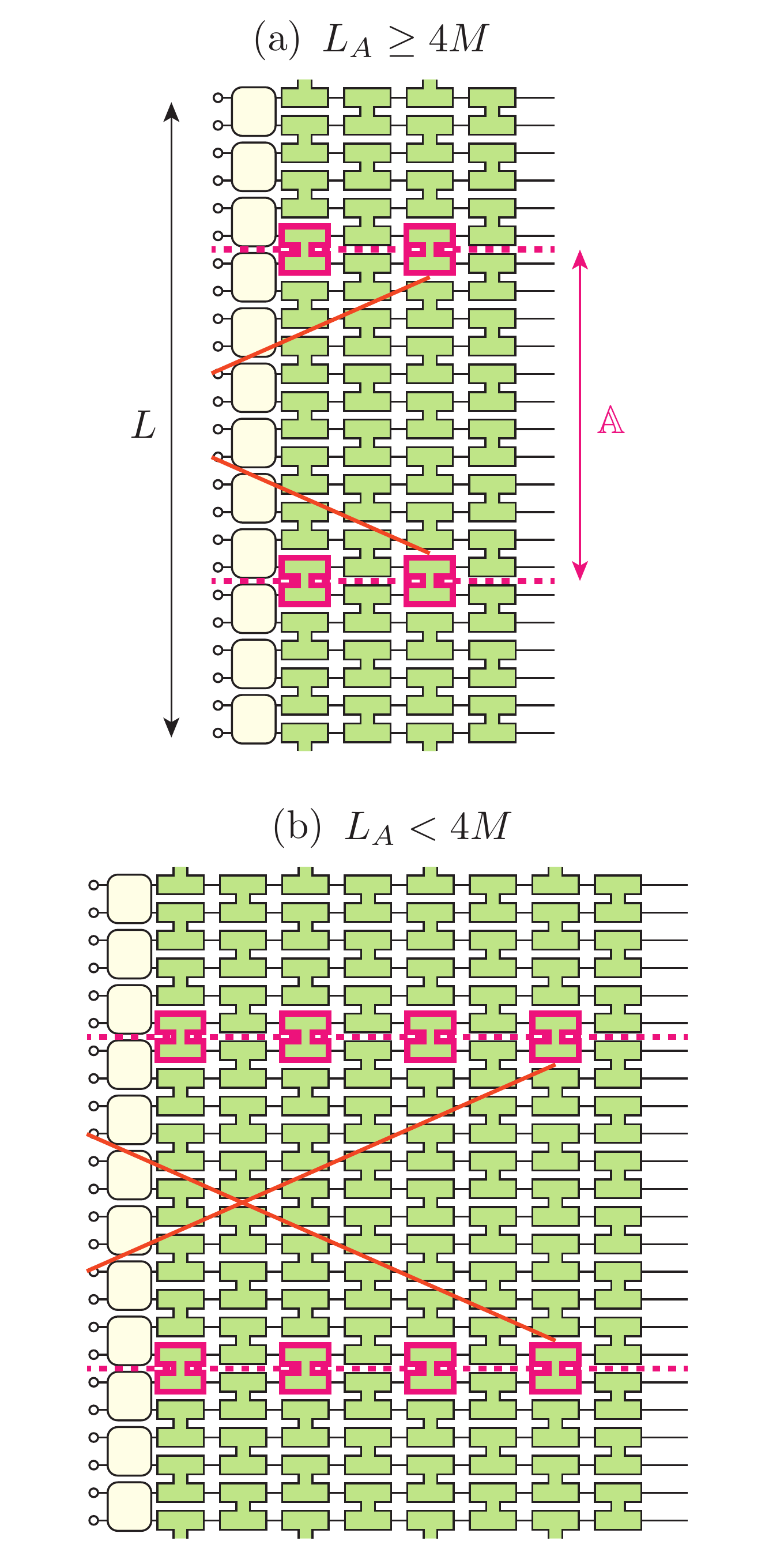}
  \caption{
    Two cases of partitioning the system: (a) 
    subsystem $\mathbb{A}$ is larger than or equal to $4M$ and
    (b) subsystem $\mathbb{A}$ is smaller than $4M$.
    Red dashed lines indicate the partitioning boundaries of the two subsystems,
    and green blocks with red outer frame indicate the local time-evolution
    operators crossing the partitioning boundaries. Orange solid lines denote halves of  
    the causality cones, centered at each partitioning boundary, within which 
    the boundary effect can be propagated via the local time-evolution
    operators. The difference between cases (a) and (b)
    is captured by whether or not these two causality cones overlap.
    In the examples shown here, $L=24$, $L_A=12$, and $M=2$ in (a) and $M=4$ in (b).
  }
  \label{fig:div}
\end{figure}

Next, we shall examine the entanglement property of the
DQAP ansatz $\vert \psi_M (\mbox{\boldmath{$\theta$}}) \rangle$.
Figure~\ref{fig:ent} shows the entanglement entropy $S_{\mathbb{A}}$
of the optimized DQAP ansatz $\vert \psi_M (\mbox{\boldmath{$\theta$}}) \rangle$
as a function of $M$ for several different system sizes $L$ under APBCs. 
Here, the variational parameters $\mbox{\boldmath{$\theta$}}$ are optimized 
for each $M$, and the bipartition is assumed to be half of the system,
namely,
\begin{equation}
  \mathbb{A} = \{ 1,2,\cdots, L/2 \}, 
\end{equation}
with the size $L_A$ of subsystem $\mathbb{A}$ being $L/2$. 
For the bipartitioning, we consider only the case  
where the system is divided into the two subsystems by not breaking any local bonding state 
in the initial state $\vert \psi_{\rm i}\rangle$, as shown schematically in Fig.~\ref{fig:div}.

As shown in Fig.~\ref{fig:ent}, we find that the entanglement entropy $S_{\mathbb{A}}$ for $4M \leq L_A$
is independent of the system size $L$ and falls into a smooth ``universal" curve of $M$.  
On the other hand, the entanglement entropy $S_{\mathbb{A}}$ starts to deviate from this universal curve for $4M > L_A$ 
and approaches the exact value at $M=L/4$ for APBCs [$M=(L-2)/4$ for PBCs]. 
These features are captured schematically in Fig.~\ref{fig:div}. 
The partitioning effect can propagate via the local time-evolution operators into the inside of subsystem $\mathbb{A}$ 
up to $2M$ lattice spaces (also taking into account the entanglement of a local bonding state in $|\psi_{\rm i}\rangle$) 
from each boundary of the partitioning, and thus this maximum propagation limit forms a causality-cone line structure centered at each 
partitioning boundary (see Fig.~\ref{fig:div}). 
The two causality cones cross each other when $4M > L_A$, and this is when  
the entanglement entropy $S_{\mathbb{A}}$ deviates from the universal curve found in Fig.~\ref{fig:ent}.

Let us discuss the results for $4M \le L_A$. 
In this case, we find that the entanglement entropy $S_{\mathbb{A}}$ 
is exactly the same for different system sizes $L$ and thus different sizes $L_A$ of subsystem $\mathbb{A}$. 
This implies that the entanglement entropy $S_{\mathbb{A}}$ is independent of the size $L_A$ of subsystem $\mathbb{A}$, 
as long as the partitioning boundaries are separated long enough. 
In other words, the entanglement carried by the DQAP ansatz $\vert \psi_M (\mbox{\boldmath{$\theta$}}) \rangle$ 
with a finite $M$ is bounded, as in the matrix product states with a finite bond dimension~\cite{Nishino1996,Pollmann2009,Pirvu2012}.

So far, we have assumed that the size $L_{\bar A}$ of the complement $\bar{\mathbb{A}}$ 
of subsystem $\mathbb{A}$ is the same as the size $L_A$ of subsystem $\mathbb{A}$. 
However, we should note that the results of the entanglement entropy $S_{\mathbb{A}}$ for $4M\le L_A$ shown in Fig.~\ref{fig:ent}
remain exactly the same even when we enlarge the size of $\bar{\mathbb{A}}$, provided that $L_A \le L_{\bar A}$. 
Thus, the entanglement entropy $S_{\mathbb{A}}$ for $4M\le L_A$ and $L_A \le L_{\bar A}$ 
is determined solely by the number $M$ of layers in the DQAP ansatz $\vert \psi_M (\mbox{\boldmath{$\theta$}}) \rangle$. 
Note also that, considering $S_{\mathbb{A}}=S_{\bar{\mathbb{A}}}$, the smaller subsystem determines the value 
of $M$ until which the entanglement entropy follows the universal curve.

We have performed similar calculations for the systems under PBCs and found that, independently of the system size $L$, 
the entanglement entropy $S_{\mathbb{A}}$
of the optimized DQAP ansatz $\vert \psi_M (\mbox{\boldmath{$\theta$}}) \rangle$ 
for the system 
under PBCs is exactly the same as that 
of the optimized DQAP ansatz $\vert \psi_M (\mbox{\boldmath{$\theta$}}) \rangle$ 
for the system under APBCs, provided that (i) $4M\le L_A$, (ii) $L_A\le L_{\bar A}$, and 
(iii) the partitioning boundaries do not break any local bonding state in the initial state $|\psi_{\rm i}\rangle$. 
Namely, the entanglement entropy $S_{\mathbb{A}}$ falls into the universal curve of $M$ shown in Fig.~\ref{fig:ent},  
independent of the system size and the boundary conditions, as long as the three conditions above are satisfied. 
This is similar to the observation of the optimized variational energy $E_M(L)$, where 
$\Delta \varepsilon = E_M(L)/L - \varepsilon_{\infty}$ falls into the universal curve of $M$, as shown in Fig.~\ref{fig:energy}(b), 
independent of system size and boundary conditions, as long as $M< \lceil (L-2)/4\rceil$.

Let us now explore how these two quantities 
approach asymptotically in the limit of $M\to \infty$, 
which thus requires us to take the limit of $L\to\infty$
as well under the condition that $4M\le L_A$ or $M< \lceil (L-2)/4\rceil$. 
To this end, here we calculate the exponents $\delta_{S}(M)$ and $\delta_{E}(M)$ by the following formulas:
\begin{equation}
  \delta_{S}(M) = 3\frac{  S_{M+1} - S_{M} }{\ln (M+1) - \ln (M)}
  \label{eq:exponent:entropy}
\end{equation}
and
\begin{equation}
  \delta_{E}(M) = \frac{1}{2} \frac{\ln \Delta \varepsilon_{M+1} - \ln \Delta \varepsilon_{M}}{\ln (M+1) - \ln (M)}.
  \label{eq:exponent:energy}
\end{equation}
Here, $S_{M}$ is the entanglement entropy $S_{\mathbb{A}}$ of the optimized DQAP ansatz 
$\vert \psi_M(\mbox{\boldmath{$\theta$}}) \rangle$ with $4M \leq L_A$ and $L_A \leq L_{\bar A}$, 
and $\Delta \varepsilon_{M}=E_M(L)/L - \varepsilon_{\infty} $ is the energy difference from the exact ground-state energy 
$\varepsilon_{\infty}$ per site in the thermodynamic limit, where the variational energy $E_M(L)$ is evaluated for 
the optimized DQAP ansatz $\vert \psi_M(\mbox{\boldmath{$\theta$}}) \rangle$ with
$M < L/4$ under APBCs.

It is known~\cite{Calabrese_2004} that the entanglement entropy $S_{\rm exact}(L_A)$ of the exact ground state 
for the size $L_A$ of subsystem $\mathbb{A}$ is given as 
\begin{equation}
  S_{\rm exact}(L_A) \approx \frac{1}{3} \ln L_A. \label{eq:exact:entropy}
\end{equation}
Equation~(\ref{eq:exponent:entropy}) is motivated by the assumption
that $L_A$ is replaced as $L_A \sim M^{\delta_S}$,
indicating that a finite $M$ introduces a finite correlation length, 
as does a finite bond dimension in the matrix product states~\cite{Nishino1996,Pollmann2009,Pirvu2012}.
Similarly, it is also known~\cite{Bioete_1986,Affleck_1986,Granet_2019} that the finite size correction
to the exact ground-state energy per site is given as 
\begin{equation}
  \Delta \varepsilon_{\rm exact} (L) = \frac{C}{L^2} + O(L^{-4}), \label{eq:exact:energy}
\end{equation}
where $C$ is a system size independent constant and $\Delta \varepsilon_{\rm exact}(L)$ is the energy difference
between the exact ground-state energy per site for the system size $L$ and
that in the thermodynamic limit. 
Equation~(\ref{eq:exponent:energy}) is thus motivated 
by assuming that $L \sim M^{\delta_E}$.

\begin{figure}[htbp]
  \includegraphics[width=\hsize]{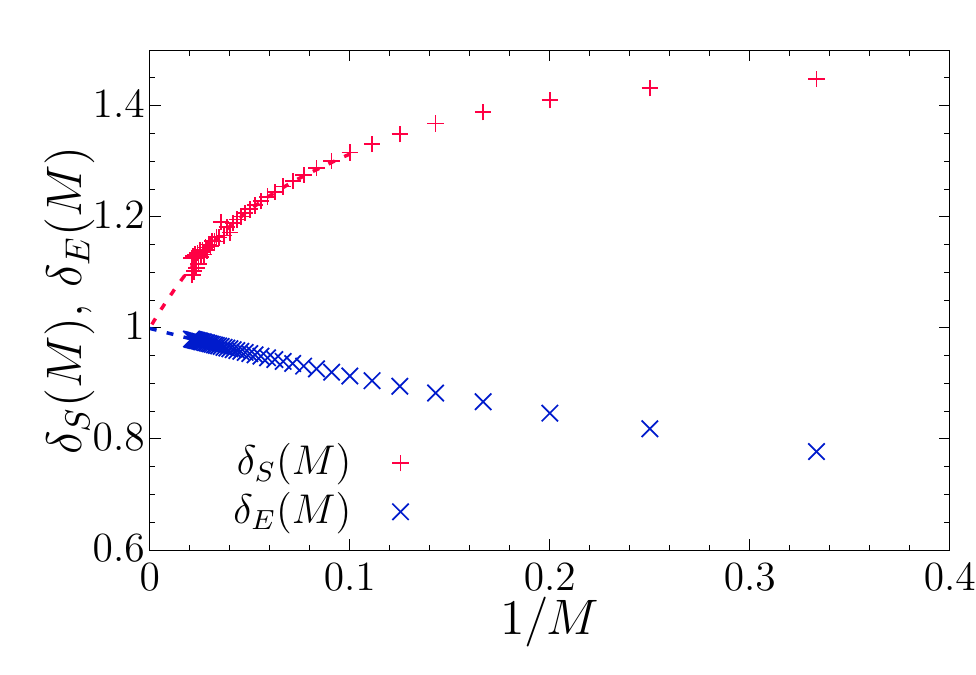}
  \caption{
  Exponents $\delta_S(M)$ and $\delta_E(M)$ for the optimized DQAP ansatz
  $\vert \psi_M (\mbox{\boldmath{$\theta$}}) \rangle$ with $4M \leq L_A$ and $L_A \leq L_{\bar A}$ 
  for $\delta_S(M)$ and with $M<L/4$ for $\delta_E(M)$.
  The dashed lines are guide for the eye.
  }
  \label{fig:exponent}
\end{figure}

We find in Fig.~\ref{fig:exponent} that
these two exponents $\delta_S(M)$ and $\delta_E(M)$
approach one in the limit of $M \to \infty$. 
Thus, the asymptotic behaviors of the entanglement entropy $S_{\mathbb{A}}$ and 
the energy deviation of the variational energy $E_M(L)/L$  
can be simply described by the expressions in Eqs.~(\ref{eq:exact:entropy}) and (\ref{eq:exact:energy}), respectively,  
with the $L_A$ and $L$ dependence replaced by $M$, i.e., 
\begin{equation}
S_M\approx \frac{1}{3}\ln M
\end{equation}
and
\begin{equation}
\Delta\varepsilon_{M} \sim M^{-2}.
\end{equation}

Finally, we discuss the relation to the evolution of the
single-particle orbitals in $\vert \psi_M (\mbox{\boldmath{$\theta$}}) \rangle$ 
via the local time-evolution operators. 
As discussed in Sec.~\ref{sec:res:psi}, the spatial extent $d_M$ of a single-particle orbital in the DQAP ansatz 
$\vert \psi_M (\mbox{\boldmath{$\theta$}}) \rangle$ with $M$ layers of the local time-evolution operators 
is $d_M = 4M + 2$. 
Therefore, when $4M \leq L_A$, for which we find that the entanglement entropy $S_{\mathbb{A}}$ of 
$\vert \psi_M (\mbox{\boldmath{$\theta$}}) \rangle$ is independent of $L$ (see Fig.~\ref{fig:ent}), 
a single-particle orbital in $\vert \psi_M (\mbox{\boldmath{$\theta$}}) \rangle$ 
may cross one of the partitioning boundaries, but 
it cannot cross the other partitioning boundary. 
Conversely, when $4M > L_A$,
a single-particle orbital (but not necessarily all the single-particle orbitals)
can cross both partitioning boundaries of the subsystems. 
Since the entanglement entropy of the free-fermion system
is determined by the hybridization between the two subsystems, 
the observation here suggests that 
the contribution of each partitioning boundary 
to the entanglement entropy $S_{\mathbb{A}}$ 
is indeed separable in the case for $4M \leq L_A$, 
i.e., $S_{\mathbb{A}}\sim S_{\partial\mathbb{A}_{\rm I}} + S_{\partial\mathbb{A}_{\rm II}}$, where 
$\partial\mathbb{A}_{\rm I (II)}$ is the partitioning boundary and 
$S_{\partial\mathbb{A}_{\rm I (II)}}$ implies the entanglement entropy from the boundary $\partial\mathbb{A}_{\rm I (II)}$ 
(see Appendix~\ref{sec:separation}).

\subsection{\label{sec:res:minf}Mutual information}

\begin{figure}
  \includegraphics[width=\hsize]{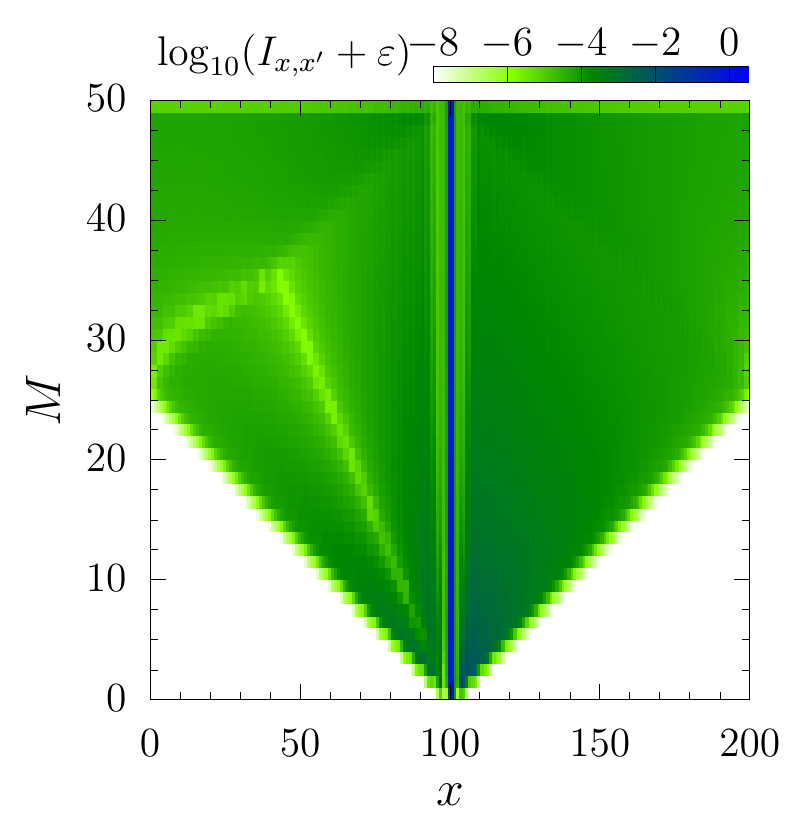}
  \caption{Intensity plot of mutual information $I_{x,x^{\prime}}$ of 
  the DQAP ansatz $\vert \psi_M (\mbox{\boldmath{$\theta$}}) \rangle$ with 
  the variational parameters $\mbox{\boldmath{$\theta$}}$ optimized for each $M$ 
  to minimize the variational energy. 
  The calculations are for $L=200$ under APBCs at half filling i.e, $N=L/2$. 
  We set $x^\prime = L/2$ in $I_{x,x^{\prime}}$. 
  For clarity, we add a small constant $\varepsilon = 10^{-8}$ in drawing the intensity plot 
  of $\log_{10}(I_{x,x'}+\varepsilon)$.
  }
  \label{fig:minf}
\end{figure}

We shall now examine the evolution of the mutual information 
$I_{x,x^{\prime}}$ defined in Eq.~(\ref{eq:2p:minf}) 
for the DQAP ansatz $\vert \psi_M (\mbox{\boldmath{$\theta$}}) \rangle$ 
with increasing the number $M$ of layers of the local time-evolution operators. 
As discussed in Sec.~\ref{sec:ent}, $I_{x,x^{\prime}}$ is a measure to quantify the entanglement 
between sites $x$ and $x^{\prime}$ for a quantum state. 
Figure \ref{fig:minf} shows the results for the DQAP ansatz $\vert \psi_M (\mbox{\boldmath{$\theta$}}) \rangle$ 
with the variational parameters $\mbox{\boldmath{$\theta$}}$ 
optimized for each $M$ to minimize the variational energy.

\begin{figure}
  \includegraphics[width=\hsize]{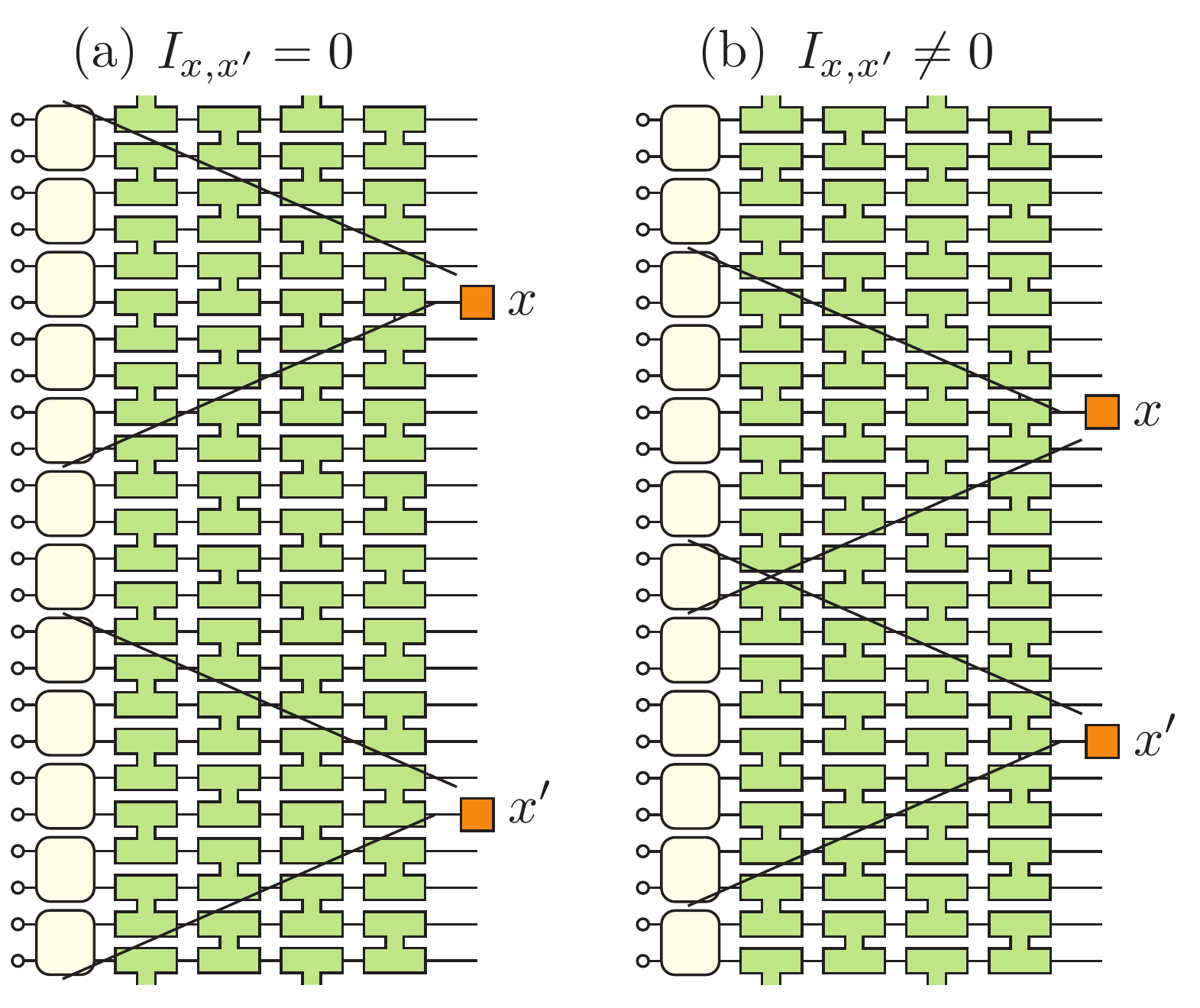}
  \caption{Schematic representations of two cases where the DQAP ansatz gives 
    (a) $I_{x,x^{\prime}}=0$, i.e., no entanglement between sites $x$ and $x^\prime$, 
    and (b) $I_{x,x^{\prime}} \neq 0$, i.e., quantum entanglement developed between sites $x$ and $x^\prime$. 
    Black sold lines indicate causality cones within which quantum entanglement can be propagated  
    via the local time-evolution operators from sites $x$ and $x^\prime$.
    }
  \label{fig:minf:overlap}
\end{figure}

We find in Fig.~\ref{fig:minf} that the mutual information
$I_{x,x^{\prime}}$ is exactly zero, implying no entanglement, when 
$ \vert x - x^{\prime} \vert > 4M + 1$. This is generally the case for any system size $L$. 
As illustrated in Fig.~\ref{fig:minf:overlap}, 
this entanglement feature reflects the causality-cone-like structure of the propagation of quantum entanglement 
via the local time-evolution operators in the DQAP ansatz, which limits the propagation speed  
set by the Lieb-Robinson bound.  
Two causality cones for the propagation of quantum entanglement from sites $x$ and $x^\prime$ are indicated 
in Fig.~\ref{fig:minf:overlap}. All the local unitary time-evolution operators inside the causality cones are connected 
to the origin of the cone (i.e., $x$ or $x^\prime$), while those outside the causality cones are essentially disconnected.  
When these two causality cones 
do not overlap to each other, the mutual information $I_{x,x^{\prime}}$ is zero. 
On the other hand, if these two causality cones overlap, we obtain $I_{x,x^{\prime}} \neq 0$. 
We should also note that although the mutual information $I_{x,x^{\prime}}$ becomes nonzero 
for all values of $x$ and $x'$
once the number $M$ of layers of the local time-evolution operators satisfies $L/2 \le 4M+1$ (see Fig.~\ref{fig:minf}), 
more layers of the local time-evolution operators are required 
for the DQAP ansatz $\vert \psi_M (\mbox{\boldmath{$\theta$}}) \rangle$ to represent the exact ground state of the system, 
as discussed in Sec.~\ref{sec:res:ene}.

The feature of the mutual information found here can also be understood on the basis of the single-particle orbitals 
in the DQAP ansatz $\vert \psi_M (\mbox{\boldmath{$\theta$}}) \rangle$ discussed in Sec.~\ref{sec:res:psi}. 
As already described in Sec.~\ref{sec:ent},
the mutual information $I_{x,x^{\prime}}$ for the free-fermion system 
is fully determined by the one-particle density matrix ${\bm D}_{ \{x,x^{\prime}\} }$
given in Eq.~(\ref{eq:2p:dmat}). 
Since $\langle \psi_M (\mbox{\boldmath{$\theta$}}) \vert \hat{c}_x^{\dagger} \hat{c}_x \vert \psi_M (\mbox{\boldmath{$\theta$}}) \rangle = 0.5$ 
in our system, $I_{x,x^{\prime}}$ is determined solely by the off-diagonal element 
$\langle \psi_M (\mbox{\boldmath{$\theta$}}) \vert \hat{c}_{x}^{\dagger} \hat{c}_{x^{\prime}} \vert \psi_M (\mbox{\boldmath{$\theta$}}) \rangle$. 
For example, as discussed in Sec.~\ref{sec:ent}, $I_{x,x^{\prime}} = 0$ when  
$\langle \psi_M (\mbox{\boldmath{$\theta$}}) \vert \hat{c}_{x}^{\dagger} \hat{c}_{x^{\prime}} \vert \psi_M (\mbox{\boldmath{$\theta$}}) \rangle=0$.

Let us now evaluate $\langle \psi_M (\mbox{\boldmath{$\theta$}}) \vert \hat{c}_{x}^{\dagger} \hat{c}_{x^{\prime}} \vert \psi_M (\mbox{\boldmath{$\theta$}}) \rangle$ using Eq.~(\ref{eq:ff:expectation}). 
Considering that $\mbox{\boldmath{$\Psi$}}_M^{\dagger} \mbox{\boldmath{$\Psi$}}_M =  {\bf I}_{N}$,
we obtain that 
\begin{equation}
  \langle \psi_M(\mbox{\boldmath{$\theta$}}) \vert
  \hat{c}_{x}^{\dagger} \hat{c}_{x^{\prime}} \vert
  \psi_M (\mbox{\boldmath{$\theta$}}) \rangle
  =
  \sum_{n=1}^N
  [ \mbox{\boldmath{$\Psi$}}_M ]_{xn}^{\ast}
  [ \mbox{\boldmath{$\Psi$}}_M ]_{x^{\prime}n}.
  \label{eq:explicit:cdagc}
\end{equation}
Because of the construction of $\vert \psi_M (\mbox{\boldmath{$\theta$}}) \rangle$ described in Sec.~\ref{sec:res:psi}, 
the $n$th single-particle orbital $[ \mbox{\boldmath{$\Psi$}}_M ]_{xn}$ in $\vert \psi_M (\mbox{\boldmath{$\theta$}}) \rangle$ 
is finite only in the region of sites $x$ where $2n-2M-1 \leq x \leq 2n+2M$. 
Here, site $2n - 2M-1$ ($2n + 2M$) should be read as 
${\rm mod}(2n-2M-2, L)+1$ [${\rm mod}(2n+2M-1, L)+1$]. 
Therefore, 
$[ \mbox{\boldmath{$\Psi$}}_M ]_{xn}^{\ast}
[ \mbox{\boldmath{$\Psi$}}_M ]_{x^{\prime}n} = 0$ 
when $\vert x-x^\prime \vert > 4M+1$,  
which thus also explains that $I_{x,x^{\prime}}=0$ when $\vert x-x^{\prime} \vert > 4M+1$ found in Fig.~\ref{fig:minf}. 
Although $I_{x,x^{\prime}}$ becomes finite for all distances $\vert x-x^{\prime} \vert$ when 
$M$ satisfies $4M\ge L/2-1$,  
it is not sufficient for the DQAP ansatz $\vert \psi_M (\mbox{\boldmath{$\theta$}}) \rangle$
to represent the exact ground state 
since the single-particle orbital $[ \mbox{\boldmath{$\Psi$}}_M ]_{xn}$ does not extend over the entire region of the system 
until $d_M=4M+2\ge L$.

\subsection{\label{sec:res:opt}Optimized variational parameters}

We shall now discuss the optimized variational parameters in the DQAP ansatz 
$\vert \psi_M (\mbox{\boldmath{$\theta$}}) \rangle$. 
The natural gradient method described in Sec.~\ref{sec:opt} is employed without any difficulty 
to optimize the variational parameters in the DQAP ansatz 
$\vert \psi_M (\mbox{\boldmath{$\theta$}}) \rangle$, 
which becomes the exact ground state of the system for $M = L/4$ under APBCs and $M = (L-2)/4$ under PBCs 
at half filling. 
However, we find that the optimized variational parameters are not unique and many different sets of optimized variational 
parameters give the same energy,   
as discussed in Appendix~\ref{sec:random}.

\begin{figure}
  \includegraphics[width=0.9\hsize]{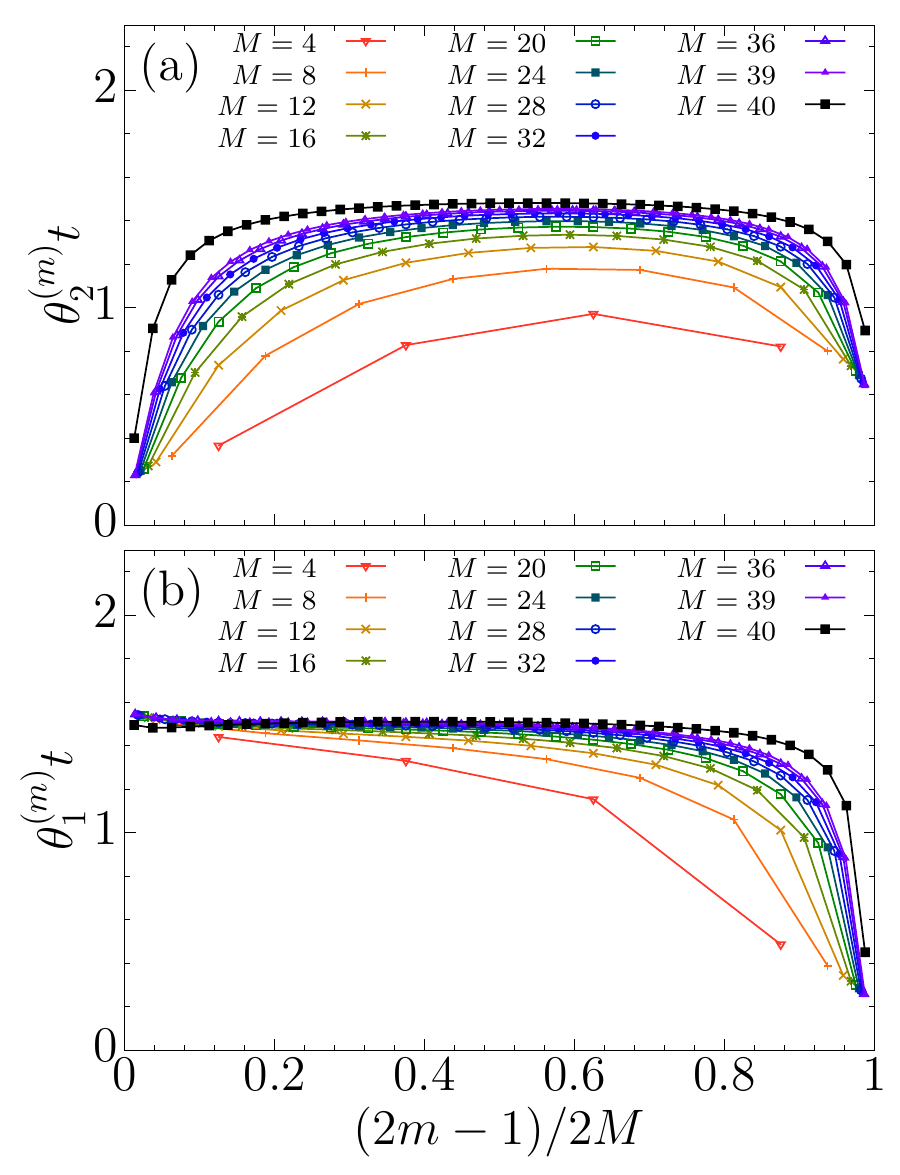}
  \caption{
  Optimized variational parameters $\mbox{\boldmath{$\theta$}} = \{ \theta_1^{(m)}, \theta_2^{(m)} \}$ 
  in the DQAP ansatz $\vert \psi_M (\mbox{\boldmath{$\theta$}}) \rangle$ for various values of $M$. 
  The variational  parameters are optimized for each $M$ to minimize the variational energy of the system 
  with $L=160$ under APBCs at half filling, i.e., $N=L/2$. In this case, the DQAP ansatz 
  $\vert \psi_M (\mbox{\boldmath{$\theta$}}) \rangle$ represents the exact ground state at $M=L/4$. 
  }
  \label{fig:prms:M}
\end{figure}

Among many sets of optimum solutions for the variational parameters, 
we find a series of systematic solutions by gradually increasing $M$ for a fixed system size $L$.
Such a series is obtained as follows.
We first start with a small value of $M$, for which the optimized variational parameters 
$\{ {\theta_p^{(1)}}^*, {\theta_p^{(2)}}^*, \cdots, {\theta_p^{(M)}}^* \}_M$
in $\vert \psi_M (\mbox{\boldmath{$\theta$}}) \rangle$ can be uniquely determined for $p=1,2$. 
Then, we use these optimized variational parameters as the initial parameters 
$\{ {\theta_p^{(1)}}, {\theta_p^{(2)}}, \cdots, {\theta_p^{(M+1)}} \}_{M+1}$
for $\vert \psi_{M+1} (\mbox{\boldmath{$\theta$}}) \rangle$, i.e., 
$\{ {\theta_p^{(1)}}, \cdots, {\theta_p^{(M/2)}}, {\theta_p^{(M/2+1)}}, {\theta_p^{(M/2)+2}},\cdots {\theta_p^{(M+1)}} \}_{M+1}
\leftarrow \{ {\theta_p^{(1)}}^{*}, \cdots, {\theta_p^{(M/2)}}^*, \frac{1}{2}({\theta_p^{(M/2)}}^{*}+{\theta_p^{(M/2+1)}}^{*}),
{\theta_p^{(M/2+1)}}^{*},\ \cdots, {\theta_p^{(M)}}^{*} \}_M$
and optimize the variational parameters in $\vert \psi_{M+1} (\mbox{\boldmath{$\theta$}}) \rangle$.
Here, we assumed that $M$ is even. When $M$ is odd, $M/2$ should be replaced with $(M-1)/2$. 
With iteratively increasing $M$ by one in this procedure, 
we finally obtain the series of the optimized variational parameters systematically, as shown in Fig.~\ref{fig:prms:M}.

The characteristic features  of the optimized variational parameters are summarized as follows. 
First, $\theta_2^{(m)}$ 
monotonically (rather almost linearly) increases with $m$, 
while $\theta_1^{(m)}$ remains almost constant, when $(2m-1)/2M$ is small. 
This dependence of the parameters on $m$ remarkably resembles 
the linear scheduling of the scheduling function $s_{\rm i}(\tau)$ and $s_{\rm f}(\tau)$ for 
the quantum adiabatic approximation given in Eq.~(\ref{eq:theta_lin}). 
Second, the optimized variational parameters $\theta_1^{(m)}$ and $\theta_2^{(m)}$ are both almost constant in the intermediate region of 
$(2m-1)/2M$. This is a part of the reason why the optimization procedure of the variational parameters described above is successful. 
Third, both parameters $\theta_1^{(m)}$ and $\theta_2^{(m)}$ finally decrease with increasing $m$ when 
$(2m-1)/2M$ approaches one. 
This might be understood because at the last stage of the process,
it would be better for the DQAP to be determined by the time-evolution operator 
${\rm e}^{-{\rm i}\hat{\mathcal{H}}_{\rm f}t}$ of the final system $\hat{\mathcal{H}}_{\rm f}$ as in the continuous time 
quantum adiabatic process. 
To this end, the parameters $\theta_1^{(m)}$ and $\theta_2^{(m)}$ should be small to reduce the Suzuki-Trotter decomposition 
error due to the discretization of time~\cite{Trotter1959,Suzuki1976}.

Notice also that there is an abrupt change of the optimized variational parameters between $M=L/4-1$ and $M=L/4$ 
(see the results for $M=39$ and $M=40$ in Fig.~\ref{fig:prms:M}). 
As already described in Sec.~\ref{sec:res:ene}, 
the optimized DQAP ansatz $\vert \psi_M (\mbox{\boldmath{$\theta$}}) \rangle$ with $M=L/4$ represents 
the exact ground state of the system under APBCs. 
This abrupt change of the optimized variational parameters is associated with that of the variational energy 
found in Fig.~\ref{fig:energy}. 
We also notice in Fig.~\ref{fig:prms:M} that the optimized variational parameters in 
$\vert \psi_M (\mbox{\boldmath{$\theta$}}) \rangle$ with $M<L/4$ for a given system size $L$ 
converges systematically to those with $M=L/4-1$ as $M$ increases, which are different 
from the optimized variational parameters in $\vert \psi_M (\mbox{\boldmath{$\theta$}}) \rangle$ with $M=L/4$. 
Furthermore, we find that the optimized variational parameters in $\vert \psi_M (\mbox{\boldmath{$\theta$}}) \rangle$ 
with the given number $M$ of layers
remain unchanged for different system sizes $L$, as long as $M<L/4$, which is associated with the observation 
that the variational energy per site, $E_M(L)/L$, is independent of $L$ when $M<L/4$, as shown in Fig.~\ref{fig:energy}(b). 
We should also note that the optimized variational parameters in $\vert \psi_M (\mbox{\boldmath{$\theta$}}) \rangle$ 
for the system under PBCs
are exactly the same as those in $\vert \psi_M (\mbox{\boldmath{$\theta$}}) \rangle$ for the system under APBCs,  
e.g., shown in Fig.~\ref{fig:prms:M}, 
independent of the system size $L$, 
as long as $M< \lceil (L-2)/4\rceil$ (also see Sec.~\ref{sec:res:eent}).

\begin{figure}
  \includegraphics[width=0.9\hsize]{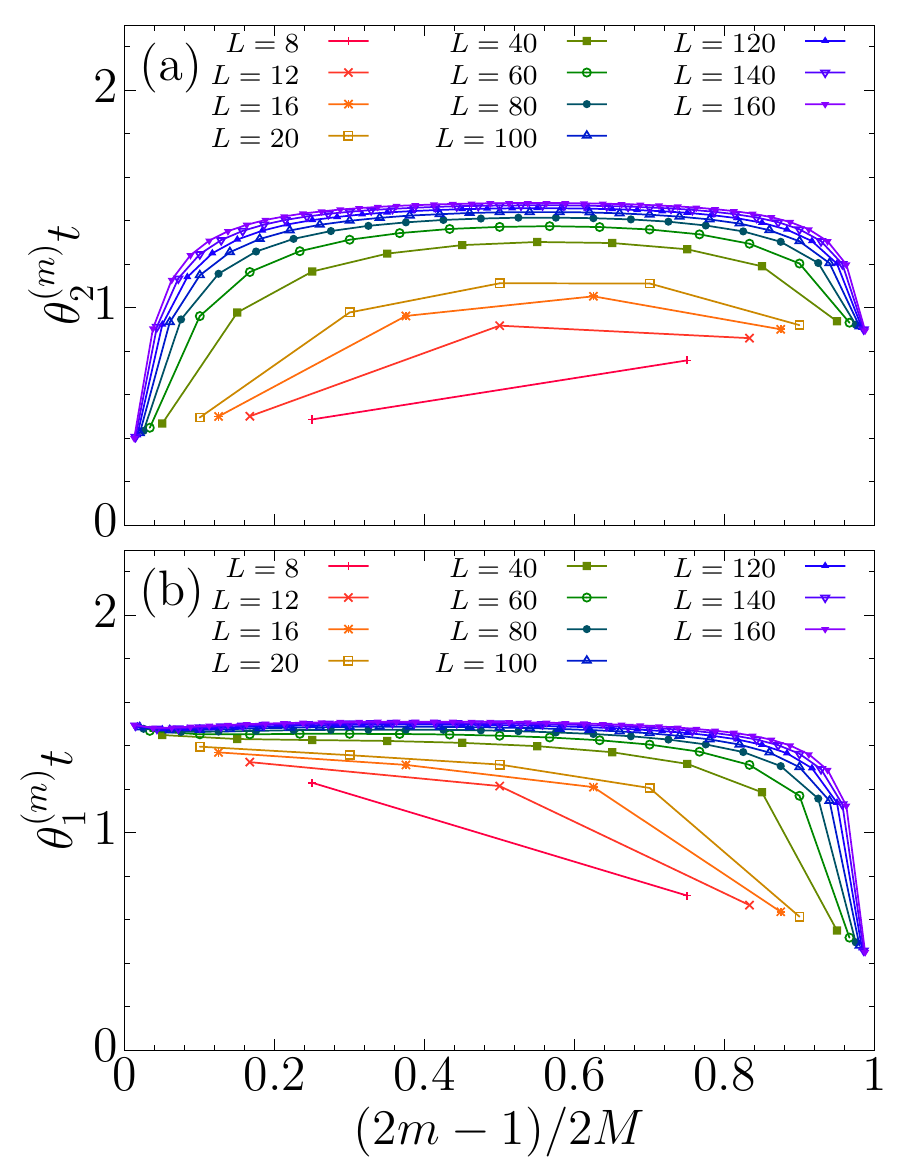}
  \caption{
  Optimized variational parameters $\mbox{\boldmath{$\theta$}} = \{ \theta_1^{(m)}, \theta_2^{(m)} \}$ 
  in the DQAP ansatz $\vert \psi_M (\mbox{\boldmath{$\theta$}}) \rangle$ with $M=L/4$ for various system sizes $L$. 
  The variational  parameters are optimized for each system size $L$ to minimize the variational energy 
  under APBCs at half filling, i.e., $N=L/2$, and thus  
  the DQAP ansatz $\vert \psi_M (\mbox{\boldmath{$\theta$}}) \rangle$ 
  here represents the exact ground state.  
  }
  \label{fig:prms:L}
\end{figure}

Figure~\ref{fig:prms:L} summarizes the optimized variational parameters in the DQAP ansatz 
$\vert \psi_M (\mbox{\boldmath{$\theta$}}) \rangle$ with $M = L/4$ for different system sizes $L$ under APBCs, 
for which $\vert \psi_M (\mbox{\boldmath{$\theta$}}) \rangle$ represents the exact ground state. 
We find that the optimized variational parameters 
$\mbox{\boldmath{$\theta$}} = \{ \theta_1^{(1)}, \theta_2^{(1)}, \cdots, \theta_1^{(M)}, \theta_2^{(M)} \}$ 
converge asymptotically to a smooth function of $m$ for each $\theta_p^{(m)}$ ($p=1,2$) with increasing the system size $L$.

\begin{figure}[htbp]
  \includegraphics[width=\hsize]{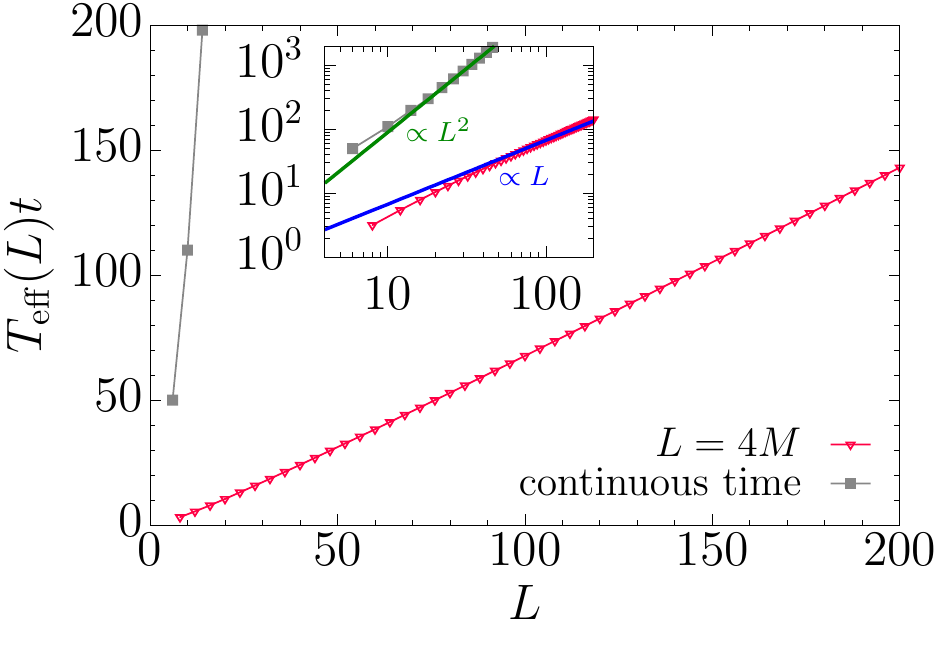}
  \caption{
  Effective total evolution time $T_{\rm eff}(L)$ of the DQAP in which the exact ground state of the final Hamiltonian 
  is obtained (red triangles). 
  The DQAP ansatz $\vert \psi_M(\mbox{\boldmath{$\theta$}})\rangle$ with $M=L/4$ is optimized for the system size $L$ 
  under APBCs, thus representing the exact ground state. 
  For comparison, 
  the total evolution time necessary to obtain the ground state of the final Hamiltonian within a given accuracy 
  in the continuous-time quantum adiabatic process with a linear scheduling (for details, see Appendix~\ref{sec:conttime}) is 
  also shown by gray squares. 
  Inset: Same plot but in the logarithmic scale. 
  For comparison, functions proportional to $L$ and $L^2$ are also plotted by blue and green solid lines, respectively.
  }
  \label{fig:times}
\end{figure}

Let us now
examine the effective total evolution time $T_{\rm eff}(L)$ of the DQAP given by 
\begin{equation}
  T_{\rm eff}(L) = \sum_{p=1}^2 \sum_{m=1}^M \theta_{p}^{(m)},
\end{equation}
where the variational parameters $\theta_p^{(m)}$ ($p=1,2$) in the DQAP ansatz 
$\vert \psi_M (\mbox{\boldmath{$\theta$}}) \rangle$ are optimized for the system size $L$ with $M=L/4$ under APBCs, 
thus representing the exact ground state, and are already shown in Fig.~\ref{fig:prms:L}. 
It is highly interesting to find in Fig.~\ref{fig:times} that the effective total evolution time $T_{\rm eff}(L)$ is 
almost perfectly proportional to the system size $L$. 
According to the quantum adiabatic theorem, 
the evolution time necessary to successfully converge to the ground state of the final Hamiltonian 
in the continuous time quantum adiabatic process 
is inversely proportional to the square of the minimum energy gap during the process~\cite{Born1928}. 
For the model studied here, the minimum gap appears at the final Hamiltonian,
i.e., the free-fermion model, and thus it is $\sim 1/L$,
suggesting that, according to the adiabatic theorem,
the evolution time to successfully obtain the final state within a given accuracy
is proportional to $L^2$, as shown in Fig.~\ref{fig:times} 
(and also see Fig.~\ref{fig:conttime} in Appendix~\ref{sec:conttime}).
The quadratic speed up found here in the DQAP ansatz resembles that
in the adiabatic quantum Grover search algorithm~\cite{Albash2018}
with the optimum scheduling function~\cite{Roland2002,Morita2008}.

  One of the methods to find a optimum scheduling function in the continuous-time quantum adiabatic process 
  is the QAB~\cite{Rezakhani2009},
  in which an optimum path of the quantum adiabatic process is determined 
  by solving the Euler-Lagrange equation derived so as to minimize 
  the total transition probability during the evolution. The outline of this theory is described 
  in Appendix~\ref{sec:qab}. 
  To apply this theory, one has to assume adiabaticity of the intermediate state:
  a quantum state remains to be the ground state of the instantaneous Hamiltonian during the time evolution, 
  at least, approximately.
  On the other hand, the variational parameters $\mbox{\boldmath{$\theta$}}$ 
  of the DQAP ansatz $\vert \psi_M(\mbox{\boldmath{$\theta$}})\rangle$ are determined
  so as to mimizize the expectation value of the final Hamiltonian, and thus there is no 
  guarantee that the optimized DQAP ansatz follows the quantum adiabatic dynamics, 
  although the DQAP ansatz $\vert \psi_M(\mbox{\boldmath{$\theta$}})\rangle$ 
  itself is motivated by the quantum adiabatic process. 
  In the rest of this section, we shall show numerically that adiabaticity in the sense described above
  is indeed not satisfied in the optimized DQAP ansatz $\vert \psi_M(\mbox{\boldmath{$\theta$}})\rangle$.

  For this purpose, here we determine an optimum scheduling function $\chi_m^{\ast}$ to maximize 
  the overlap between the intermediate states of the optimized DQAP ansatz $\vert \psi_M(\mbox{\boldmath{$\theta$}})\rangle$ 
  representing the exact ground state of the final Hamiltonian 
  and the exact ground state of the instantaneous Hamiltonian, i.e.,  
  \begin{equation}
    ( \chi_m^{\ast},\alpha_m^{\ast} ) = \underset{\chi,\alpha}{\text{arg max}} \left[ F_{m}(\chi,\alpha) \right].
  \end{equation}
  Here, $F_m(\chi,\alpha)$ is the state overlap given by
  \begin{equation}
    F_m(\chi,\alpha) = \vert \langle \chi \vert \phi^{(m)}_M (\mbox{\boldmath{$\theta$}},\alpha) \rangle \vert^2,
  \end{equation}
  where $\vert \chi \rangle$ is the ground state of the following Hamiltonian: 
  \begin{equation}
    \hat{\mathcal{H}}(\chi) = \hat{\mathcal{V}}_1 + \chi \hat{\mathcal{V}}_2
    \label{eq:ham:1d:2}
  \end{equation}
  with $N$ fermions at half filling and $\hat{\mathcal{V}}_1$ and $\hat{\mathcal{V}}_2$ being given in 
  Eqs.~(\ref{eq:1d:v:bond1}) and (\ref{eq:1d:v:bond2}), respectively, and 
  \begin{align}
    \vert \phi^{(m)}_M (\mbox{\boldmath{$\theta$}},\alpha) \rangle = &
          {\rm e}^{-{\rm i}\theta_1^{(m)} \alpha \hat{\mathcal{V}}_1}
          {\rm e}^{-{\rm i}\theta_2^{(m)} \hat{\mathcal{V}}_2} \nonumber \\
          & \times 
          \prod_{k=m-1}^1 {\rm e}^{-{\rm i}\theta_1^{(k)} \hat{\mathcal{V}}_1} {\rm e}^{-{\rm i}\theta_2^{(k)} \hat{\mathcal{V}}_2}
          \vert \psi_{\rm i} \rangle .
          \label{eq:phi_m}
  \end{align}
  is the $m$th intermediate state ($m=0,1,2,\cdots,M$) of the DQAP ansatz $\vert \psi_M (\mbox{\boldmath{$\theta$}}) \rangle$ 
  with the variational parameters $\mbox{\boldmath{$\theta$}} = \{ \theta_1^{(1)}, \theta_2^{(1)}, \cdots, \theta_1^{(M)}, \theta_2^{(M)} \}$ 
  optimized for $M=L/4$ under APBCs and $M=(L-2)/4$ under PBCs at half filling, thus representing the exact ground state of the 
  one-dimensional 
  free-fermion system described by the Hamiltonian in Eq,~(\ref{eq:1d:ham}), i.e., the Hamiltonian $\hat{\mathcal{H}}(\chi=1)$ 
  in Eq.~(\ref{eq:ham:1d:2}). 
  Here, we introduce an additional parameter $\alpha$ to increase the state overlap 
  $F_m(\chi,\alpha)$ and define $\vert \phi^{(0)}_M (\mbox{\boldmath{$\theta$}},\alpha) \rangle = \vert \psi_{\rm i} \rangle$.

  \begin{figure*}
    \includegraphics[width=\hsize]{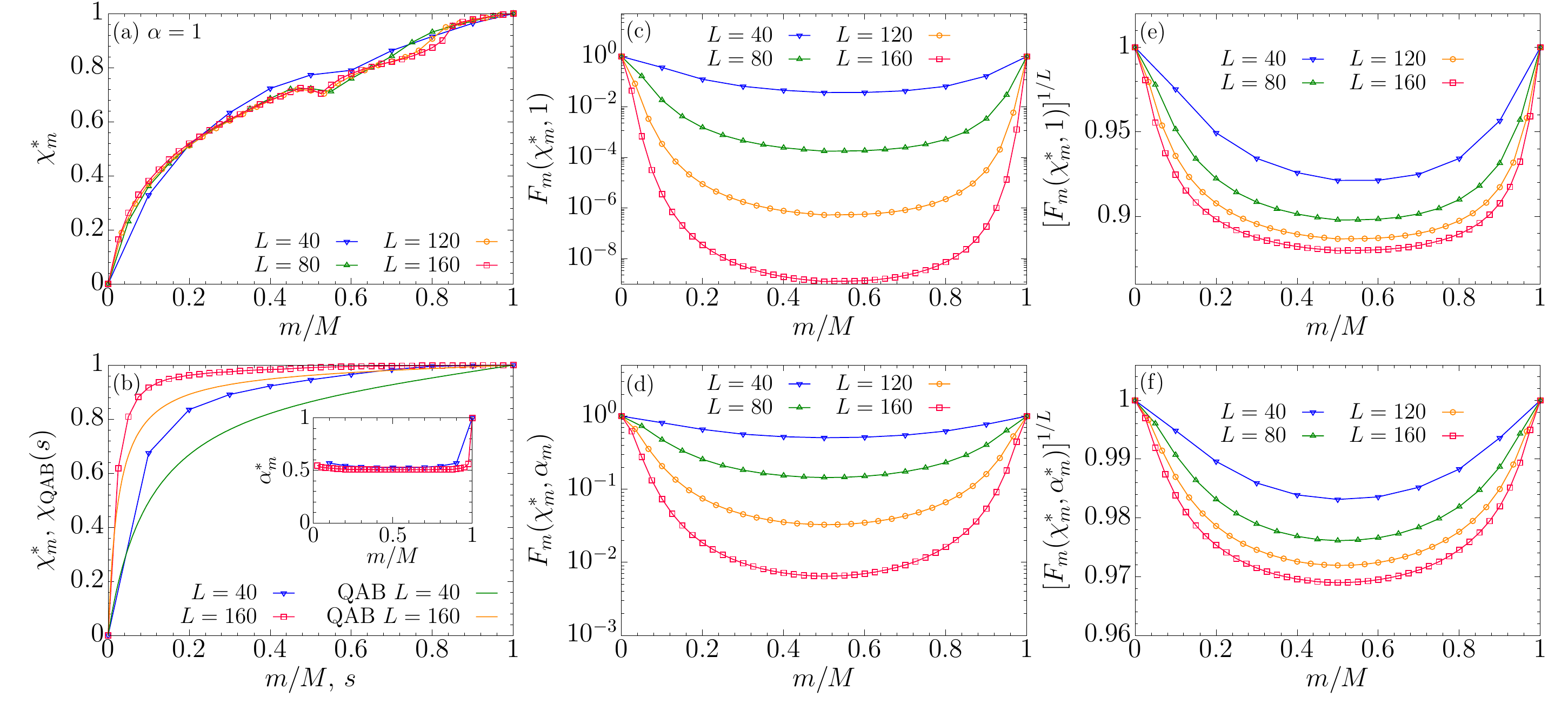}
    \caption{(a), (b) Optimum scheduling function $\chi_m^{\ast}$, (c), (d) the state overlap 
    $F_m(\chi_m,\alpha)$, and (e), (f) the state overlap per site $\left[F_m(\chi_m,\alpha)\right]^{1/L}$ 
    for the one-dimensional free-fermion systems with various system sizes $L$ and $M=L/4$ under APBCs at half filling. 
    The parameter $\alpha$ is set to be 1 in (a), (c), and (e), while it is optimized to maximize the state overlap 
    in (b), (d), and (f). The optimized $\alpha$ for each intermediate state, $\alpha_m^*$, is shown in the inset of (b).  
    For comparison, the optimum scheduling function $\chi_{\rm QAB}(s)$ obtained
    by quantum adiabatic brachistochrone (QAB) (for details, see Appendix~\ref{sec:qab}) is also shown by solid lines in (b).
    }
    \label{fig:schedule}
  \end{figure*}

  Figures~\ref{fig:schedule}(a) and \ref{fig:schedule}(c) show the optimum scheduling function $\chi_m^{\ast}$ and 
  the corresponding state overlap $F_m(\chi_m^{\ast},\alpha=1)$, respectively, when the parameter $\alpha$ in 
  Eq.~(\ref{eq:phi_m}) is set to be 1. 
  Although it is slightly concave, the optimum scheduling function $\chi_m^*$ in Fig.~\ref{fig:schedule}(a)
  is somewhat closer to a linear  
  function of $m$, which is expected for the linear scheduling.  
  However, the state overlap $F_m(\chi_m^*,1)$ becomes exponentially small with increasing 
  system size $L$, especially for $m\approx M/2$. Indeed, when $m$ is away from 0 and $M$ (i.e., the initial and final states), 
  the state overlap per site 
  $\left[F_m(\chi_m^{\ast},\alpha=1)\right]^{1/L}$ shown in Fig.~\ref{fig:schedule}(e) seems to converge to a value less than 1 in 
  the limit of $L\to\infty$. Therefore, the intermediate states of the optimized DQAP ansatz are rather far from the ground state of 
  the instantaneous Hamiltonian, suggesting that the discretized time evolution of the optimized DQAP ansatz is 
  much more different from a quantum adiabatic evolution, but closer to a quantum diabatic evolution. 
  This quantum diabatic-like evolution, instead of a quantum adiabatic evolution, of the optimized DQAP 
  ansatz could explain the quadratic speedup of the effective total evolution time $T_{\rm eff}(L)$ found in Fig.~\ref{fig:times}, 
  but certainly more systematic analysis is highly required and is left for a future study.
  
  This feature does not alter even when the parameter $\alpha$ is also optimized to maximize the state overlap. 
  As shown in Fig.~\ref{fig:schedule}(d), 
  the optimization of $\alpha$ increases the state overlap $F_m(\chi_m^{\ast},\alpha^*)$ significantly in orders of magnitude.
  More interestingly, the optimized values of $\alpha$ are neither 0 nor 1, but 
  approximately 0.5 for all values of $m$ except for $m=M$, as shown in the inset of Fig.~\ref{fig:schedule}(b). 
  However, the state overlap $F_m(\chi_m^{\ast},\alpha^*)$ still decreases exponentially with increasing the system size $L$ 
  for the intermediate values of $m$, and the state overlap per site shown in Fig.~\ref{fig:schedule}(f) seems to converge 
  to a value less than 1 in the limit of $L\to\infty$ for these values of $m$.

  Although it is no longer appropriate to identify the optimum scheduling function $\chi_m^{\ast}$ determined here 
  with an effective scheduling function of the quantum adiabatic process for the DQAP ansatz 
  $\vert \psi_M (\mbox{\boldmath{$\theta$}}) \rangle$, 
  it is highly interesting to compare $\chi_m^{\ast}$ with the optimum scheduling function
  $\chi_{\rm QAB}(s)$ obtained by the QAB~\cite{Rezakhani2009}, which is outlined in Appendix~\ref{sec:qab}. 
  As shown in Fig.~\ref{fig:schedule}(b), we find that, assuming $m/M$ is a similar quantity to the normalized time $s$ over the 
  total evolution time, the optimum scheduling function $\chi_m^{\ast}$, 
  optimized along with the parameter $\alpha$, is rather similar to the optimum scheduling function $\chi_{\rm QAB}(s)$ 
  obtained by the QAB, which is clearly distinct from the scheduling function expected for the linear scheduling.

\section{Imaginary-time evolution: comparison with the DQAP ansatz}\label{sec:res:imag}

Let us now consider an ansatz inspired by the imaginary-time evolution, instead of the real-time evolution 
as in the DQAP ansatz $\vert \psi_M (\mbox{\boldmath{$\theta$}}) \rangle$ discussed in the previous sections.
The imaginary-time counterpart $\vert \varphi_M (\mbox{\boldmath{$\tau$}}) \rangle $ 
of the DQAP ansatz $\vert \psi_M (\mbox{\boldmath{$\theta$}}) \rangle$ 
defined in Eq.~(\ref{eq:1d:dqap}) for the free-fermion system is given by
\begin{equation}
  \vert \varphi_M (\mbox{\boldmath{$\tau$}}) \rangle =
  \prod_{m=M}^1 (
       {\rm e}^{- \tau_1^{(m)} \hat{\mathcal{V}}_1}
       {\rm e}^{- \tau_2^{(m)} \hat{\mathcal{V}}_2}
       )
   \vert \psi_{\rm i} \rangle, 
   \label{eq:psi_im}
\end{equation}
where the initial state $\vert \psi_{\rm i} \rangle$ is a product state of the local bonding states given in Eq.~(\ref{eq:1d:init}), 
i.e., the ground state of $\hat{\mathcal{V}}_{1}$,  
and the imaginary-time steps $\mbox{\boldmath{$\tau$}} = \{ \tau_1^{(1)}, \tau_2^{(1)}, \cdots, \tau_1^{(M)}, \tau_2^{(M)} \}$ 
are considered as real variational parameters that are determined so as to minimize the variational energy~\cite{Yanagisawa1998}.
As in the DQAP ansatz $\vert \psi_M (\mbox{\boldmath{$\theta$}}) \rangle$, $\vert \varphi_M (\mbox{\boldmath{$\tau$}}) \rangle$ 
is constructed by repeatedly applying the local but now imaginary-time evolution operators ${\rm e}^{- \tau_1^{(m)} \hat{\mathcal{V}}_1}$ 
and ${\rm e}^{- \tau_2^{(m)} \hat{\mathcal{V}}_2}$. Since these imaginary-time evolution operators are not unitary, 
$\vert \varphi_M (\mbox{\boldmath{$\tau$}}) \rangle$ is no longer normalized.

We can follow exactly the same analysis in Sec.~\ref{sec:method} for the 
discretized imaginary-time evolution ansatz $\vert \varphi_M (\mbox{\boldmath{$\tau$}}) \rangle$ in Eq.~(\ref{eq:psi_im}) 
and obtain that
\begin{equation}
  \vert \varphi_M (\mbox{\boldmath{$\tau$}}) \rangle
  = \prod_{n=1}^N [ \hat{\bm c}^{\dagger} {\bm G}_M ]_{n} \vert 0 \rangle,
  \label{eq:psi_im:matform}
\end{equation}
where ${\bm G}_M$ is the $L \times N$ matrix given by
\begin{equation}
  {\bm G}_M = \prod_{m=M}^1 (
  {\rm e}^{-\tau_1^{(m)} {\bm V}_1} {\rm e}^{-\tau_2^{(m)} {\bm V}_2} )
  \mbox{\boldmath{$\Psi$}}_{\rm i}. 
\end{equation}
Here, the $L\times L$ matrices ${\bm V}_1$ and ${\bm V}_2$ are defined in Eqs.~(\ref{eq:matv1}) and (\ref{eq:matv2}), respectively, 
and the $L\times N$ matrix $\mbox{\boldmath{$\Psi$}}_{\rm i}$ is given in Eq.~(\ref{eq:matint}) 
for the number $N$ of fermions with $N=L/2$. Note also that 
${\rm e}^{-\tau {\bm V}_1}$ and ${\rm e}^{-\tau {\bm V}_2}$ are the same block diagonal matrices of 
${\rm e}^{-{\rm i}\theta {\bm V}_1}$ and ${\rm e}^{-{\rm i}\theta {\bm V}_2}$ in Eqs.~(\ref{eq:mat_v1}) and (\ref{eq:mat_v2}), respectively, 
except that $\theta$ is replaced with $-{\rm i}\tau$. 
It is now apparent that the imaginary-time evolved state $\vert \varphi_M (\mbox{\boldmath{$\tau$}}) \rangle$ from a state initially 
prepared as a single Slater determinant state $\vert \psi_{\rm i} \rangle$ can still be represented as a single Slater determinant state, 
in which each single-particle orbital is given by each column vector of ${\bm G}_M$. However, note that the imaginary-time evolved 
single-particle orbitals are neither normalized nor orthogonal to each other, i.e., ${\bm G}_M^{\dagger} {\bm G}_M \neq {\bf I}_{N}$, 
even though the initial single-particle orbitals 
are orthonormalized, i.e., $ \mbox{\boldmath{$\Psi$}}_{\rm i}^{\dagger} \mbox{\boldmath{$\Psi$}}_{\rm i} = {\bf I}_{N}$.

The natural gradient method described in Sec.~\ref{sec:opt} is straightforwardly extended 
to optimize the variational parameters $\mbox{\boldmath{$\tau$}}=\{\tau_1^{(m)},\tau_2^{(m)}\}$ 
in $\vert \varphi_M (\mbox{\boldmath{$\tau$}}) \rangle$
by replacing ${\bm S}$ and ${\bm f}$ in Eqs.~(\ref{eq:s_kk}) and (\ref{eq:f_kk}), respectively, with 
\begin{equation}
  \begin{split}
    [ {\bm S} ]_{kk^{\prime}} = &
    {\rm tr} [ {\bm F} (\partial_k {\bm G}_M ) (\partial_{k^{\prime}} {\bm G}_M) ] \\
    & - {\rm tr} [ {\bm F} (\partial_k {\bm G}_M^{\dagger} ) {\bm G}_M
      {\bm F} {\bm G}_M^{\dagger} ( \partial_{k^{\prime}} {\bm G}_M )]
  \end{split}
\end{equation}
and
\begin{equation}
  \begin{split}
    [ {\bm f} ]_{k} = & {\rm tr} [ {\bm F} ( \partial_k {\bm G}_M^{\dagger} ) {\bm T} {\bm G}_M ] \\
    & - {\rm tr} [ {\bm F} ( \partial_k {\bm G}_M^{\dagger} ) {\bm G}_M {\bm F} {\bm G}_M^{\dagger} {\bm T} {\bm G}_M ], 
  \end{split}
\end{equation}
where ${\bm F} = ( {\bm G}_M^{\dagger} {\bm G}_M )^{-1}$ and 
$k,k'=1,2,\cdots,2M$ labeling the variational parameters $\{\tau_1^{(m)},\tau_2^{(m)}\}_{m=1}^{M}$ as 
$\{ \tau_1, \tau_2, \cdots, \tau_k, \cdots , \tau_{2M-1}, \tau_{2M} \} = \{ \tau_1^{(1)}, \tau_2^{(1)}, \cdots, \tau_{p}^{(m)}, \cdots, \tau_{1}^{(M)}, \tau_{2}^{(M)} \}$.
$\partial_k {\bm G}_M$ is an $L\times N$ matrix defined as the first derivative of ${\bm G}_M$ with respect to the 
$k$th variational parameter $\tau_k$, i.e., 
\begin{equation}
  \partial_k {\bm G}_M = - (\prod_{l=2M}^{k+1} {\rm e}^{-\tau_l {\bm W}_l} )
  {\bm W}_k ( \prod_{l=k}^{1} {\rm e}^{-\tau_l {\bm W}_l} ) \mbox{\boldmath{$\Psi$}}_{\rm i}
\end{equation}
where 
$\{ {\bm W}_1, {\bm W}_2, {\bm W}_3, \cdots, {\bm W}_{2M} \} = \{ {\bm V}_2, {\bm V}_1, {\bm V}_2, \cdots, {\bm V}_1 \} $.

\begin{figure}
  \includegraphics[width=\hsize]{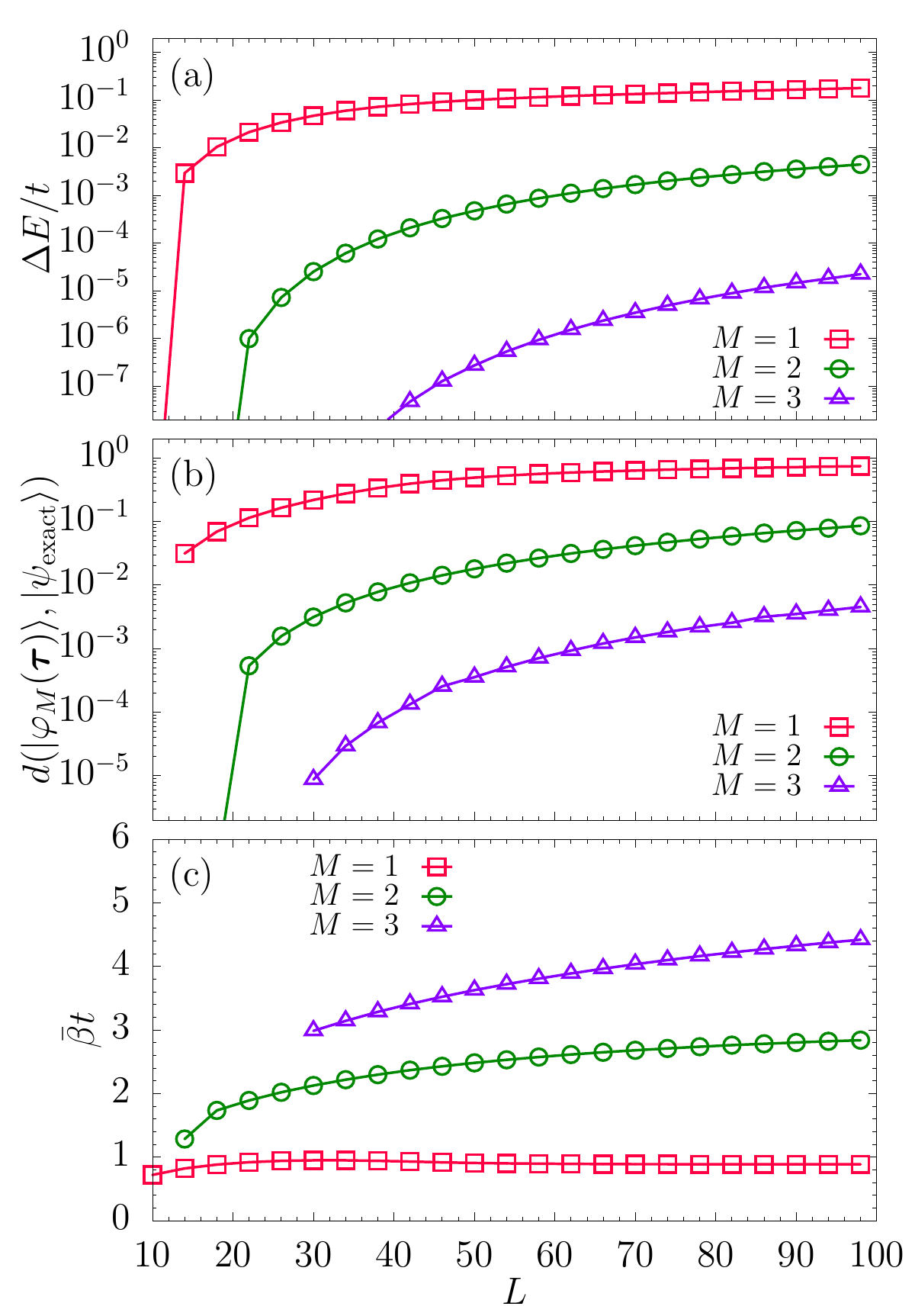}
  \caption{
    (a) Energy difference $\Delta E$ between the variational energy and the exact energy,  
    (b) distance 
    $d(\vert \varphi_M(\mbox{\boldmath{$\tau$}}) \rangle, \vert \psi_{\rm exact} \rangle)$, 
    and (c) effective total evolution time $\bar{\beta}(L)$
    for three values of $M$ as a function of the system size $L$. 
    The variational parameters in the discretized imaginary-time evolution ansatz 
    $\vert \varphi_M (\mbox{\boldmath{$\tau$}}) \rangle$ are optimized for each $M$ to 
    minimize the variational energy of the system under PBCs at half filling, i.e., $N=L/2$, satisfying 
    the closed shell condition. 
  }
  \label{fig:imag:energy}
\end{figure}

Figures~\ref{fig:imag:energy}(a) and \ref{fig:imag:energy}(b), respectively,
show an error of the optimized variational energy
\begin{equation}
  \Delta E =
  \frac{\langle \varphi_M(\mbox{\boldmath{$\tau$}}) \vert \hat{\mathcal{H}}
    \vert \varphi_M(\mbox{\boldmath{$\tau$}}) \rangle}
       {\langle \varphi_M(\mbox{\boldmath{$\tau$}}) \vert
        \varphi_M(\mbox{\boldmath{$\tau$}}) \rangle}
       - E_{\rm exact}(L)
\end{equation}
and a distance defined through the fidelity 
\begin{equation}
  d(\vert \varphi_M(\mbox{\boldmath{$\tau$}}) \rangle, \vert \psi_{\rm exact} \rangle) = \sqrt{1 -
    \frac{\vert \langle \varphi_M(\mbox{\boldmath{$\tau$}}) \vert \psi_{\rm exact} \rangle \vert^2}
    {\langle \varphi_M(\mbox{\boldmath{$\tau$}}) \vert \varphi_M (\mbox{\boldmath{$\tau$}}) \rangle }}
\end{equation}
as a function of the system size $L$ for three values of $M$. 
Here, $E_{\rm exact}(L)$ and $\vert \psi_{\rm exact} \rangle$ denote
the exact ground state energy and the normalized exact ground state of the system with the system size $L$, respectively. 
Although the system considered here is at the critical point
where the expectation value of $\hat{c}_{x}^{\dagger} \hat{c}_{x^{\prime}}$
decays algebraically with the distance $\vert x - x^{\prime} \vert$,
we find that the variational energy of the discretized imaginary-time evolution ansatz 
$\vert \varphi_M (\mbox{\boldmath{$\tau$}}) \rangle$ converges exponentially faster to the exact energy 
with increasing $M$~\cite{Zanca2017}.
This is in sharp contrast to the results for the discretized real-time evolution ansatz 
$\vert \psi_M (\mbox{\boldmath{$\theta$}}) \rangle$ shown in Fig.~\ref{fig:energy}. 
For example, for $L < 100$, we obtain the variational energy with accuracy 
$\Delta E$ as large as $2\times10^{-5}$ or less and the distance 
smaller than $10^{-2}$ by using only $M=3$. 
We should note that the exponentially fast convergence of the discretized imaginary-time evolution ansatz 
has also been reported for the transverse-field Ising model 
even at the critical point~\cite{Beach2019}.

We shall now discuss how the efficiency of the discretized imaginary-time evolution 
ansatz $\vert \varphi_M (\mbox{\boldmath{$\tau$}}) \rangle$ occurs. 
First, we should recognize that, although the discretized imaginary-time evolution 
ansatz $\vert \varphi_M (\mbox{\boldmath{$\tau$}}) \rangle$ is composed
of the local imaginary-time evolution operators, there is no limit of speed for 
propagating quantum entanglement via the local imaginary-time evolution operators 
because these local operators are non-unitary. 
This is indeed easily understood if we evaluate the local expectation value 
$\langle \varphi_M(\mbox{\boldmath{$\tau$}}) \vert \hat{c}_{x}^{\dagger} \hat{c}_{x+1} \vert \varphi_M (\mbox{\boldmath{$\tau$}}) \rangle$. 
In this case, we have to treat all local imaginary-time evolution operators 
in $\vert \varphi_M (\mbox{\boldmath{$\tau$}}) \rangle$, no matter how far a local imaginary-time evolution operator 
${\rm e}^{t\tau_p^{(m)} ( \hat{c}_{y}^{\dagger} \hat{c}_{y+1} + \hat{c}_{y+1}^{\dagger} \hat{c}_{y} ) }$ acting at sites $y$ and 
$y+1$ is distant from site $x$. 
Because of the non-unitarity,  
there is no cancellation of local imaginary-time evolution operators on the left and right sides of the local expectation value 
$\langle \varphi_M(\mbox{\boldmath{$\tau$}}) \vert \hat{c}_{x}^{\dagger} \hat{c}_{x+1} \vert \varphi_M (\mbox{\boldmath{$\tau$}}) \rangle$, 
implying that there is no causality-cone-like structure for the propagation of quantum entanglement illustrated in Fig.~\ref{fig:causality}.

Second, for the free-fermion system, we can understand the efficiency of 
the discretized imaginary-time evolution ansatz $\vert \varphi_M (\mbox{\boldmath{$\tau$}}) \rangle$ in terms of 
the imaginary-time evolution of the single-particle orbitals 
in ${\bm G}_M$.
For this purpose, we introduce the following $L\times N$ matrix: 
\begin{equation}
  {\bm g}_m = \prod_{m^{\prime}=m}^{1}(
    {\rm e}^{-\tau_1^{(m)} {\bm V}_1} {\rm e}^{-\tau_2^{(m)} {\bm V}_2} )
    \mbox{\boldmath{$\Psi$}}_{\rm i}
    \label{eq:imag:gmat}
\end{equation}
for $m=0,1,\cdots,M$ with ${\bm g}_0 = \mbox{\boldmath{$\Psi$}}_{\rm i}$ and ${\bm g}_M = {\bm G}_M$ 
to represent the single-particle orbitals at an intermediate imaginary time. 
Here, the variational parameters $\mbox{\boldmath{$\tau$}} = \{ \tau_1^{(1)}, \tau_2^{(1)}, \cdots, \tau_1^{(M)}, \tau_2^{(M)} \}$ 
are optimized for $\vert \varphi_M (\mbox{\boldmath{$\tau$}}) \rangle$ to minimize the variational energy and these parameters 
are used to define the matrix ${\bm g}_m$ in Eq.~(\ref{eq:imag:gmat}). 
Note that ${\bm g}_m$ is an analog to $\mbox{\boldmath{$\Phi$}}_m$ introduced in Eq.~(\ref{eq:phim}) for 
the DQAP ansatz $\vert \psi_M (\mbox{\boldmath{$\theta$}}) \rangle$.

\begin{figure}
  \includegraphics[width=\hsize]{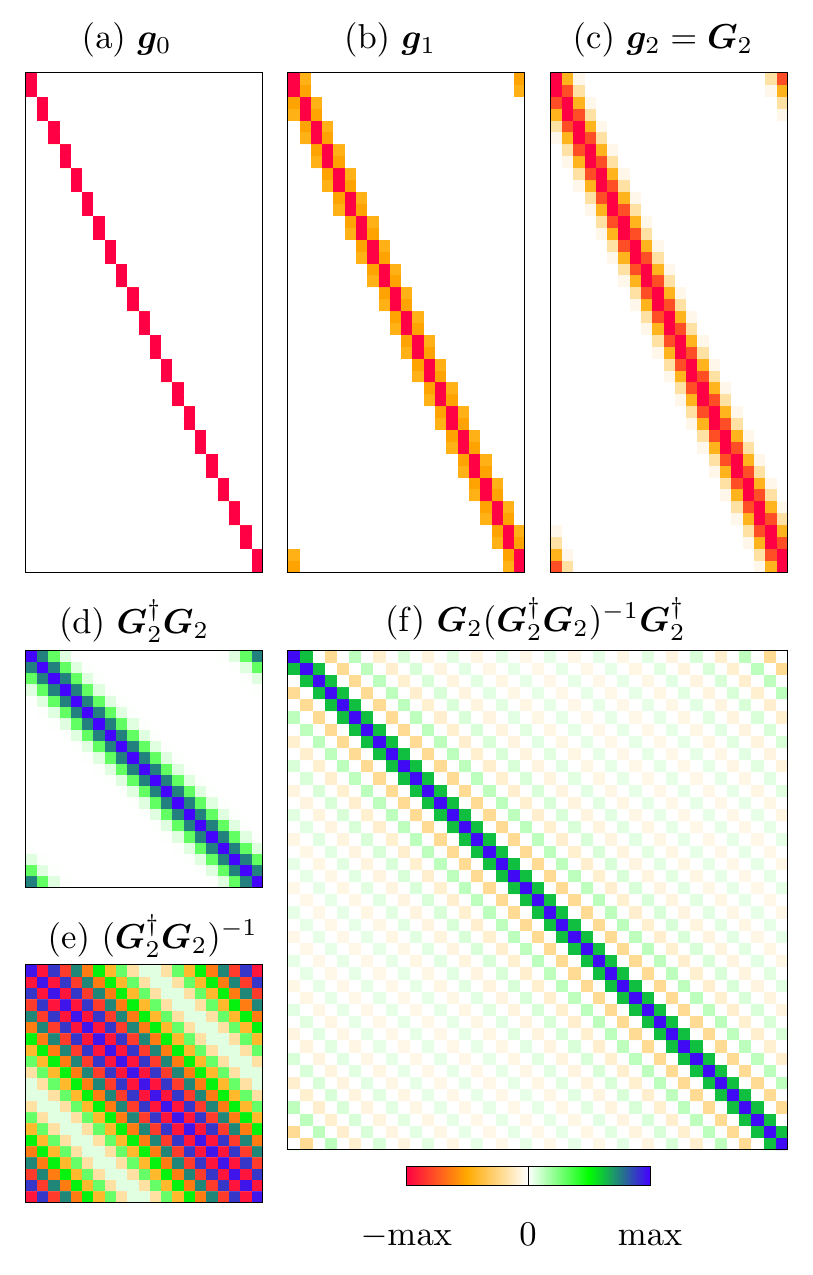}
  \caption{
  Intensity plot of matrix elements for 
  (a) ${\bm g}_0 = \mbox{\boldmath{$\Psi$}}_{\rm i}$, (b) ${\bm g}_1$,
  (c) ${\bm g}_2 = {\bm G}_2$, (d) ${\bm G}_2^{\dagger} {\bm G}_2$, 
  (e) $({\bm G}_2^{\dagger} {\bm G}_2)^{-1}$, and  
  (f) ${\bm G}_2 ({\bm G}_2^{\dagger} {\bm G}_2)^{-1} {\bm G}_2^{\dagger}$
  in the discretized  imaginary-time evolution ansatz $\vert \varphi_M(\mbox{\boldmath{$\tau$}}) \rangle$ with $M=2$.
  The matrix elements are all real.
  The variational parameters in $\vert \varphi_M(\mbox{\boldmath{$\tau$}}) \rangle$ are optimized for 
  $L=42$ under PBCs at half filling, i.e., $N=L/2$. 
  In (a)--(c), the vertical axis corresponds to the site index $x$, while the horizontal axis corresponds to the single-particle 
  orbital index $n$. 
  In (d) and (e), both vertical and horizontal axes correspond to the single-particle orbital index $n$. 
  In (f), both vertical and horizontal axes correspond to the site index $x$.
  }
  \label{fig:imag:orbital}
\end{figure}

Figures~\ref{fig:imag:orbital}(a)--\ref{fig:imag:orbital}(c) show the discretized imaginary-time evolution of 
all matrix elements in three matrices ${\bm g}_0$, ${\bm g}_1$, and ${\bm g}_2={\bm G}_2$ when $M=2$. 
As in the case of a single-particle orbital in $\mbox{\boldmath{$\Phi$}}_m$, 
we can readily find that a single-particle orbital in ${\bm g}_m$ extends spatially four lattice spaces every time 
the imaginary time $m$ increases 
by one: the spatial extent $\bar{d}_m$ (in unit of lattice constant) of a single-particle orbital in ${\bm g}_m$ is 
${\bar d}_m = 4m+2$, exactly the same as the spatial extent $d_m$ of a single-particle orbital in $\mbox{\boldmath{$\Phi$}}_m$ 
for the real-time evolution (see Sec.~\ref{sec:res:psi}). This is simply because the imaginary- and real-time evolutions are both 
governed by the spatially local evolution operators.  
Therefore, as in the case of the real-time evolution ansatz 
$\vert \psi_M (\mbox{\boldmath{$\theta$}}) \rangle$, 
the single-particle orbitals in ${\bm g}_m$ can extend over the entire system only when 
$\bar{d}_m$ reaches to the system size $L$.

However, unlike the case of the real-time evolution ansatz, 
the single-particle orbitals in ${\bm g}_m$ are not orthonormalized, i.e., 
${\bm g}_m^{\dagger} {\bm g}_m \neq {\bf I}_{N}$ for $m>0$, 
as shown in Fig.~\ref{fig:imag:orbital}(d). 
As a result, $({\bm g}_m^{\dagger}{\bm g}_m)^{-1}$ becomes non-local in the sense that  
$[({\bm g}_m^{\dagger}{\bm g}_m)^{-1}]_{xx^\prime}\ne0$ even when sites $x$ and $x^\prime$ are 
distant from each other [see Fig.~\ref{fig:imag:orbital}(e)], 
although ${\bm g}_m^{\dagger} {\bm g}_m$ might be local. 
This has a significant consequence when we evaluate the expectation value 
$\langle \varphi_M(\mbox{\boldmath{$\tau$}}) \vert \hat{c}_x^{\dagger} \hat{c}_{x^{\prime}} \vert \varphi_M (\mbox{\boldmath{$\tau$}}) \rangle$, 
taking also into account the normalization of $\vert \varphi_M (\mbox{\boldmath{$\tau$}}) \rangle$. 
Using Eq.~(\ref{eq:ff:expectation}), we can show that 
\begin{align}
  \frac{\langle \varphi_M(\mbox{\boldmath{$\tau$}}) \vert \hat{c}_x^{\dagger} \hat{c}_{x^{\prime}} \vert \varphi_M (\mbox{\boldmath{$\tau$}}) \rangle}
       {\langle \varphi_M(\mbox{\boldmath{$\tau$}}) \vert \varphi_M (\mbox{\boldmath{$\tau$}}) \rangle}
       = &{\rm tr} \left[ {\bm G}_M ({\bm G}_M^{\dagger} {\bm G}_M)^{-1} {\bm G}_M^{\dagger} \mbox{\boldmath{$\delta$}}_{xx^{\prime}} \right] \nonumber\\
       = &
           \left[ {\bm G}_M ({\bm G}_M^{\dagger} {\bm G}_M)^{-1} {\bm G}_M^{\dagger} \right]_{x^{\prime}x}.
       \label{eq:cc_im}
\end{align}
Because $({\bm G}_M^{\dagger}{\bm G}_M)^{-1}$ is non-local, 
${\bm G}_M ({\bm G}_M^{\dagger} {\bm G}_M)^{-1} {\bm G}_M^{\dagger}$ is also non-local even for $M\ll L$, 
as shown in Fig.~\ref{fig:imag:orbital}(f), implying that the expectation value of $\hat{c}_x^{\dagger} \hat{c}_{x^{\prime}}$ is 
non-zero even when sites $x$ and $x^\prime$ are far apart.  
This should be contrasted with the case of the the DQAP ansatz $\vert \psi_M (\mbox{\boldmath{$\theta$}}) \rangle$, where 
the corresponding expectation value is given as 
\begin{equation}
  \langle \psi_M(\mbox{\boldmath{$\theta$}}) \vert \hat{c}_x^{\dagger} \hat{c}_{x^{\prime}} \vert \psi_M (\mbox{\boldmath{$\theta$}}) \rangle
  = {\rm tr} \left[  \mbox{\boldmath{$\Psi$}}_M   \mbox{\boldmath{$\Psi$}}_M^{\dagger} \mbox{\boldmath{$\delta$}}_{xx^{\prime}} \right]
    =  \left[
      \mbox{\boldmath{$\Psi$}}_M   \mbox{\boldmath{$\Psi$}}_M^{\dagger}
      \right]_{x^{\prime}x}
\end{equation}
because the real-time evolution ansatz $\vert \psi_M (\mbox{\boldmath{$\theta$}}) \rangle$ is normalized, i.e., 
$ \mbox{\boldmath{$\Psi$}}_{\rm M}^{\dagger} \mbox{\boldmath{$\Psi$}}_{\rm M} = {\bf I}_{N}$, and thus 
the expectation value of $\hat{c}_x^{\dagger} \hat{c}_{x^{\prime}}$ is zero when sites $x$ and $x^\prime$ are far apart, 
provided that $M$ is not large enough, as discussed in Sec.~\ref{sec:res:minf}.

\begin{figure}[htbp]
  \includegraphics[width=\hsize]{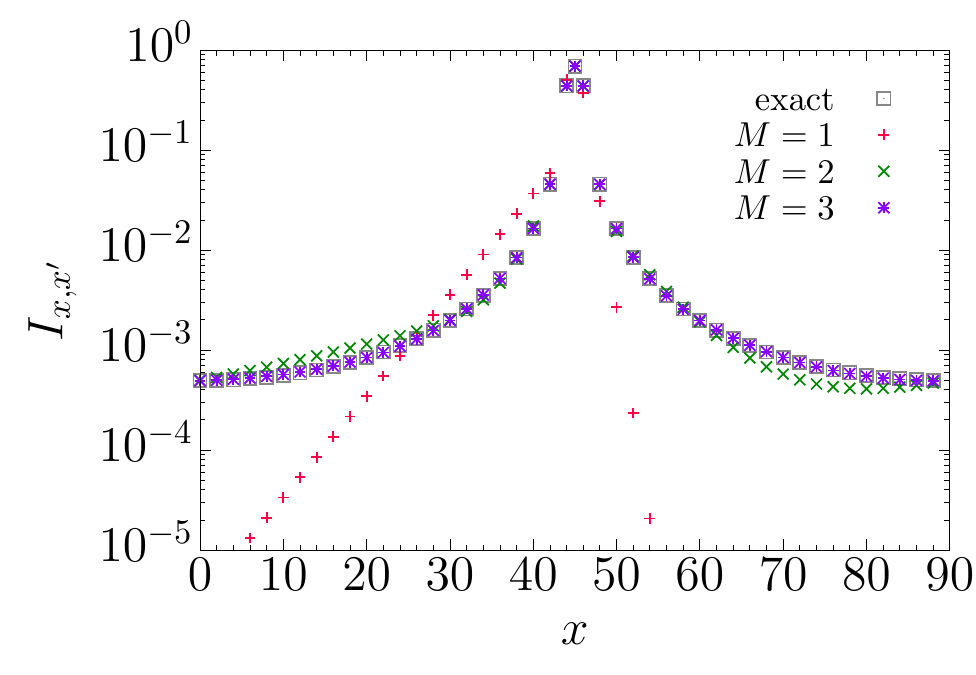}
  \caption{Mutual information $I_{x,x^{\prime}}$ of the
    discretized imaginary-time evolution ansatz $\vert \varphi_M(\mbox{\boldmath{$\tau$}}) \rangle$
    with the variational parameters $\mbox{\boldmath{$\tau$}}$ optimized for each $M$ to minimize the variational energy. 
    The calculations are for 
    $L=90$ under PBCs at half filling, i.e., $N=L/2$.
    Gray squares indicate the mutual information $I_{x,x^{\prime}}$ of the exact ground state.
    We set $x^{\prime} = 45$. 
    Note that $I_{x,x^{\prime}}$ exhibits an oscillatory behavior and is exactly zero at site $x$ on the same sublattice of 
    site $x'$ except for $x=x^{\prime}$.
  }
  \label{fig:imag:minfo}
\end{figure}

Consequently, the discretized imaginary-time evolution ansatz $\vert \varphi_M (\mbox{\boldmath{$\tau$}}) \rangle$ 
acquires the global correlation with the extremely small number of $M$. 
Figure~\ref{fig:imag:minfo}
shows the mutual information $I_{x,x^{\prime}}$ of the optimized $\vert \varphi_M (\mbox{\boldmath{$\tau$}}) \rangle$ with 
$M=1,2,3$. We find that, although $I_{x,x^{\prime}}$ for $M=1$
shows exponential decay as a function of distance $|x-x'|$,
it is drastically improved with increasing $M$ and
the mutual information $I_{x,x^{\prime}}$ evaluated for $M=2$ already almost coincides with the exact value, 
despite that there are only four variational parameters for $M=2$. 
We should also emphasize that no causality-cone-like structure is observed in the mutual information $I_{x,x^{\prime}}$ 
of the discretized imaginary-time evolution ansatz $\vert \varphi_M (\mbox{\boldmath{$\tau$}}) \rangle$, which is in sharp contrast 
to the results for the discretized real-time evolution ansatz 
$\vert \psi_M (\mbox{\boldmath{$\theta$}}) \rangle$ shown in Fig.~\ref{fig:minf}.

Finally, we show in Fig.~\ref{fig:imag:energy}(c) 
the effective total evolution time $\bar{\beta}(L)$ 
of the imaginary-time evolution in $\vert \varphi_M (\mbox{\boldmath{$\tau$}}) \rangle$ defined by
\begin{equation}
  \bar{\beta}(L) = \frac{1}{2} \sum_{m=1}^M \sum_{p=1}^2 \tau_{p}^{(m)},
\end{equation}
where the variational parameters $\{\tau_p^{(m)} \}$ are optimized for each $M$ to minimize the variational energy. 
First, it is noticed that $\bar{\beta}(L)$ exhibits 
the system size dependence, which is different from that found for the discretized real-time evolution ansatz 
$\vert \psi_M (\mbox{\boldmath{$\theta$}}) \rangle$ shown in Fig.~\ref{fig:times}. 
It is also important to note that $\bar{\beta}(L)$ is not proportional to system size $L$. 
On one hand, one would expect that a large $\tau_p^{(m)}$ is preferable 
to reach the ground state fast, i.e., with a fewer number of variational parameters, in a standard sense of the imaginary-time evolution. 
On the other hand, a large $\tau_p^{(m)}$ might introduce bias in approximating 
the continuous imaginary-time evolution by the discretized evolution via the Suzuki-Trotter decomposition~\cite{Trotter1959,Suzuki1976}. 
Therefore, the optimized solution should be determined by compromising these two factors. 
The discretized imaginary-time evolution ansatz $\vert \psi_M (\mbox{\boldmath{$\theta$}}) \rangle$ 
finds the best solution available within a given value of $M$.

\section{\label{sec:summary}Summary and discussion}

As a quantum-classical hybrid algorithm to generate a desired quantum state in a quantum circuit, 
we have studied the DQAP ansatz $\vert \psi_M (\mbox{\boldmath{$\theta$}}) \rangle$ to 
represent the ground state of the one-dimensional free-fermion system.  
The DQAP ansatz $\vert \psi_M (\mbox{\boldmath{$\theta$}}) \rangle$ considered here is 
inspired by the QAOA and is composed of $M$ layers of 
two elementary sets of local time-evolution operators acting on neighboring sites (i.e., qubits),
as illustrated in Fig.~\ref{fig:1d:ansatz}. 
By numerically optimizing the variational parameters 
$\mbox{\boldmath{$\theta$}} = \{ \theta_1^{(1)}, \theta_2^{(1)}, \cdots, \theta_1^{(M)}, \theta_2^{(M)} \}$ 
so as to minimize the variational energy, we have found that 
the exact ground state can be attained by 
the DQAP ansatz $\vert \psi_M (\mbox{\boldmath{$\theta$}}) \rangle$ 
with the number $M_B$ of layers 
as large as $(L-2)/4$ for PBCs and $L/4$ for APBCs, i.e., the minimum number of $M$ set by the Lieb-Robinson 
bound for the propagation of quantum entanglement via the local time-evolution operators (see Fig.~\ref{fig:causality}).  
Our results thus suggest that the DQAP ansatz $\vert \psi_M (\mbox{\boldmath{$\theta$}}) \rangle$ is the ideal ansatz to 
represent the exact ground state based on the quantum adiabatic process. 
Indeed, in the DQAP scheme, the exact ground state is prepared by the shallowest quantum circuit with linear depth, 
containing ${\mathcal O}(L^2)$ single-qubit and CNOT gates, where $L$ is the number of sites in the system,  
i.e., the number of qubits.

We have also found that the optimized DQAP ansatz $\vert \psi_M (\mbox{\boldmath{$\theta$}}) \rangle$ with $M$ less than 
$M_B$ exhibits another series of the states that are independent of system size $L$. We have shown that 
the entanglement entropy $S_{\mathbb A}$ of subsystem $\mathbb{A}$ and the variational energy $E_M(L)/L$ per site 
evaluated for these states with $4M\le L_A$ and $M<M_B$, respectively, fall into smooth universal functions of $M$, 
independently of the system size $L$ and the boundary conditions. 
This implies that the entanglement acquired by the DQAP ansatz 
$\vert \psi_M (\mbox{\boldmath{$\theta$}}) \rangle$ with a finite $M$ is bounded,
as in the case of the matrix product states 
with a finite bond dimension~\cite{PerezGarcia2007,Verstraete2008}.
Moreover, we have found that 
the entanglement entropy $S_{\mathbb A}$ and the energy difference between the variational energy and the exact one 
$\Delta \varepsilon= E_M(L)/L - \varepsilon_{\infty} $
behave asymptotically as $S_{\mathbb A}\approx\frac{1}{3}\ln M$
and $\Delta \varepsilon \sim M^{-2}$, respectively.

We have also analyzed the evolution of the single-particle orbitals in the DQAP ansatz 
$\vert \psi_M (\mbox{\boldmath{$\theta$}}) \rangle$ 
via the local time-evolution operators 
and the mutual information of $\vert \psi_M (\mbox{\boldmath{$\theta$}}) \rangle$ to explore how quantum entanglement is 
developed in the quantum state. The latter quantity also reveals the causality-cone-like structure of the propagation of 
quantum entanglement via the local time-evolution operators.  
Furthermore, we have found that the optimized variational parameters 
$\mbox{\boldmath{$\theta$}} = \{ \theta_1^{(1)}, \theta_2^{(1)}, \cdots, \theta_1^{(m)}, \theta_2^{(m)}, \cdots, \theta_1^{(M)}, \theta_2^{(M)} \}$ 
in the DQAP ansatz $\vert \psi_M (\mbox{\boldmath{$\theta$}}) \rangle$ converge to a smooth function of $m$ for each $\theta_p^{(m)}$ 
($p=1,2$), which is quite different from the linear scheduling functions expected when the quantum adiabatic 
process is naively discretized in time. 
We have also found that the effective total evolution time $T_{\rm eff}(L)$ of the optimized variational parameters 
$\mbox{\boldmath{$\theta$}}$ in the DQAP ansatz $\vert \psi_M (\mbox{\boldmath{$\theta$}}) \rangle$ with $M= M_B$, 
thus representing the exact ground state, 
scales linearly with the 
system size $L$, as opposed to $L^2$ expected in the continuous-time quantum adiabatic process
with the linear scheduling.
Moreover, we have  found that 
the intermediate state in the optimized DQAP ansatz representing the exact ground state of the final Hamiltonian 
has an exponentially small state overlap with the ground state of the instantaneous Hamiltonian,
implying that the discretized time evolution of the DQAP ansatz is far from a quantum adiabatic evolution but rather close to 
a quantum diabatic evolution, although the DQAP ansatz itself is motivated by a quantum adiabatic process. 
Nonetheless, we have also estimated the optimum scheduling function by maximizing the state overlap between the 
intermediate state of the DQAP ansatz and the ground state of the instantaneous Hamiltonian and found that it is rather 
similar to the optimum scheduling function obtained by the QAB, 
if the additional parameter $\alpha$ is also optimized. 
The quantum diabatic like evolution of the optimized DQAP ansatz could be
responsible for the quadratic speedup of the effective total evolution time $T_{\rm eff}(T)$
of the optimized DQAP ansatz.

We have also explored the discretized imaginary-time evolution ansatz $\vert \varphi_M (\mbox{\boldmath{$\tau$}}) \rangle$, 
an imaginary-time counterpart of the DQAP ansatz $\vert \psi_M (\mbox{\boldmath{$\theta$}}) \rangle$, for the same
free-fermion system. 
Similarly to the DQAP ansatz $\vert \psi_M (\mbox{\boldmath{$\theta$}}) \rangle$, the discretized imaginary-time evolution ansatz 
$\vert \varphi_M (\mbox{\boldmath{$\tau$}}) \rangle$ is composed of $M$ layers of two elementary sets of local 
imaginary-time evolution operators acting on neighboring sites. 
We have found that the convergence of $\vert \varphi_M (\mbox{\boldmath{$\tau$}}) \rangle$ to the exact ground 
state is exponentially fast in terms of the number $M$ of layers, as compared to that of the DQAP ansatz 
$\vert \psi_M (\mbox{\boldmath{$\theta$}}) \rangle$, although both ansatze are composed of the local evolution operators. 
This difference is attributed to the fact that the imaginary-time evolution operator is not unitary and thus there is no limit of speed 
for the propagation of quantum entanglement via the local non-unitary imaginary-time evolution operators.  
In particular, for the free-fermion system, 
we can show directly that the expectation value  
$\langle \varphi_M(\mbox{\boldmath{$\tau$}}) \vert \hat{c}_x^{\dagger} \hat{c}_{x^{\prime}} \vert \varphi_M (\mbox{\boldmath{$\tau$}}) \rangle / \langle \varphi_M(\mbox{\boldmath{$\tau$}}) \vert \varphi_M (\mbox{\boldmath{$\tau$}}) \rangle$ is non-local 
even when $M\ll L$, regardless of the distance $|x-x^\prime|$, because the discretized imaginary-time evolution ansatz 
$\vert \varphi_M (\mbox{\boldmath{$\tau$}}) \rangle$ is not normalized, i.e., $ \langle \varphi_M(\mbox{\boldmath{$\tau$}}) \vert \varphi_M (\mbox{\boldmath{$\tau$}}) \rangle \ne 1$. 
This is in sharp contrast to the case of the DQAP ansatz $\vert \psi_M (\mbox{\boldmath{$\theta$}}) \rangle$, in which the 
corresponding expectation value 
$\langle \psi_M(\mbox{\boldmath{$\theta$}}) \vert \hat{c}_x^{\dagger} \hat{c}_{x^{\prime}} \vert \psi_M (\mbox{\boldmath{$\theta$}}) \rangle$ is zero when the two causality cones set by the Lieb-Robinson bound formed from the origins at sites $x$ and $x^\prime$ do not 
overlap (see Fig.~\ref{fig:minf:overlap}). 
Our result thus implies that if the local non-unitary imaginary-time evolution operator can be implemented in a quantum circuit by 
using ${\mathcal O}(1)$ local single- and two-qubit unitary gates, one can prepare the ground state in this scheme 
by a much shallower quantum circuit with depth ${\mathcal O}(1)$. 
However, it is challenging to represent a local non-unitary operator by  ${\mathcal O}(1)$ local single- and two-qubit unitary gates, 
especially for a quantum state at criticality.

The free-fermion system considered here is at the critical point 
where the correlation function 
$\langle  \hat{c}_{x}^{\dagger} \hat{c}_{x^{\prime}} \rangle$ decays algebraically with $|x-x^\prime|$ and thus 
the correlation is extended over the entire system. 
In a critical system, it is intuitively understood that at least $L^2$  
local two-qubit unitary gates are required in a quantum circuit to represent the quantum entanglement of the state 
for the system with $L$ sites. 
Therefore, also in this sense, the DQAP ansatz $\vert \psi_M (\mbox{\boldmath{$\theta$}}) \rangle$ is an ideally compact 
form to represent the ground state of this system.  
However, it is not trivial for more general cases such as an interacting fermion system. 
It is thus valuable to consider a possible improvement in the quantum adiabatic process, for example, 
by introducing navigation proposed in the VanQver algorithm~\cite{Matsuura_2020}, 
for reducing the complexity of quantum processes.
It is also an interesting extension to introduce a non-unitary process by inserting measurements 
during the quantum adiabatic process~\cite{Yaodong2018,Cao2019}.

We have also found that the natural gradient method can optimize 
the variational parameters in 
the DQAP ansatz $\vert \psi_M (\mbox{\boldmath{$\theta$}}) \rangle$ 
without any difficulty. 
Even with randomly chosen initial variational parameters, the optimization method can eventually 
find sets of optimized variational parameters to converge to the lowest variational energy (see Appendix~\ref{sec:random}), 
implying that there is no problem such as the barren plateaus phenomena~\cite{McClean:2018aa}.   
However, this could be due to the fact that for the free-fermion system, the independent matrix elements in 
the DQAP ansatz $\vert \psi_M (\mbox{\boldmath{$\theta$}}) \rangle$ can be significantly reduced 
(see discussion in Sec.~\ref{sec:res:psi}). Therefore, it is desirable to examine the case for an interacting fermion system, 
for example, and this is left for a future study.

The focus in this paper was limited to the free-fermion system, 
where a time-evolved $N$-fermion state can still be described by a single Slater determinant state, 
and therefore 
any quantum advantage is expected in simulating this system on a quantum computer. 
However, this system is one of the ideal systems to test the operations of NISQ devices 
because the quantum state described by the DQAP ansatz $\vert \psi_M (\mbox{\boldmath{$\theta$}}) \rangle$ 
is highly entangled but can still be treated in large systems 
on a classical computer. 

\section*{Acknowledgement}
We are grateful to Sandro Sorella and Giuseppe E. Santoro for their valuable comments and inputs on the subjects 
studied in this paper.
We are also thankful to Hiroshi Ueda for
his insightful comments on the relation to the matrix product states.
Parts of the numerical simulations have been done on the HOKUSAI supercomputer
at RIKEN (Project ID: No. G20015). 
This work was supported by JST PRESTO (No. JPMJPR191B),
Grant-in-Aid for Research Activity start-up
(No. JP19K23433), and Grant-in-Aid for Scientific Research (B) 
(No. JP18H01183) from MEXT, Japan.

\appendix

\section{\label{sec:ff}Derivation for the free-fermion formulas}

In this Appendix, we derive Eqs.~(\ref{eq:tsps:psi}) and (\ref{eq:psi_im:matform}) 
by following Ref.~\cite{PhysRevB.41.11352}. 
To this end, let us first notice the following commutation relation:
\begin{equation}
  [ \hat{\mathcal{W}}_k,\ \hat{\bm c}^{\dagger} ] = \hat{\bm c}^{\dagger} {\bm W}_k,
  \label{eq:comm:w:c}
\end{equation}
namely, 
\begin{equation}
  [ \hat{\mathcal{W}}_k,\ \hat{c}_x^{\dagger} ] 
  = \sum_{x^{\prime}=1}^L \hat{c}_{x^{\prime}}^{\dagger} \left[ {\bm W}_k \right]_{x^{\prime}x},
\end{equation}
where $\hat{\mathcal{W}}_k$ and ${\bm W}_k$ are respectively a single-particle operator
and the corresponding $L \times L$ matrix defined in Eq.~(\ref{eq:ff:op:w}) with $k=1,2,3,\cdots$. 
Note that the matrix ${\bm W}_k$ in Eq.~(\ref{eq:ff:op:w}) is assumed to be Hermitian, 
but Eq.~(\ref{eq:comm:w:c}) is satisfied for any $L \times L$ matrix ${\bm W}_k$ and the argument given below in this 
Appendix is generally correct for any matrix ${\bm W}_k$.

Rearranging the terms in Eq.~(\ref{eq:comm:w:c}), 
we obtain that 
\begin{equation}
  \hat{\mathcal{W}}_k \hat{\bm c}^{\dagger} = \hat{\bm c}^{\dagger}
  ({\bm W}_k + \hat{\mathcal{W}}_k ).
  \label{eq:wc:linear}
\end{equation}
By sequentially using Eq.~(\ref{eq:wc:linear}), 
we can find that 
\begin{equation}
  ( \hat{\mathcal{W}}_k )^l \hat{\bm c}^{\dagger} =
  \hat{\bm c}^{\dagger}
  ({\bm W}_k + \hat{\mathcal{W}}_k )^l
\end{equation}
for integer $l\geq0$. 
This formula can be extended
to a general function of a matrix 
and we can readily show that 
\begin{equation}
  {\rm e}^{- z_k \hat{\mathcal{W}}_k } \hat{\bm c}^{\dagger} =
  \hat{\bm c}^{\dagger}
  {\rm e}^{- z_k ({\bm W}_k + \hat{\mathcal{W}}_k ) }
\end{equation}
for any complex number $z_k$.  
Notice here that the single-particle operator $\hat{\mathcal{W}}_k$ is 
a scalar in the matrix-vector multiplication 
and the $L\times L$ matrix ${\bm W}_k$ is a $c$-number in the operator space. 
Therefore, $\hat{\mathcal{W}}_k$ and ${\bm W}_k$ commute with each other and thus 
\begin{equation}
  {\rm e}^{- z_k \hat{\mathcal{W}}_k } \hat{\bm c}^{\dagger} =
  \hat{\bm c}^{\dagger}
      {\rm e}^{- z_k {\bm W}_k} {\rm e}^{- z_k \hat{\mathcal{W}}_k }.
  \label{eq:ff:lemma}
\end{equation}
By multiplying an $L \times N$ matrix $\mbox{\boldmath{$\Phi$}}_{k-1}$ from the right side,
we obtain that 
\begin{align}
  {\rm e}^{- z_k \hat{\mathcal{W}}_k } \hat{\bm c}^{\dagger} \mbox{\boldmath{$\Phi$}}_{k-1} =
  \hat{\bm c}^{\dagger}
  \mbox{\boldmath{$\Phi$}}_{k} {\rm e}^{- z_k \hat{\mathcal{W}}_k }
  \label{eq:ff:lemma:2}
\end{align}
with
\begin{equation}
  \mbox{\boldmath{$\Phi$}}_{k} \equiv {\rm e}^{- z_k {\bm W}_k} \mbox{\boldmath{$\Phi$}}_{k-1}
\end{equation}
and $\mbox{\boldmath{$\Phi$}}_0 = \mbox{\boldmath{$\Psi$}}_0$, where $\mbox{\boldmath{$\Psi$}}_0$ is an $L \times N$ matrix. 

Let us now introduce the initial state $\vert \psi_0 \rangle$ with $N$ fermions as 
\begin{equation}
  \vert \psi_0 \rangle = \prod_{n=1}^{N} \left[ \hat{\bm c}^{\dagger} \mbox{\boldmath{$\Psi$}}_0 \right]_n \vert 0 \rangle, 
\end{equation}
where $\vert 0\rangle$ is the vacuum of fermions. 
Using Eq.~(\ref{eq:ff:lemma:2}),
we find that
\begin{align}
  \vert \phi_1 \rangle \equiv & 
  {\rm e}^{- z_1 \hat{\mathcal{W}}_1 } \vert \psi_0 \rangle \nonumber \\
  = &
  {\rm e}^{- z_1 \hat{\mathcal{W}}_1 }
  \left[ \hat{\bm c}^{\dagger} \mbox{\boldmath{$\Phi$}}_0 \right]_1
  \left[ \hat{\bm c}^{\dagger} \mbox{\boldmath{$\Phi$}}_0 \right]_2
  \cdots
  \left[ \hat{\bm c}^{\dagger} \mbox{\boldmath{$\Phi$}}_0 \right]_N
  \vert 0 \rangle \nonumber \\
  = &
  \left[ \hat{\bm c}^{\dagger} \mbox{\boldmath{$\Phi$}}_1 \right]_1
  {\rm e}^{- z_1 \hat{\mathcal{W}}_1 }
  \left[ \hat{\bm c}^{\dagger} \mbox{\boldmath{$\Phi$}}_0 \right]_2
  \cdots
  \left[ \hat{\bm c}^{\dagger} \mbox{\boldmath{$\Phi$}}_0 \right]_N
  \vert 0 \rangle \nonumber \\
  = & 
  \left[ \hat{\bm c}^{\dagger} \mbox{\boldmath{$\Phi$}}_1 \right]_1
  \left[ \hat{\bm c}^{\dagger} \mbox{\boldmath{$\Phi$}}_1 \right]_2
       {\rm e}^{- z_1 \hat{\mathcal{W}}_1 } 
  \cdots
  \left[ \hat{\bm c}^{\dagger} \mbox{\boldmath{$\Phi$}}_0 \right]_N
  \vert 0 \rangle \nonumber \\
  \vdots \nonumber \\
  = & 
  \left[ \hat{\bm c}^{\dagger} \mbox{\boldmath{$\Phi$}}_1 \right]_1
  \left[ \hat{\bm c}^{\dagger} \mbox{\boldmath{$\Phi$}}_1 \right]_2
  \cdots
  \left[ \hat{\bm c}^{\dagger} \mbox{\boldmath{$\Phi$}}_1 \right]_N
       {\rm e}^{- z_1 \hat{\mathcal{W}}_1 }
  \vert 0 \rangle \nonumber \\
  = & \prod_{n=1}^N
  \left[ \hat{\bm c}^{\dagger} \mbox{\boldmath{$\Phi$}}_1 \right]_n
  \vert 0 \rangle,
\end{align}
and thus we obtain that 
\begin{equation}
  \vert \phi_1 \rangle = \prod_{n=1}^N \left[ \hat{\bm c}^{\dagger} \mbox{\boldmath{$\Phi$}}_1 \right]_n \vert 0 \rangle
  \label{eq:ff:evolv:w:1}
\end{equation}
with
\begin{equation}
  \mbox{\boldmath{$\Phi$}}_1 =
  {\rm e}^{-z_1 {\bm W}_1}
  \mbox{\boldmath{$\Phi$}}_0.
\end{equation}
By repeating the same procedure, we finally arrive at the following formula:  
\begin{align}
  \vert \phi_k \rangle \equiv & {\rm e}^{-z_k \hat{\mathcal{W}}_k}
        {\rm e}^{-z_{k-1} \hat{\mathcal{W}}_{k-1}}
        \cdots
        {\rm e}^{-z_1 \hat{\mathcal{W}}_1}
        \vert \psi_0 \rangle \nonumber \\
= & \prod_{n=1}^N \left[ \hat{\bm c}^{\dagger} \mbox{\boldmath{$\Phi$}}_k \right]_n \vert 0 \rangle
  \label{eq:ff:evolv:w}
\end{align}
with
\begin{equation}
  \mbox{\boldmath{$\Phi$}}_k =
       {\rm e}^{-z_k {\bm W}_k}
       {\rm e}^{-z_{k-1} {\bm W}_{k-1}}
       \cdots
       {\rm e}^{-z_{1} {\bm W}_{1}}       
       \mbox{\boldmath{$\Phi$}}_{0}.
  \label{eq:ff:evolv:w:mat}
\end{equation}
Equations~(\ref{eq:ff:evolv:w}) and (\ref{eq:ff:evolv:w:mat})
yield Eq.~(\ref{eq:tsps:psi}) with $z_k = {\rm i} \theta_k$ and $k=1,2,\cdots,K$. 
Equation~(\ref{eq:psi_im:matform}) is also shown similarly 
with $z_{k} = \tau_k$.

\section{\label{sec:random}Robustness of optimization}

In this Appendix, we show that the optimization of the variational parameters $\mbox{\boldmath{$\theta$}}$ 
in the DQAP ansatz $\vert \psi_M (\mbox{\boldmath{$\theta$}}) \rangle$ is robust for the one-dimensional free-fermion system 
described in Eq.~(\ref{eq:1d:ham}).
To this end, we start the optimization iteration with randomly initialized variational parameters and examine how the variational 
parameters as well as the variational energy are eventually optimized.

\begin{figure}
  \includegraphics[width=\hsize]{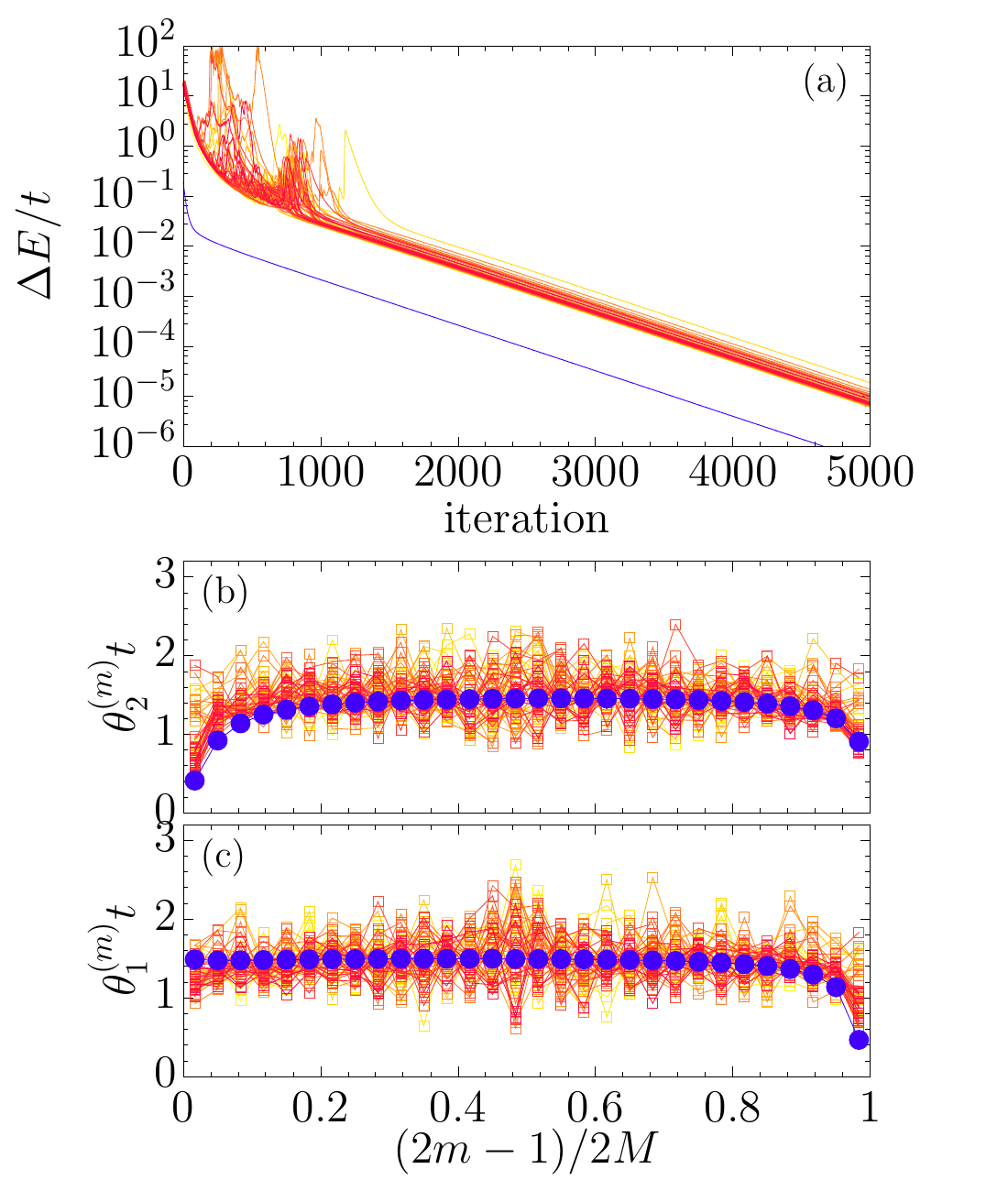}
  \caption{
  (a) Error of the variational energy, $\Delta E$, as a function of the number of optimization iterations, 
    and (b), (c) optimized variational parameters $\theta_2^{(m)}$ and $\theta_1^{(m)}$, started with 50 different sets of 
    initial variational parameters that are chosen randomly within $0 \leq \theta_p^{(m)} \leq 0.01/t$ for $m=1,2,\cdots,M$ and $p=1,2$. 
    The variational parameters in the DQAP ansatz $\vert \psi_M (\mbox{\boldmath{$\theta$}}) \rangle$ with $M=L/4$ 
    are optimized for $L=120$ under APBCs at half-filling, i.e., $N=L/4$, and thus 
    $\vert \psi_M (\mbox{\boldmath{$\theta$}}) \rangle$ can represent the exact ground state of the system. 
    Colors varying from yellow to red indicate the results for 50 different sets of initial parameters.  
     Blue line in (a) and blue circles in (b) and (c) denote the results for the DQAP ansatz 
     $\vert \psi_M (\mbox{\boldmath{$\theta$}}) \rangle$
     with the variational parameters optimized in the procedure 
     described in Sec.~\ref{sec:res:opt}. 
  }
  \label{fig:random}
\end{figure}

Figure~\ref{fig:random}(a) shows the convergence of the variational energy 
as a function of the number of optimization iterations starting with 50 different sets of initial variational parameters that are  
chosen randomly within $0 \leq \theta_p^{(m)} \leq 0.01/t$ for $m=1,2,\cdots,M$ and $p=1,2$. 
The vertical axis in Fig.~\ref{fig:random}(a) is the energy difference $\Delta E$ between the variational energy at a give iteration 
and the exact energy. 
We set $\delta \beta = 0.01$ for the learning rate in Eq.~(\ref{eq:update}) to optimize the variational parameters 
in the DQAP ansatz $\vert \psi_M (\mbox{\boldmath{$\theta$}}) \rangle$ with $M=L/4$ 
for $L=120$ under APBCs at half filling, and thus  
the DQAP ansatz $\vert \psi_M (\mbox{\boldmath{$\theta$}}) \rangle$ can represent the exact ground state. 

As shown in Fig.~\ref{fig:random}(a), 
while they are somewhat scattered in the beginning of the iterations, 
all cases studied with 50 different sets of randomly chosen initial variational parameters finally converge 
to the exact energy exponentially with increasing the number of iterations. 
However, the convergence of the variational energy for the DQAP ansatz $\vert \psi_M (\mbox{\boldmath{$\theta$}}) \rangle$ 
optimized systematically in the procedure described in Sec.~\ref{sec:res:opt} is 
at least one order of magnitude better than that for the DQAP ansatz $\vert \psi_M (\mbox{\boldmath{$\theta$}}) \rangle$ 
optimized with the randomly chosen initial variational parameters.  

More interestingly, we find in Figs.~\ref{fig:random}(b) and \ref{fig:random}(c) that the optimized sets of variational parameters 
are all different, although these sets can reproduce the exact ground state energy. 
Note here that $\theta_p^{(m)}$ has redundancy with a period of $\pi / t$ [see Eqs.~(\ref{eq:mat_v1}) and (\ref{eq:mat_v2})]. 
However, even if we take this redundancy into account, these sets are clearly different. 
Consequently, the single-particle orbitals in the optimized DQAP ansatz $\vert \psi_M (\mbox{\boldmath{$\theta$}}) \rangle$ 
have different shapes, depending on the variational parameters $\mbox{\boldmath{$\theta$}}_p^{(m)}$.  
However, this does not alter the conclusion discussed in Sec.~\ref{sec:res:psi}, qualitatively.

\section{\label{sec:separation}Boundary contribution to entanglement entropy}

As described in Sec.~\ref{sec:res:eent}, the entanglement entropy $S_{\mathbb{A}}$ of the optimized 
DQAP ansatz $\vert \psi_M (\mbox{\boldmath{$\theta$}}) \rangle$ is determined by the number $M$ of layers 
in the DQAP ansatz, independently of system size $L$ and boundary conditions, provided that 
$4M \leq L_A$ and $L_A \leq L_{\bar{A}}$, where $L_A$ ($L_{\bar A}$) is the size of subsystem $\mathbb{A}$ 
(complement of subsystem $\mathbb{A}$) and $L=L_A+L_{\bar A}$. This finding suggests that the 
entanglement entropy $S_{\mathbb{A}}$ is separable to the contributions from the partitioning boundaries 
$\partial\mathbb{A}_{\rm I}$ and $\partial\mathbb{A}_{\rm II}$ between 
the two subsystems, i.e., $S_{\mathbb{A}}\sim S_{\partial\mathbb{A}_{\rm I}} + S_{\partial\mathbb{A}_{\rm II}}$, 
where 
$S_{\partial\mathbb{A}_{\rm I (II)}}$ implies the entanglement entropy from the boundary $\partial\mathbb{A}_{\rm I (II)}$. 
Note that the partitioning boundaries are assumed not to break any local bonding state in the initial state $|\psi_{\rm i}\rangle$ 
(see Fig.~\ref{fig:div}). 
Here, in this Appendix, we discuss more details of this point through the one-particle density matrix ${\bm D}_A$ 
defined in Eq.~(\ref{eq:D_A}).

Let us first introduce the one-particle density matrix ${\bm D}_{\mathbb{U}}$ of the whole system 
${\mathbb{U}}={\mathbb{A}} + \bar{\mathbb{A}}$ as 
\begin{alignat}{1}
  {\bm D}_{\mathbb{U}} &=
  \langle \psi_M (\mbox{\boldmath{$\theta$}}) \vert \hat{\bm c}^{\ast} \hat{\bm c}^t \vert \psi_M (\mbox{\boldmath{$\theta$}}) \rangle \\
  & = \mbox{\boldmath{$\Psi$}}_M^* \mbox{\boldmath{$\Psi$}}_M^{t}, \label{eq:dupp}
\end{alignat}
where $\hat{\bm c}^{\ast}$ ($\hat{\bm c}^t$) is 
the matrix transpose of $\hat{\bm c}^{\dagger}$ ($\hat{\bm c}$) in Eqs.~(\ref{eq:vec:c}) 
and $\mbox{\boldmath{$\Psi$}}_M$ is given in Eq.~(\ref{eq:1d:psi}). 
Note that we have used Eq.~(\ref{eq:ff:expectation}) to derived Eq.~(\ref{eq:dupp}).
Then, we can readily show that
\begin{equation}
  {\bm D}_{\mathbb{U}}^2 = {\bm D}_{\mathbb{U}}
  \label{eq:idem}
\end{equation}
because $\mbox{\boldmath{$\Psi$}}_M^{\dagger} \mbox{\boldmath{$\Psi$}}_M = {\bf I}_{N}$. 
This implies that the eigenvalues
of ${\bm D}_{\mathbb{U}}$ are either 0 or 1. 
In the following, we assume that $\mathbb{A} = \{ 1, 2, 3, \cdots, L_A \}$ and 
$\bar{\mathbb{A}} = \{ L_A+1, L_A+2, L_A+3, \cdots, L \}$, for simplicity.

Let us now write ${\bm D}_{\mathbb{U}}$ as
\begin{equation}
  {\bm D}_{\mathbb{U}} = \left(
  \begin{array}{cc}
    {\bm D}_{\mathbb{A}\mathbb{A}} & {\bm D}_{\mathbb{A}\bar{\mathbb{A}}} \\
    {\bm D}_{\bar{\mathbb{A}}\mathbb{A}} & {\bm D}_{\bar{\mathbb{A}}\bar{\mathbb{A}}} \\
  \end{array}
  \right),
\end{equation}
where ${\bm D}_{\mathbb{A}\mathbb{A}} $ is an $L_A \times L_A$ matrix, 
corresponding to the one-particle density matrix ${\bm D}_{\mathbb{A}}$ of subsystem $\mathbb{A}$ 
defined in Eq.~(\ref{eq:D_A}), and 
${\bm D}_{\bar{\mathbb{A}}\bar{\mathbb{A}}}$ is an $L_{\bar{A}} \times L_{\bar{A}}$ matrix.
Due to the idempotence of ${\bm D}_{\mathbb{U}}$ in Eq.~(\ref{eq:idem}), we find that 
\begin{equation}
  {\bm D}_{\mathbb{A}\mathbb{A}}^2 +
  {\bm D}_{\mathbb{A}\bar{\mathbb{A}}}
  {\bm D}_{\bar{\mathbb{A}}\mathbb{A}}
  = {\bm D}_{\mathbb{A}\mathbb{A}}.
\end{equation}
Considering the spatial extent $d_M = 4 M + 2$ of the single-particle orbitals in the 
DQAP ansatz $\vert \psi_M (\mbox{\boldmath{$\theta$}}) \rangle$ discussed in Sec.~\ref{sec:res:psi},  
we can show that, in general,
\begin{equation}
  {\rm rank} [{\bm D}_{{\mathbb{A}}\bar{\mathbb{A}}}] =   {\rm rank} [{\bm D}_{\bar{\mathbb{A}}\mathbb{A}}] = {\rm min}( L_A, 4M ),
  \label{eq:rank}
\end{equation}
irrespectively of the values of the variational parameters in the 
DQAP ansatz $\vert \psi_M (\mbox{\boldmath{$\theta$}}) \rangle$.
Here, the spatial extent $d_M = 4 M + 2$ of the single-particle orbitals in $\mbox{\boldmath{$\Psi$}}_M$ suggests
that ${\bm D}_{\mathbb{U}} = \mbox{\boldmath{$\Psi$}}_M^{\ast} \mbox{\boldmath{$\Psi$}}^t_M$
is a band-like matrix with $8M+2$ non-zero elements in each row and each column, while 
${\bm D}_{\mathbb{A}\bar{\mathbb{A}}}$ has an opposite matrix structure with non-zero elements appearing at the upper right and 
lower left corners, and the apparent rank of ${\bm D}_{\mathbb{A}\bar{\mathbb{A}}}$ is $8M$. 
For example, when $L=16$, $N=8$, and $M=1$, the matrix structures of 
$\mbox{\boldmath{$\Psi$}}_M^{\ast} \mbox{\boldmath{$\Psi$}}^t_M = {\bm D}_{\mathbb{U}}$ 
are schematically given as  
\begin{eqnarray}
&\left(
\begin{array}{cccccccc}
 * & * & 0 & 0 & 0 & 0 & 0 & * \\
 * & * & 0 & 0 & 0 & 0 & 0 & * \\
 * & * & * & 0 & 0 & 0 & 0 & 0 \\
  * & * & * & 0 & 0 & 0 & 0 & 0 \\
 0 & * & * & * & 0 & 0 & 0 & 0 \\
 0 & * & * & * & 0 & 0 & 0 & 0 \\
 0 & 0 & * & * & * & 0 & 0 & 0 \\
 0 & 0 & * & * & * & 0 & 0 & 0 \\
 0 & 0 & 0 & * & * & * & 0 & 0 \\
 0 & 0 & 0 & * & * & * & 0 & 0 \\
 0 & 0 & 0 & 0 & * & * & * & 0 \\
 0 & 0 & 0 & 0 & * & * & * & 0 \\
 0 & 0 & 0 & 0 & 0 & * & * & * \\
 0 & 0 & 0 & 0 & 0 & * & * & * \\
 * & 0 & 0 & 0 & 0 & 0 & * & * \\
 * & 0 & 0 & 0 & 0 & 0 & * & * \\
\end{array}
\right) \nonumber \\
&\times 
\left(
\begin{array}{cccccccccccccccc}
 * & * & * & * & 0 & 0 & 0 & 0 & 0 & 0 & 0 & 0 & 0 & 0 & * & * \\
 * & * & * & * & * & * & 0 & 0 & 0 & 0 & 0 & 0 & 0 & 0 & 0 & 0 \\
 0 & 0 & * & * & * & * & * & * & 0 & 0 & 0 & 0 & 0 & 0 & 0 & 0 \\
 0 & 0 & 0 & 0 & * & * & * & * & * & * & 0 & 0 & 0 & 0 & 0 & 0 \\
 0 & 0 & 0 & 0 & 0 & 0 & * & * & * & * & * & * & 0 & 0 & 0 & 0 \\
 0 & 0 & 0 & 0 & 0 & 0 & 0 & 0 & * & * & * & * & * & * & 0 & 0 \\
 0 & 0 & 0 & 0 & 0 & 0 & 0 & 0 & 0 & 0 & * & * & * & * & * & * \\
 * & * & 0 & 0 & 0 & 0 & 0 & 0 & 0 & 0 & 0 & 0 & * & * & * & * \\
\end{array}
\right) \nonumber \\
&=
  \left(
  \begin{array}{cccccccc|cccccccc}
  * & * & * & * & * & * & 0 & 0 & 0 & 0 & 0 & 0 & * & * & * & *  \\
  * & * & * & * & * & * & 0 & 0 & 0 & 0 & 0 & 0 & * & * & * & *  \\
  * & * & * & * & * & * & * & * & 0 & 0 & 0 & 0 & 0 & 0 & * & *  \\
  * & * & * & * & * & * & * & * & 0 & 0 & 0 & 0 & 0 & 0 & * & *  \\
  * & * & * & * & * & * & * & * & * & * & 0 & 0 & 0 & 0 & 0 & 0  \\
  * & * & * & * & * & * & * & * & * & * & 0 & 0 & 0 & 0 & 0 & 0  \\
  0 & 0 & * & * & * & * & * & * & * & * & * & * & 0 & 0 & 0 & 0  \\
  0 & 0 & * & * & * & * & * & * & * & * & * & * & 0 & 0 & 0 & 0  \\ \hline
  0 & 0 & 0 & 0 & * & * & * & * & * & * & * & * & * & * & 0 & 0  \\
  0 & 0 & 0 & 0 & * & * & * & * & * & * & * & * & * & * & 0 & 0  \\
  0 & 0 & 0 & 0 & 0 & 0 & * & * & * & * & * & * & * & * & * & *  \\
  0 & 0 & 0 & 0 & 0 & 0 & * & * & * & * & * & * & * & * & * & *  \\
 * & * & 0 & 0 & 0 & 0 & 0 & 0 & * & * & * & * & * & * & * & *  \\
 * & * & 0 & 0 & 0 & 0 & 0 & 0 & * & * & * & * & * & * & * & *  \\
 * & * & * & * & 0 & 0 & 0 & 0 & 0 & 0 & * & * & * & * & * & *  \\
 * & * & * & * & 0 & 0 & 0 & 0 & 0 & 0 & * & * & * & * & * & *  \\
  \end{array}
  \right), \label{eq:matrix_da}
\end{eqnarray}
where $*$ indicates a non-zero element and ${\bm D}_{\mathbb{A}\bar{\mathbb{A}}}$ is the upper right quadrant of 
the matrix in the right hand side. 
However, due to the characteristic structure of $\mbox{\boldmath{$\Psi$}}_M$, a single-particle orbital 
being extended spatially by two lattice spaces in each spatial direction every time the local time-evolution operators 
are applied, 
we find that the Gaussian elimination eliminates 
a half of the non-zero row (or column) vectors in ${\bm D}_{\mathbb{A}\bar{\mathbb{A}}}$, 
implying that the non-zero row (or column) vectors in ${\bm D}_{\mathbb{A}\bar{\mathbb{A}}}$ are 
linearly dependent and only the half of them are linearly independent, which leads to Eq.~(\ref{eq:rank}).
Because ${\bm D}_{\bar{\mathbb{A}}\mathbb{A}}  = {\bm D}_{\mathbb{A}\bar{\mathbb{A}}}^{\dagger}$ and thus 
${\rm rank} [{\bm D}_{{\mathbb{A}}\bar{\mathbb{A}}} {\bm D}_{\bar{\mathbb{A}}\mathbb{A}} ] = {\rm rank} [{\bm D}_{{\mathbb{A}}\bar{\mathbb{A}}}] $, 
we finally obtain that
\begin{equation}
  {\rm rank} [ {\bm D}_{\mathbb{A}\mathbb{A}}^2 - {\bm D}_{\mathbb{A}\mathbb{A}} ] =
  {\rm min} ( L_A, 4M ).
  \label{eq:rankD}
\end{equation}

\begin{figure}
  \includegraphics[width=\hsize]{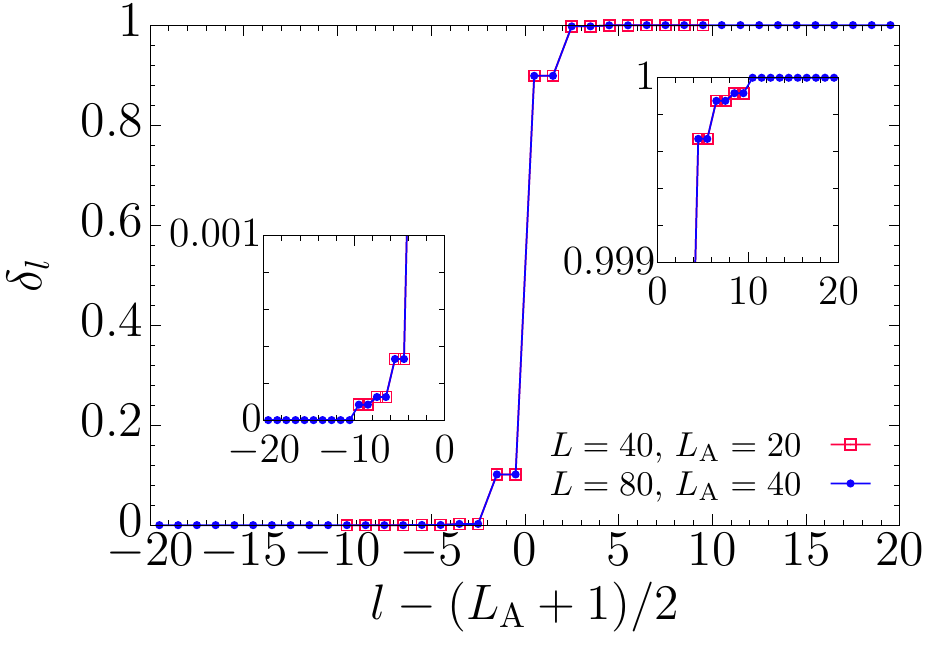}
  \caption{Eigenvalues $\delta_l$ of the one-particle density matrix ${\bm D}_{\mathbb{A}}$
    for the optimized DQAP ansatz $\vert \psi_M (\mbox{\boldmath{$\theta$}}) \rangle$ 
    with $M=5$. The variational parameters are optimized for $L=40$ and $80$ under APBC at half filling,
    i.e., $N=L/2$. We assume 
    that the size of subsystem ${\mathbb{A}}$ is $L_A=L/2$ and thus $4M\le L_A$.
    Insets: Enlarged plots around $\delta_l =0$ and $\delta_l = 1$. 
    Notice that 
    (i) all eigenvalues are symmetric around $1/2$,
    (ii) all eigenvalues for $L=40$ are neither 0 nor 1 , (iii) all eigenvalues (except for the eigenvalues being either 0 or 1) 
    for $L=80$ are identical to those for $L=40$, and (iv) all eigenvalues with $0<\delta_l<1$ are pairwise degenerate. 
  }
  \label{fig:deigen}
\end{figure}

Equation~(\ref{eq:rankD}) immediately implies that, for the DQAP ansatz $\vert \psi_M (\mbox{\boldmath{$\theta$}}) \rangle$ 
with $4M \le L_A$,
$L_A - 4 M$ eigenvalues of ${\bm D}_{\mathbb{A}\mathbb{A}}$ ($= {\bm D}_{\mathbb{A}}$) are either 0 or 1. 
As shown schematically in Fig.~\ref{fig:div}(a), there are $L_A/2-2M$ single-particle orbitals in the DQAP ansatz 
$\vert \psi_M (\mbox{\boldmath{$\theta$}}) \rangle$ that do not cross either side of the partitioning boundaries between the 
two subsystems and stay inside subsystem $\mathbb{A}$. 
These $L_A/2-2M$ single-particle orbitals contribute to the eigenvalues of ${\bm D}_{\mathbb{A}}$ with $\delta_l=1$. 
Due to the particle-hole symmetry, the eigenvalues of ${\bm D}_{\mathbb{A}}$
should appear symmetrically around $1/2$~\cite{Barghathi2018,Wybo2020}. Therefore, there exist 
$L_A/2 -2 M$ eigenvalues of ${\bm D}_{\mathbb{A}}$ with $\delta_l=0$,  
corresponding to the unoccupied single-particle orbitals that stay inside subsystem $\mathbb{A}$ without 
crossing the partitioning boundaries. 
The remaining $4M$ eigenvalues of ${\bm D}_{\mathbb{A}}$ are neither 0 nor 1, i.e., $0<\delta_l<1$. 
These contributions are due to the single-particle orbitals ($2M$) in the DQAP ansatz $\vert \psi_M (\mbox{\boldmath{$\theta$}}) \rangle$ 
that cross either side of the partitioning boundaries between the two subsystems and the hole counterparts ($2M$) 
due to the particle-hole symmetry.  
These eigenvalues of ${\bm D}_{\mathbb{A}}$ with $0<\delta_l<1$ contribute to the non-zero entanglement entropy $S_{\mathbb{A}}$ in 
Eq.~(\ref{eq:ent}).

\begin{figure}
  \includegraphics[width=\hsize]{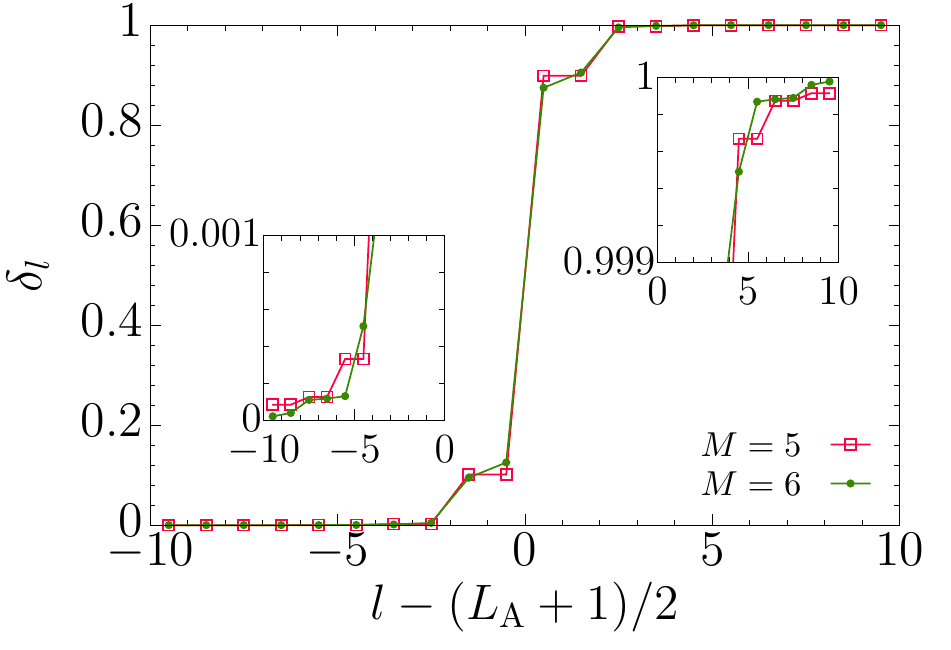}
  \caption{
  Same as Fig.~\ref{fig:deigen} but for $L=40$ with $M=6$, i.e., $4M>L_A$. 
  For comparison, the same results for $L=40$ with $M=5$ shown in Fig.~\ref{fig:deigen} are also plotted. 
  }
  \label{fig:deigen_M}
\end{figure}

Figure~\ref{fig:deigen} shows the numerical results of the eigenvalues of ${\bm D}_{\mathbb{A}}$ for the 
optimized DQAP ansatz $\vert \psi_M (\mbox{\boldmath{$\theta$}}) \rangle$ with $M=5$ and two different system 
sizes $L=40$ and 80, assuming that $L_A=L/2$ and thus $4M\le L_A$. 
First, we can notice that the eigenvalues are all symmetric around $1/2$, as expected due to the particle-hole symmetry. 
Second, we can confirm that all the eigenvalues are neither 0 nor 1 for $L=40$. Third, there are $L_A/2 - 2 M = 10$ eigenvalues 
with $\delta_l = 1$ as well as 10 eigenvalues with
$\delta_l = 0$ for $L=80$. In addition, we can find that other eigenvalues different from 0 or 1 for 
$L=80$ are identical to the eigenvalues of ${\bm D}_{\mathbb{A}}$ found for $L=40$. These eigenvalues $\delta_l$ 
with $0<\delta_l<1$ contribute 
to the entanglement entropy $S_{\mathbb{A}}$, and therefore this finding is in good agreement with the result that 
the entanglement entropy $S_{\mathbb{A}}$ of the optimized DQAP ansatz $\vert \psi_M (\mbox{\boldmath{$\theta$}}) \rangle$ 
with $4M\le L_A$ is independent of system size $L$ (see Fig.~\ref{fig:ent}). 
Fourth, the eigenvalues $\delta_l$ with $0<\delta_l<1$ are pairwise degenerate. 
Considering that only these $4M$ eigenvalues $\delta_l$ with $0<\delta_l<1$ 
contribute to the entanglement entropy $S_{\mathbb{A}}$ and these eigenvalues correspond to the single-particle orbitals 
in the DQAP ansatz $\vert \psi_M (\mbox{\boldmath{$\theta$}}) \rangle$ that cross either side of the partitioning boundaries and the hole 
counterparts, 
we interpret the pairwise degeneracy as a fingerprint that the contribution to the entanglement entropy $S_{\mathbb{A}}$ 
can be separated from 
the two partitioning boundaries $\partial \mathbb{A}_{\rm I}$ and $\partial \mathbb{A}_{\rm II}$, i.e, 
$S_{\mathbb{A}}\sim S_{\partial\mathbb{A}_{\rm I}} + S_{\partial\mathbb{A}_{\rm II}}$. 
Indeed, as shown in Fig.~\ref{fig:deigen_M}, the pairwise degeneracy disappears once $4M>L_A$~\cite{Chang2019}.

\section{Evolution time in a continuous-time quantum adiabatic process}\label{sec:conttime}

To make a comparison with the effective total evolution time $T_{\rm eff}(L)$ of the DQAP discussed in 
Fig.~\ref{fig:times},
here we estimate the total evolution time necessary to obtain the ground state within a given accuracy in the 
continuous-time quantum adiabatic process with a linear scheduling.

According to the quantum adiabatic theorem~\cite{Morita2007}, 
the transition amplitude at time $\tau$ to the excited state $\vert \phi_{\alpha} (\tau) \rangle$ 
($\alpha=1,2,\cdots$) is generally given by 
\begin{equation}
  \vert \langle \phi_{\alpha} (\tau) \vert \Psi(\tau) \rangle \vert \sim
  \frac{\vert \langle \phi_{\alpha}(\tau) \vert \partial_{\tau} \hat{\mathcal{H}}(\tau) \vert \phi_0 (\tau) \rangle \vert}
       {( \Omega_{\alpha} (\tau) - \Omega_0(\tau))^2}, 
       \label{eq:qat}
\end{equation}
where $\vert \phi_{\alpha} (\tau) \rangle$ is the $\alpha$th eigenstate of the instantaneous Hamiltonian $\hat{\mathcal{H}}(\tau)$ 
at time $\tau$ with the eigenvalue $\Omega_{\alpha}(\tau)$, and 
$\vert \phi_0 (\tau) \rangle$ is the ground state of $\hat{\mathcal{H}}(\tau)$ with its eigenvalue 
$\Omega_0(\tau)\,(<\Omega_\alpha(\tau))$. 
$\vert \Psi (\tau) \rangle$ is the time-evolving state at time $\tau$ via the time-dependent Schr\"odinger equation 
from the initial state $\vert \Psi(0) \rangle = \vert \phi_0 (0) \rangle$ at time $\tau=0$. 
$\partial_{\tau} \hat{\mathcal{H}}(\tau)$ indicates the time derivative of $\hat{\mathcal{H}}(\tau)$.

For the one-dimensional free-fermion system in Eq.~(\ref{eq:1d:ham}), the time-dependent Hamiltonian $\hat{\mathcal{H}}(\tau)$ 
is given as
\begin{equation}
  \hat{\mathcal{H}}(\tau) = \hat{\mathcal{V}}_1 + (\tau/T) \hat{\mathcal{V}}_2, 
  \label{eq:ham_lin}
\end{equation}
where $\hat{\mathcal{V}}_1$ and $\hat{\mathcal{V}}_2$ are defined in Eqs.~(\ref{eq:1d:v:bond1}) and (\ref{eq:1d:v:bond2}), respectively, 
and the linear scheduling is assumed with the total evolution time $T$, i.e., 
the initial time $\tau_{\rm i}=0$ and the final time $\tau_{\rm f}=T$ 
in Eqs.~(\ref{eq:lin_sche}).
In this case, the derivative of the time-dependent Hamiltonian $\hat{\mathcal{H}}(\tau)$ is simply 
\begin{equation}
  \partial_{\tau} \hat{\mathcal{H}}(\tau) = \frac{1}{T} \hat{\mathcal{V}}_2. 
\end{equation}
An upper bound of the numerator in Eq.~(\ref{eq:qat})
is given by the operator norm~\cite{Duan2020}, i.e., 
\begin{align}
  & \vert \langle \phi_\alpha(\tau) \vert \partial_{\tau} \hat{\mathcal{H}}(\tau)  \vert \phi_0 (\tau) \rangle \vert \nonumber \\
  \leq &  \underset{\vert \psi \rangle \text{ with } \langle \psi \vert \psi \rangle = 1}{\rm max}
  \vert \langle \psi \vert \partial_{\tau} \hat{\mathcal{H}}(\tau) \vert \psi \rangle \vert
  = \frac{L t}{2T}.
  \label{eq:cont:over}
\end{align}
The last equality follows from the fact that $\hat{\mathcal{V}}_2$ is a direct sum of $L/2$
two-qubit $XY$ Hamiltonians acting on every distinct pairs of adjacent qubits.
However, as shown in the following, we find that
the upper bound given in Eq.~(\ref{eq:cont:over}) 
overestimates by a factor of $O(L)$.

Let us first represent $\hat{\mathcal{H}}(\tau)$ in the matrix form
\begin{equation}
  \hat{\mathcal{H}}(\tau) = \hat{\bm c}^{\dagger} {\bm T}(\tau) \hat{\bm c}
\end{equation}
with
\begin{equation}
  {\bm T}(\tau) = {\bm V}_1 + (\tau/T) {\bm V}_2,
\end{equation}
where $\hat{\bm c}^{\dagger}$ and $\hat{\bm c}$ are defined in Eqs.~(\ref{eq:vec:c}). 
We then diagonalize the $L\times L$ matrix ${\bm T}(\tau)$ by a unitary matrix ${\bm U}(\tau)$ as 
\begin{equation}
  {\bm T}(\tau) = {\bm U}^\dagger(\tau) {\bm E}(\tau) {\bm U}(\tau), 
\end{equation}
where ${\bm E}(\tau) = {\rm diag}(E_1 (\tau), E_2 (\tau), \cdots, E_L(\tau) )$
is the diagonal matrix with the diagonal elements being the eigenvalues of the matrix ${\bm T}(\tau)$.
The Hamiltonian $\hat{\mathcal{H}}(\tau)$ is then represented as
\begin{equation}
  \hat{\mathcal{H}}(\tau) = \hat{\bm a}^{\dagger}(\tau) {\bm E}(\tau) \hat{\bm a}(\tau)
  = \sum_{n=1}^L E_n(\tau) \hat{a}_{n}^{\dagger}(\tau) \hat{a}_{n}(\tau),
\end{equation}
where $\hat{\bm a}^{\dagger}(\tau)$ and $\hat{\bm a}(\tau)$ are the fermion operators given by
\begin{equation}
  \hat{\bm a}^{\dagger}(\tau) = \hat{\bm c}^{\dagger} {\bm U}(\tau), \
  \hat{\bm a}(\tau) = {\bm U}^{\dagger}(\tau) \hat{\bm c}. 
\end{equation}
It is important to notice here that, since the Hamiltonian $\hat{\mathcal{H}}(\tau)$ is non-local for $\tau > 0$,
the operator $\hat{a}_{n}^\dag(\tau)$ representing the $n$th single-particle orbital is no longer local 
and the weight at each site can be approximated by $1/\sqrt{L}$, i.e., 
$\vert U_{xn}(\tau) \vert \sim 1/\sqrt{L}$, where
$\left[ {\bm U}(\tau) \right]_{xn} = U_{xn}(\tau)$.

The ground state $\vert \phi_0(\tau) \rangle$ of the Hamiltonian $\hat{\mathcal{H}}(\tau)$ at time $\tau$ 
with $N$ fermions is given as
\begin{equation}
  \vert \phi_0 (\tau) \rangle = \prod_{n=1}^N a_{n}^{\dagger}(\tau) \vert 0 \rangle 
\end{equation}
with $\Omega_0(\tau)=\sum_{n=1}^N E_n(\tau)$, 
assuming that $E_1(\tau)\le E_2(\tau)\le \cdots \le E_L(\tau)$. 
Since $\partial_{\tau} \hat{\mathcal{H}}(\tau)$ is a single-particle operator,
the possible excitations are restricted to particle-hole excitations, i.e.,  
\begin{equation}
  \vert \phi_{\alpha} (\tau) \rangle = \hat{a}_{n_1}^{\dagger}(\tau) \hat{a}_{n_2}(\tau) \vert \phi_{0} (\tau) \rangle, 
\end{equation}
where $n_1 > N$ and $n_2 \leq N$.
We then find that
\begin{align}
  & \langle \phi_{\alpha} (\tau) \vert \partial_{\tau} \hat{\mathcal{H}} \vert \phi_0 (\tau) \rangle \nonumber \\
  & = \frac{1}{T} \sum_{x=1}^{L/2} [ {\bm V}_2 ]_{2x,2x+1} U_{2x,n_1}^{\ast}(\tau) U_{2x+1,n_2}(\tau) \nonumber \\
  & + \frac{1}{T} \sum_{x=1}^{L/2} [ {\bm V}_2 ]_{2x+1,2x} U_{2x+1,n_1}^{\ast}(\tau) U_{2x,n_2} (\tau)
\end{align}
and hence
\begin{equation}
  \vert \langle \phi_{\alpha} (\tau) \vert \partial_{\tau} \hat{\mathcal{H}} \vert \phi_0 (\tau) \rangle \vert
  \sim \frac{L^0t}{T},
\end{equation}
where $[{\bm V}_2]_{2x,2x+1}$ for $x=L/2$ is assumed to be $[{\bm V}_2]_{L,1}$. 
Noting that the minimum of the spectral gap $\Omega_{\alpha}(\tau) - \Omega_0(\tau)$ 
is realized at $\tau=\tau_f$ and is proportional to $t/L$ for the one-dimensional free-fermion system, 
one can estimate the maximum of the transition amplitude as
\begin{equation}
  \left. \frac{
  \vert \langle \phi_{\alpha}(\tau) \vert \partial_{\tau} \hat{\mathcal{H}}(\tau)
  \vert \phi_{0} (\tau) \rangle \vert}
  {(\Omega_{\alpha}(\tau) - \Omega_0(\tau))^2} \right|_{\tau=\tau_{\rm f}}
  \sim \frac{L^2}{tT}. 
  \label{eq:tanaly}
\end{equation}
Therefore, the evolution time $T$ has to be at least as large as $T \sim L^2$ 
for the time-evolving state $\vert \Psi(\tau) \rangle$ to follow the quantum dynamics adiabatically.

Next, we confirm this analysis directly by numerically solving the time-dependent Schr\"odinger equation 
in Eqs.~(\ref{eq:qap}) and (\ref{eq:ut}). 
In the numerical simulation,
the time evolution operator $\hat{\mathcal{U}}(\tau,\tau_{\rm i})$ 
is treated as
\begin{equation}
  \hat{\mathcal{U}}(\tau,\tau_{\rm i}) = \prod_{m=M}^{1} \hat{\mathcal{U}}(\tau_{m},\tau_{m-1})
\end{equation}
and 
\begin{equation}
\hat{\mathcal{U}}(\tau_{m},\tau_{m-1}) =\mathcal{T}_\tau {\rm e}^{-{\rm i}  \int_{\tau_{m-1}}^{\tau_m}\hat{\mathcal{H}}(\tau)d\tau}, 
\end{equation}
where $\tau_m$ ($m=0,1,2,\cdots,M$) is the discretized time 
\begin{equation}
  \tau_m = m \delta \tau_M
\end{equation}
with the small time step
\begin{equation}
   \delta \tau_M = T/M 
\end{equation}
and $\mathcal{T}_\tau$ is the time-ordered operator. 
Following the Magnus expansion~\cite{Magnus1954},
the time evolution operator $\hat{\mathcal{U}}(\tau_{m},\tau_{m-1})$ for the small time step $\delta \tau_M$ can be expressed as
\begin{equation}
  \hat{\mathcal{U}} (\tau_{m},\tau_{m-1}) = \exp
  \left[
    \hat{\mathcal{F}}(\tau_m,\tau_{m-1})
    \right],
\end{equation}
where $\hat{\mathcal{F}}(\tau_m,\tau_{m-1})$ 
is expanded in 
the order of the perturbation, 
\begin{equation}
  \hat{\mathcal{F}}(\tau_m,\tau_{m-1}) = \sum_{\mu = 1} \hat{\mathcal{F}}_{\mu}(\tau_{m},\tau_{m-1}), 
  \label{eq:fexp}
\end{equation}
and the $\mu$th order term $ \hat{\mathcal{F}}_{\mu}(\tau_{m},\tau_{m-1})$ is anti-Hermitian, 
i.e., $\left[ \hat{\mathcal{F}}_{\mu}(\tau_{m},\tau_{m-1})\right]^\dag = - \hat{\mathcal{F}}_{\mu}(\tau_{m},\tau_{m-1})$. 
Therefore, unitarity of the time evolution operator $\hat{\mathcal{U}} (\tau_{m},\tau_{m-1})$ is guaranteed 
even when the expansion of $\hat{\mathcal{F}}(\tau_m,\tau_{m-1})$ in Eq.~(\ref{eq:fexp}) is terminated at a finite order. 
 
The first and second order terms in $\hat{\mathcal{F}}(\tau_{m},\tau_{m-1})$ are given respectively as 
\begin{equation}
  \hat{\mathcal{F}}_1(\tau_{m},\tau_{m-1}) = - \frac{{\rm i}\delta \tau_M}{2} ( \hat{\mathcal{H}}(\tau_m) + \hat{\mathcal{H}}(\tau_{m-1}) ) 
\end{equation}
and 
\begin{equation}
  \hat{\mathcal{F}}_2(\tau_{m},\tau_{m-1}) = - \frac{({\rm i}\delta \tau_M)^2}{6}
  [ \hat{\mathcal{H}}(\tau_m), \hat{\mathcal{H}}(\tau_{m-1}) ].
\end{equation}
Note that for any single-particle operators
\begin{equation}
  \hat{\mathcal{A}} = \hat{\bm c}^{\dagger} {\bm A} \hat{\bm c}
\end{equation}
and
\begin{equation}
  \hat{\mathcal{B}} = \hat{\bm c}^{\dagger} {\bm B} \hat{\bm c} 
\end{equation}
with $\bm A$ and $\bm B$ being $L\times L$ matrices, 
the commutator of these single-particle operators,
\begin{equation}
  [ \hat{\mathcal{A}}, \hat{\mathcal{B}} ] = \hat{\bm c}^{\dagger}
   \left[ {\bm A}, {\bm B} \right]  \hat{\bm c},
\end{equation}
is still a single-particle operator. 
Therefore, as in the case of the free-fermion system described by the time-dependent Hamiltonian $\hat{\mathcal{H}}(\tau)$ 
with the linear scheduling in Eq.~(\ref{eq:ham_lin}), 
all order terms $\hat{\mathcal{F}}_{\mu}(\tau_m,\tau_{m-1})$, including the higher order terms, in the Magnus expansion
remain in the form of single-particle operators. Thus, 
we can use the algebra for free fermions without any difficulty.

\begin{figure}
  \includegraphics[width=\hsize]{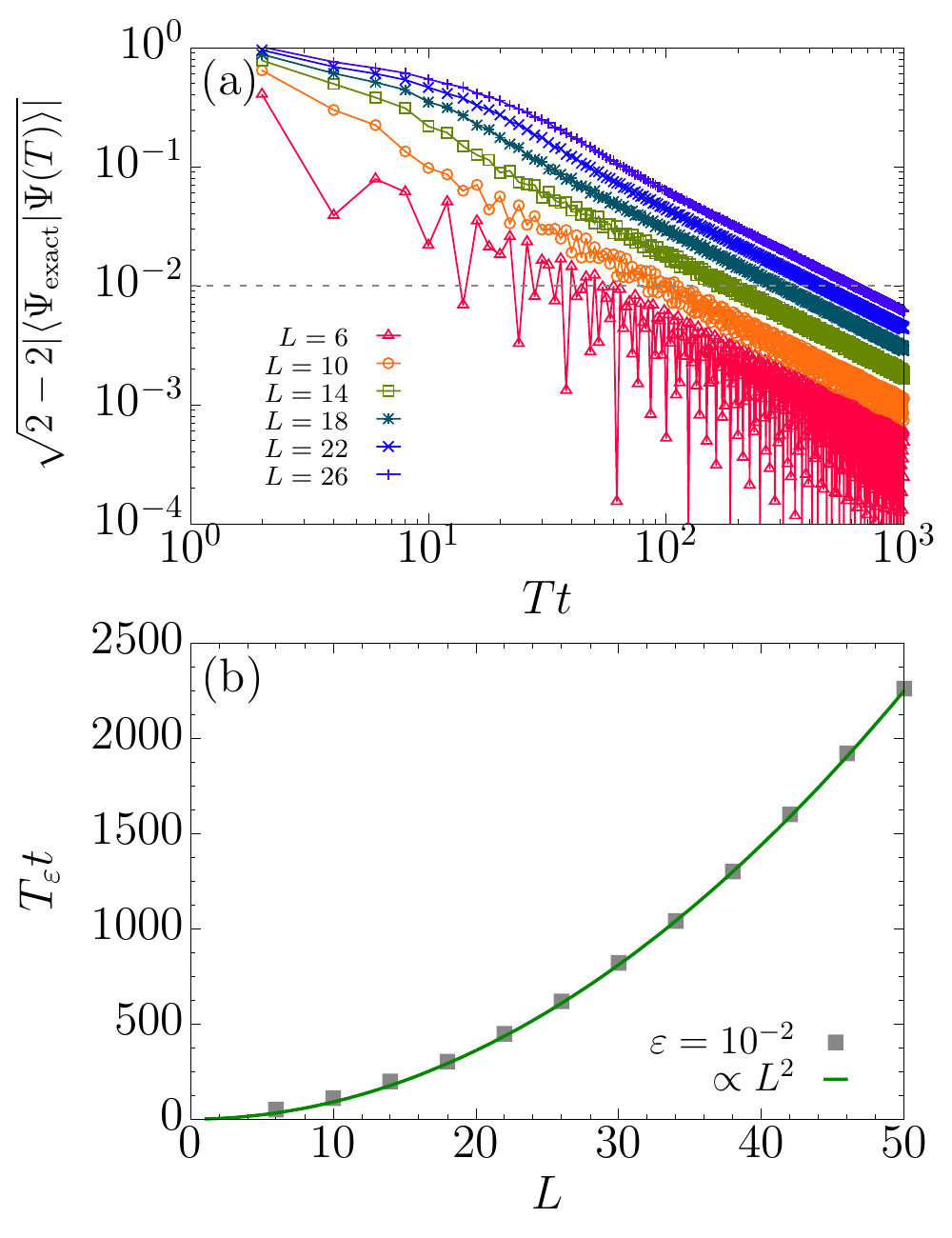}
  \caption{
  (a) Error $\varepsilon = \sqrt{2 - 2 \vert \langle \Psi_{\rm exact} \vert \Psi(T) \rangle \vert}$
    in fidelity between the final state $\vert \Psi (T) \rangle$ in the continuous-time quantum adiabatic process with the linear scheduling 
    and the exact ground state $\vert \Psi_{\rm exact} \rangle$ of the final Hamiltonian. 
    The results are obtained by the numerical simulation for the one-dimensional free-fermion system described in 
    Eq.~(\ref{eq:ham_lin}) with the total evolution time $T$ and various system sizes $L$ 
    under PBCs at half filling. 
    (b) System size dependence of the total evolution time $T_{\varepsilon}$ necessary to obtain the ground state 
    of the final Hamiltonian within the accuracy of error $\varepsilon = 0.01$
    indicated by gray dashed line in (a).
    For comparison, a function proportional to $L^2$ is also plotted by green solid line. 
  }
  \label{fig:conttime}
\end{figure}

We perform the numerical simulation with keeping $\delta \tau_M = 0.01 / t$,  
for which we find that the result by using the time-evolution operator with only the first order expansion 
is essentially 
unchanged even when we use the time-evolution operator expanded up to the second order, 
suggesting that the results are well converged.
Figure~\ref{fig:conttime}(a) shows the error $\varepsilon$ in fidelity between the final state 
$\vert \Psi(T) \rangle$ and the exact ground state $\vert \Psi_{\rm exact} \rangle$ of the final Hamiltonian, 
\begin{equation}
  \varepsilon = \sqrt{ 2 - 2 \vert \langle \Psi_{\rm exact} \vert \Psi(T) \rangle \vert},
\end{equation}
plotted as a function of the total evolution time $T$. 
We find that the error decreases asymptotically as $\varepsilon\sim1/T$ for a give system size $L\, (\agt10)$.
This is understood because the leading term of the error $\varepsilon$ is 
proportional to the transition amplitude given in Eq.~(\ref{eq:tanaly}). 
Figure~\ref{fig:conttime}(b) shows the system size dependence of the total evolution time $T_{\varepsilon}$ necessary to obtain the 
ground state of the final Hamiltonian within the accuracy of error $\varepsilon = 0.01$. 
We indeed find that $T_{\varepsilon}$ is proportional to $L^2$, except for small values of $L$, which is in good agreement with 
the analytical result in Eq.~(\ref{eq:tanaly}).
We thus conclude that the total evolution time  
necessary to obtain the ground state within a given accuracy 
in the continuous-time quantum adiabatic process with the linear scheduling 
is proportional to $L^2$.  
This is in sharp contrast to the case of the DQAP where the effective total evolution time $T_{\rm eff}(L)$ necessary 
for the DQAP ansatz $\vert \psi_M(\mbox{\boldmath{$\theta$}})\rangle$ to converge to the exact ground state of the final Hamiltonian 
is proportional to $L$, as shown in Fig.~\ref{fig:times}.

\section{\label{sec:qab}Optimum scheduling by quantum adiabatic brachistochrone}

QAB~\cite{Rezakhani2009} is a method to find 
an optimum scheduling function 
for the quantum adiabatic process. 
To illustrate this method,
here we consider the system described by the following Hamiltonian: 
\begin{equation}
  \hat{\mathcal{H}} [ \mbox{\boldmath{$\chi$}} ] = \sum_{p=1}^P \chi_p \hat{\mathcal{V}}_p,
\end{equation}
where $\hat{\mathcal{V}}_p$ is time independent and the quantum dynamics of the system is controlled through a set of parameters 
$\mbox{\boldmath{$\chi$}} = \{ \chi_1, \chi_2, \cdots, \chi_P \}$. 
Namely, these parameters are varied as a function of time $\tau$, i.e., $\mbox{\boldmath{$\chi$}} = \mbox{\boldmath{$\chi$}}(\tau)$.
For simplicity, we reparametrize this function via a dimensionless parameter $s(\tau)$
with $s(0) = 0$ at the initial time $\tau_{\rm i}=0$ and $s(T) = 1$ at the final time $\tau_{\rm f}=T$ (i.e., the total evolution time $T$), 
for instance, by rescaling $s(\tau) = \tau/T$.

In the QAB, the functional to be optimized is 
\begin{equation}
  T [ \dot{\mbox{\boldmath{$\chi$}}}, \mbox{\boldmath{$\chi$}} ] =
  \int_0^{1} {\rm d}s \mathcal{L}[ \dot{\mbox{\boldmath{$\chi$}}}, \mbox{\boldmath{$\chi$}} ],
  \label{eq:tchi}
\end{equation}
where the Lagrangian $\mathcal{L}[ \dot{\mbox{\boldmath{$\chi$}}}, \mbox{\boldmath{$\chi$}} ]$ is given by
\begin{equation}
  \mathcal{L} [ \dot{\mbox{\boldmath{$\chi$}}}, \mbox{\boldmath{$\chi$}} ] =
  \frac{\Vert \sum_{p=1}^P \dot{\chi}_p \hat{\mathcal{V}}_p \Vert^2}{\Delta^4 [\mbox{\boldmath{$\chi$}}(s) ]}.
  \label{eq:lag}
\end{equation}
Here, $\Vert \hat{\mathcal{A}} \Vert$ denotes the Hilbert-Schmidt norm, i.e., 
$\Vert \hat{\mathcal{A}} \Vert = \sqrt{ {\rm Tr} [ \hat{\mathcal{A}}^{\dagger} \hat{\mathcal{A}} ] }$,
$\Delta [ \mbox{\boldmath{$\chi$}}(s) ]$ is the minimum gap between the ground state and the first excited state 
for the Hamiltonian $\hat{\mathcal{H}}[\mbox{\boldmath{$\chi$}}]$ at time $s$,
and $\dot{\chi}_p = \frac{{\rm d} \chi_p}{{\rm d}s}$.
Note that the Lagrangian 
\begin{equation}
\frac{ \Vert \dot{\hat{\mathcal{H}}} \Vert^2 }{ \Delta^4 (s) }
\end{equation}
corresponds to the upper bound of the transition probability at a given $s$.
Therefore, the QAB determines the optimum path for $\mbox{\boldmath{$\chi$}}$
so as to minimize the total transition probability
by solving the Euler-Lagrange equation for $T [ \dot{\mbox{\boldmath{$\chi$}}}, \mbox{\boldmath{$\chi$}} ] $ 
in Eq.~(\ref{eq:tchi}):
\begin{equation}
  \frac{{\rm d}}{{\rm d}s} \left( \frac{\partial \mathcal{L}}{\partial \dot{\chi}_p}  \right)
  - \frac{\partial \mathcal{L}}{\partial \chi_p} = 0.
  \label{eq:ele}
\end{equation}
Inserting the explicit form of the Lagrangian $\mathcal{L} [ \dot{\mbox{\boldmath{$\chi$}}}, \mbox{\boldmath{$\chi$}} ]$ given in 
Eq.~(\ref{eq:lag}) into the Euler-Lagrange equation,
we obtain the following equation: 
\begin{equation}
  \ddot{\chi}_p + \sum_{i,j} \Gamma^p_{ij} \dot{\chi}_i \dot{\chi}_j = 0,
\end{equation}
where
\begin{equation}
  \Gamma^p_{ij} = \frac{2}{\Delta} \left( [ {\bm C} ]_{ij}  \sum_{q} [{\bm C}^{-1} ]_{pq} \frac{\partial \Delta}{\partial \chi_q}
  - \delta_{pi} \frac{\partial \Delta}{\partial \chi_j}
  - \delta_{pj} \frac{\partial \Delta}{\partial \chi_i} \right) ,
\end{equation}
with
\begin{equation}
  [ {\bm C} ]_{ij} = {\rm Tr}[ \hat{\mathcal{V}}_i \hat{\mathcal{V}}_j ].
\end{equation}

\begin{figure}
  \includegraphics[width=\hsize]{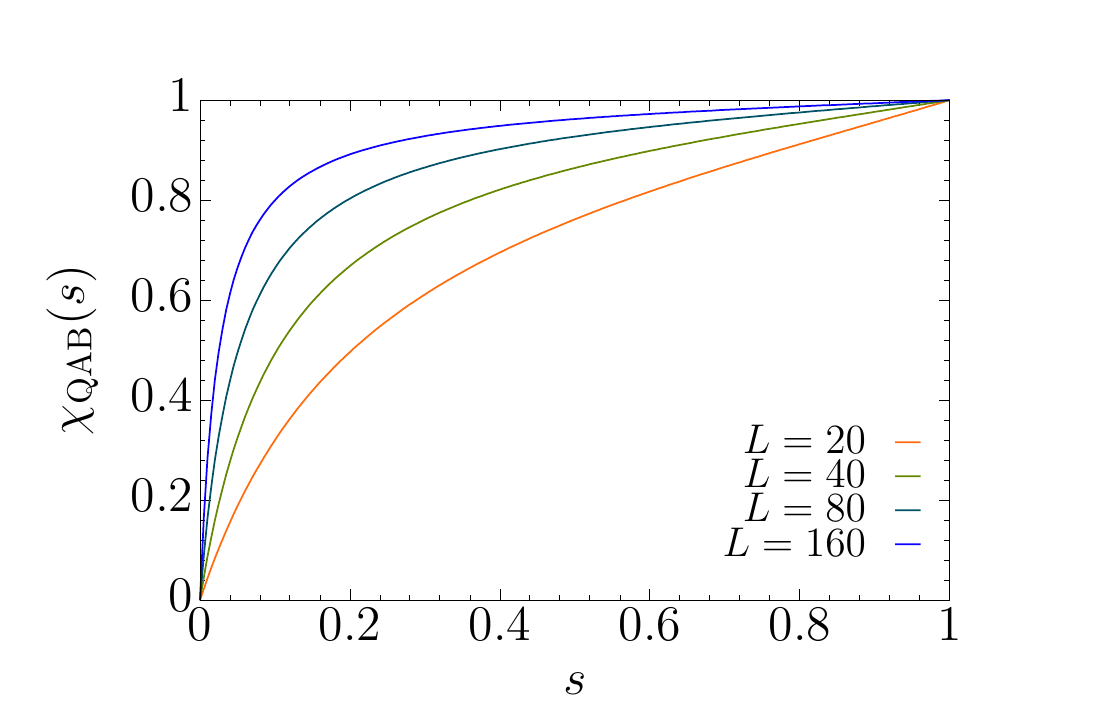}
  \caption{The optimum scheduling function $\chi_{\rm QAB}(s)$
    obtained by the quantum adiabatic brachistochrone (QAB) for the one-dimensional free-fermion system for various system 
    sizes $L$ at half filling.}
  \label{fig:qab}
\end{figure}

We shall now apply this theory to the one-dimensional free-fermion system described by the Hamiltonian in Eq.~(\ref{eq:1d:ham}).
To facilitate an analytical treatment, we consider the following parametrization:
\begin{equation}
  \hat{\mathcal{H}} [\chi] = \hat{\mathcal{V}}_1 + \chi \hat{\mathcal{V}}_2
\end{equation}
where $\hat{\mathcal{V}}_1$ and $\hat{\mathcal{V}}_2$ are given in Eqs.~(\ref{eq:1d:v:bond1}) and (\ref{eq:1d:v:bond2}), respectively.
Assuming that the system size is $L$, the minimum gap $\Delta(\chi)$ in this case is 
\begin{equation}
  \Delta (\chi) = 2 t \sqrt{ ( \chi - \cos ( 2\pi / L ) )^2 + \sin^2 ( 2 \pi / L ) }
\end{equation}
for both PBCs and APBCs with the closed shell condition at half filling. 
By inserting these into the Euler-Lagrange equation in Eq.~(\ref{eq:ele}),
we obtain the following differential equation for the parameter $\chi$:
\begin{equation}
  \ddot{\chi} - 2 \frac{ \chi - \cos(2\pi/L) }{( \chi - \cos(2\pi/L))^2 + \sin^2(2\pi/L) } \dot{\chi}^2 = 0.
\end{equation}
The solution $\chi_{\rm QAB} (s)$ of this differential equation under the conditions $\chi(0)=0$ and $\chi(1)=1$ is readily found as 
\begin{equation}
  \chi_{\rm QAB} (s) = \cos( 2\pi / L) - \sin ( 2 \pi /L ) \tan ( a s + b ) 
\end{equation}
with
\begin{equation}
  a = - \arctan \left( \frac{\sin 2 \pi/ L}{1 - \cos 2 \pi / L } \right)
\end{equation}
and
\begin{equation}
  b = \arctan \left( \frac{\cos 2 \pi/ L}{ \sin 2 \pi / L } \right).
\end{equation}
Figure~\ref{fig:qab} shows $\chi_{\rm QAB}(s)$ for several values of $L$.
The result indicates that the curvature of $\chi_{\rm QAB}(s)$ becomes flatter as one approaches the final time at $s=1$ 
because the minimum gap becomes smaller.

\bibliography{ffqc}

\begin{thebibliography}{87}
\bibitem [{pbc()}]{pbc}%
  \BibitemOpen
  \href@noop {} {}\bibinfo {note} {For $L=4n_L+2$ ($n_L$: integer) with PBC,
  which satisfies the closed shell condition, we also obtain essentially the
  same results except that now $\Delta E=0$ exactly at
  $M=(L-2)/4$.}\BibitemShut {Stop}%
\bibitem [{not()}]{note1}%
  \BibitemOpen
  \href@noop {} {}\bibinfo {note} {Note that if the initial state was composed
  simply of a product state of single sites, $\ell_{\rm i}$ ($\ell_{\rm ii}$)
  would be $4M+2$ ($4M$). Additional two sites arise because the initial state
  $ \vert \psi_{\rm i} \rangle$ considered here is a product state of local
  bonding states given in Eq.~(\ref{eq:1d:init}) and thus sites $x-2M-1$ and
  $x+2M+2$ for case (i) [sites $x-2M$ and $x+2M+1$ for case (ii)] also
  contribute to the local expectation value $\langle \psi_M
  (\mbox{\boldmath{$\theta$}}) \vert \hat{c}_{x}^{\dagger} \hat{c}_{x+1} \vert
  \psi_M (\mbox{\boldmath{$\theta$}}) \rangle$, leading to $\ell_{\rm i}=4M+4$
  ($\ell_{\rm ii}=4M+2$).}\BibitemShut {Stop}%
\end{thebibliography}

\end{document}